\documentclass[conference]{IEEEtran}
\newcommand{\paragraphb}[1]{\noindent{\bf #1} }

\usepackage{bm}
\usepackage{subcaption}
\usepackage{multirow}
\usepackage{amsmath}
\usepackage{color, xcolor}
\usepackage{booktabs}
\usepackage{makecell}
\usepackage{soul}
\usepackage{comment}
\usepackage{graphicx} 
\usepackage{adjustbox}
\usepackage{array}
\usepackage{microtype}
\usepackage{url}
\usepackage{booktabs}
\usepackage{amssymb}

\usepackage{url}
\usepackage{graphicx}
\usepackage{xspace}
\usepackage[hidelinks]{hyperref}

\usepackage{xspace}
\newcommand{\name}{\textsc{Diffence}\xspace}

\ifCLASSINFOpdf
\else
\fi
\begin{document}
\title{\name: Fencing Membership Privacy\\ With Diffusion Models}

\author{\IEEEauthorblockN{Yuefeng Peng}
	\IEEEauthorblockA{University of Massachusetts Amherst\\
		yuefengpeng@cs.umass.edu}
	\and
	\IEEEauthorblockN{Ali Naseh}
	\IEEEauthorblockA{University of Massachusetts Amherst\\
		anaseh@cs.umass.edu}
	\and
	\IEEEauthorblockN{Amir Houmansadr}
	\IEEEauthorblockA{University of Massachusetts Amherst\\
		amir@cs.umass.edu}}

\maketitle

\begin{abstract}
Deep learning models, while achieving remarkable performances across various tasks, are vulnerable to membership inference attacks (MIAs), wherein adversaries identify if a specific data point was part of the model's training set. This susceptibility raises substantial privacy concerns, especially when models are trained on sensitive datasets. 
% Current defenses against MIAs often struggle to provide robust protection without hurting model utility, and they often require retraining the model.
Although various defenses have been proposed, there is still substantial room for improvement in the privacy-utility trade-off.
In this work, we introduce a novel defense framework against MIAs by leveraging generative models. 
The key intuition of our defense is to \emph{remove the differences between member and non-member inputs}, which is exploited by MIAs, by re-generating input samples before feeding them to the target model. 
Therefore, our defense, called \name, works \emph{pre inference}, which is unlike prior defenses that are either training-time (modify the model) or post-inference time (modify the model's output). 

A unique feature of \name is that it works on input samples only, without modifying the training or inference phase of the target model. Therefore, it can be \emph{cascaded with other defense mechanisms} as we demonstrate through experiments. \name is specifically designed to preserve the model's prediction labels for each sample, thereby not affecting accuracy. Furthermore, we have empirically demonstrated that it does not reduce the usefulness of the confidence vectors.
Through extensive experimentation, we show that \name can serve as a robust plug-n-play defense mechanism, enhancing membership privacy without compromising model utility\textemdash both in terms of accuracy and the usefulness of confidence vectors\textemdash across standard and defended settings. 
For instance, \name reduces MIA attack accuracy against an undefended model by 15.8\% and attack AUC by 14.0\% on average across three datasets, all without impacting model utility. 
By integrating \name with prior defenses, we can achieve new state-of-the-art performances in the privacy-utility trade-off.
For example, when combined with the state-of-the-art SELENA defense it reduces attack accuracy by 9.3\%,  and attack AUC by 10.0\%.
\name achieves this by imposing a negligible computation overhead, adding only 57ms to the inference time per sample processed on average.

\end{abstract}


% \fi
\IEEEpeerreviewmaketitle

\section{Introduction}
% {\let\thefootnote\relax\footnotetext{This work was supported in part by the NSF grant 2131910. }}
\label{sec:intro}

\begin{figure}[ht]
\centering
\includegraphics[scale=0.2]{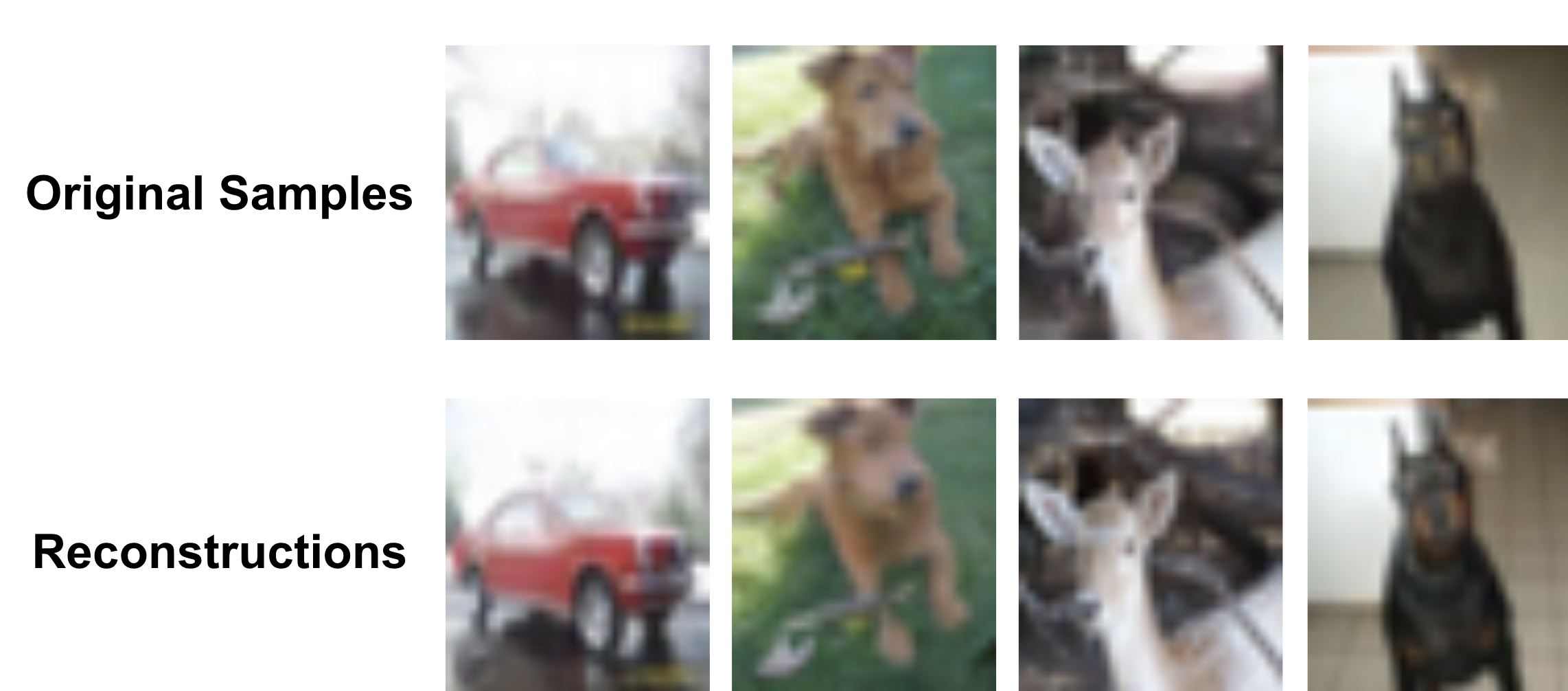}
\caption{\textbf{Examples of original samples and their reconstructions on CIFAR-10.} Original samples are successfully identified as members by the adversary, while the reconstructed samples are classified as non-members.}
\label{Fig:examples}
\end{figure}

\begin{figure*}[ht]
\centering
\includegraphics[scale=0.45]{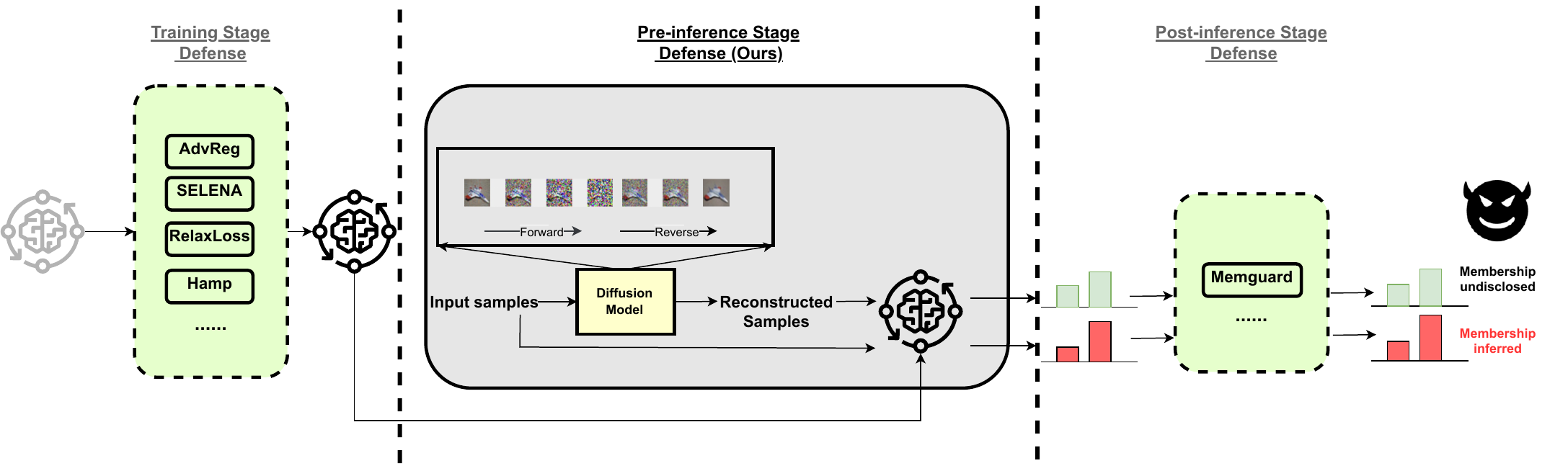}
\caption{\textbf{An illustration of different MIA defense stages and our proposed \name.} Defenses are categorized according to their implementation stages in the machine learning pipeline: the training phase, pre-inference phase, and post-inference phase. Different defense strategies can be deployed at various stages for an integrated defense approach. \name uniquely positioned in the pre-inference phase, is compatible with all other methods.}
\label{Fig: illus}
\end{figure*}

\begin{table}
\large
    \caption{\textbf{A comparison to prior works.} \checkmark means the information is required by the adversary, - otherwise.} 
    \centering
    \scalebox{0.55}{
    \begin{tabular}{m{4cm}<{\centering}m{2.2cm}<{\centering}m{2cm}<{\centering}m{2cm}<{\centering}m{3cm}<{\centering}}
    \toprule
     \textbf{Technique}& \textbf{Requires Re-training}& \textbf{Requires Additional Data}& \textbf{Impact on Model Accuracy} & \textbf{Deployment Stage} \\
    \midrule
    AdvReg~\cite{nasr2018machine}        & \checkmark  & \checkmark   & High    & Training \\
    MemGuard~\cite{jia2019memguard}      & -           & \checkmark   & None    & Post-Inference \\
    DPSGD~\cite{abadi2016deep}           & \checkmark  & -            & High    & Training \\
    SELENA~\cite{tang2022mitigating}     & \checkmark  & -            & Low     & Training \\
    RelaxLoss~\cite{chen2022relaxloss}   & \checkmark  & -            & None    & Training \\
    HAMP~\cite{chen2023overconfidence}   & \checkmark  & -            & Low     & Training \\
     \name (Scenario 1) & -& \checkmark& None&Pre-inference\\
     \name (Scenario 2) & -& -& None&Pre-inference\\
     \name (Scenario 3) & -& -& None&Pre-inference\\
    \bottomrule
    \end{tabular}}
    \label{table:compare}
\end{table}

Deep learning has made remarkable achievements in recent years. However, it has been demonstrated that deep learning models can memorize information from their training~\cite{Carlini2019secret,Song2017Machine,Leino2020Stolen}, making them susceptible to \emph{membership inference attacks (MIAs)}. Such attacks seek to infer if a specific data point was part of a model's training set~\cite{shokri2017membership}. Given that deep learning models are often trained on sensitive data, such as facial images~\cite{gurovich2019identifying} and medical records~\cite{erickson2017machine}, the potential success of MIAs poses a significant threat to privacy~\cite{shokri2017membership}.

The growing threat of MIAs has motivated the development of different defense strategies.
Existing defense mechanisms can be categorized into two main classes based on their application phase in the machine learning pipeline. 
The first category includes defenses that operate \emph{during the training phase}. 
These methods employ privacy-preserving techniques to train models in a way that reduces their memorization of training data, thereby mitigating privacy risks~\cite{abadi2016deep,nasr2018machine,tang2022mitigating,chen2022relaxloss,chen2023overconfidence}. Key approaches within this category include Differential Privacy Stochastic Gradient Descent (DPSGD)~\cite{abadi2016deep}, adversarial regularization (AdvReg)~\cite{nasr2018machine}, and knowledge distillation~\cite{tang2022mitigating}. 
The second category targets the \emph{post-inference} phase, focusing on reducing membership privacy leakage by directly rectifying the disparities in the model’s outputs between members and non-members. A representative defense in this category is MemGuard~\cite{jia2019memguard}, which modifies output confidence vectors to fool the adversary's attack model. 
% Unfortunately, existing defenses suffer from several limitations, as exemplified in Table~\ref{table:compare}. In particular, they often struggle to achieve a satisfactory privacy-utility trade-off~\cite{chen2022relaxloss,tang2022mitigating}. Additionally, some defenses often rely on extra data to support their methods. 
% \cite{nasr2018machine,jia2019memguard} 
% \hl{Despite many defenses being proposed, they either provide limited privacy protection or compromise the model's utility. There is still significant room for improvement in the privacy-utility trade-off.} 
% \hl{However, there still remains a gap between their effectiveness and the ideal defense, which should provide sufficient protection without hurting model utility. The privacy-utility trade-off still requires further improvement.}

% \ancomment{How much extra data does our approach need? (Scenario 1)- 1000 non-members}. 

\paragraphb{A new class of MIA defenses:}
In this paper, we introduce a third category of defense mechanisms that operate 
in the \emph{pre-inference} stage. 
A pre-inference defense does \emph{not} alter model outputs or the model itself, as with the two main categories introduced above.
Instead, a pre-inference defense mechanism \emph{modifies input samples} before they are sent to the target model for inference.
To our best knowledge, no prior work has developed a pre-inference defense against MIAs. 
Figure~\ref{Fig: illus} compares the three categories of MIA defense techniques, and 
Table \ref{table:compare} shows example mechanisms across these categories. 

% \paragraphb{Our contributions.} 
We introduce \name, a novel pre-inference   MIA defense that enhances privacy without sacrificing model accuracy.
\name uses diffusion models. Advanced generative models such as GANs~\cite{goodfellow2014generative} and diffusion models~\cite{dhariwal2021diffusion} have been previously employed during inference to safeguard deep learning models from adversarial attacks~\cite{Nie2022Pure}.  % This defense strategy is termed "adversarial purification." Its core principle is leveraging the generative model to recreate a sanitized version of every malicious input, thereby shielding the model from potential breaches. 
However, the application of generative models during inference to defend against privacy attacks, specifically MIA, remains largely unexplored. In this paper, we show that diffusion models can also serve as powerful tools integrated into the defense framework against MIAs, preserving membership privacy without harming model utility.

% Prior studies on MIAs have highlighted a distinct model behavior concerning training and test data, pinpointing it as a key contributor to membership privacy leakage. For instance, models often output higher confidence when predicting members. We attribute such distinct behavior to the fact that models tend to over-memorize the intricate details of the training data samples (members) during the training phase. During inference, when classifying members, models rely not just on shared category semantics but also on irrelevant nuances or noise memorized during training. In contrast, while classifying non-members, these memorized noises are absent. Inspired by adversarial purification~\cite{Nie2022Pure, wang2022guided}, we propose to deploy generative models during the classifier's inference phase, purging input samples of noises that induce overconfidence while preserving label semantics. 
% % Since non-members are excluded from the model's training, they remain unaffected post-reconstruction. By reconstructing details in members we can mitigate the model's over-reliance on them, making them more likely to be perceived as non-members.
% Consequently, whether classifying the reconstructions of both members and non-members, the model encounters novel samples it has not seen before. This effectively eliminates the differences in classification behavior between members and non-members.

\paragraphb{Intuitions behind \name:} The target model's memorization of members results in distinct behaviors when classifying seen (member) versus unseen (non-member) samples. MIAs exploit these behavioral discrepancies to distinguish between members and non-members, though different attack methods may leverage different features. In this work, we focus on black-box attacks, which typically infer membership by utilizing information from the sample outputs. He et al.~\cite{he2022membershipdoctor} categorized the information exploited in these attacks into two parts: prediction posteriors and labels. We contend that the fundamental cause of MIAs is the non-negligible gap between the model's outputs for members and non-members. Building on this, we divide the features exploitable by attacks into two categories: the train-to-test accuracy gap and the prediction distribution gap. These features, as utilized in existing attacks~\cite{shokri2017membership,salem2018ml,yeom2018privacy,nasr2018machine,song2021systematic}, are summarized in Table \ref{table:feature}. Note that some attacks which rely solely on labels~\cite{choquette2021label,li2021membership} may also indirectly exploit the prediction distribution gap through the robustness of the predicted labels. 

\begin{table}
\normalsize
    \caption{\textbf{Summary of the features and information utilized by various MIAs.} \checkmark means the specific gap is exploited by the attack, - otherwise. Regarding features, P denotes the use of prediction posteriors and L denotes the use of label.}
    \centering
    \scalebox{0.7}{
    \begin{tabular}{m{5cm}<{\centering}m{2cm}<{\centering}m{2cm}<{\centering}m{2cm}}
    \toprule
     \textbf{Attacks}&  \textbf{Train-to-test Accuracy Gap}&  \textbf{Prediction Distribution Gap} & \textbf{Features} \\
    \midrule
     NN-based attacks~\cite{salem2018ml,shokri2017membership,nasr2018machine} & \checkmark  & \checkmark  & P,L\\
     Metric-corr~\cite{yeom2018privacy}      & \checkmark  & -  &L\\
     Metric-loss~\cite{yeom2018privacy}      & \checkmark  & \checkmark &P,L \\
     Metric-conf~\cite{salem2018ml}      & -  & \checkmark &P \\
     Metric-ent~\cite{song2021systematic}      & -  & \checkmark  &P\\
     Metric-ment~\cite{salem2018ml}      & \checkmark  & \checkmark &P,L \\
     Label-only attacks~\cite{choquette2021label,li2021membership}  & \checkmark  & \checkmark &P,L \\
    \bottomrule
    \end{tabular}}
    \label{table:feature}
\end{table}

A successful MIA defense should fundamentally eliminate/shrink the two gaps described above between members and nonmembers while preserving the model's utility. \name primarily focuses on minimizing the prediction distribution gap without altering the samples' prediction labels, thus maintaining model accuracy. This is achieved by reconstructing the samples using a generative model before they are input into the target model. Through sample reconstruction, the model encounters samples in the inference stage that are not exact replicas of those observed during training, regardless of whether they are member or non-member samples, effectively reducing the discrepancy in the prediction distributions. Meanwhile, our reconstruction process retains the semantic information of the original samples, differing only in details, which allows the model's output confidence scores to accurately reflect the characteristics of the original sample. \name includes generating multiple reconstructions for each sample and implementing various strategic selection techniques based on different assumptions about the defender. These techniques aim to align the prediction distributions of members and non-members as closely as possible while maintaining the prediction label, further details are provided in Section \ref{sec:sample_s}.

\paragraphb{\name's implementation:} 
While our defense can be implemented using arbitrary generative models, we  
use diffusion models~\cite{ho2020denoising}  due to their state-of-the-art performances in generative tasks~\cite{dhariwal2021diffusion} and compatibility with our defense strategy; we therefore call our mechanism \name, i.e., a diffusion-based fence against MIAs. A diffusion model can be used to modify an original sample through forward diffusion and reverse denoising steps, creating a new sample that maintains the same semantic attributes while altering non-essential details, aligning with our defense objectives (details are discussed in Section \ref{sec:method_diff}). Figure \ref{Fig:examples} provides some example inputs and their selected reconstructions. The diffusion model employed for defense does not need to be exclusively designed for this purpose; hence, defenders can utilize off-the-shelf diffusion models. Additionally, our results demonstrate that defenders can employ diffusion models pre-trained on different datasets to effectively defend and achieve comparable performance. For instance, a diffusion model trained on ImageNet can successfully provide defense for CIFAR10.
% Additionally, the exceptional generative quality and diversity inherent in diffusion models make them particularly apt for our defense purpose. 

% For instance, during its training phase, a model could be trained using a privacy-preserving technique, such as adversarial regularization, and then, during inference, seamlessly transition to leveraging our defense mechanism.
We evaluated \name on three benchmark datasets (CIFAR-10 {\cite{krizhevsky2009learning}}, CIFAR-100 {\cite{krizhevsky2009learning}}, and SVHN {\cite{netzer2011reading}}) and two high-resolution image datasets (CelebA {\cite{liu2015deep}} and UTKFace {\cite{zhang2017age}}) using four model architectures: {ResNet18~\cite{he2016deep}}, {DenseNet121~\cite{huang2017densely}}, VGG16 {\cite{simonyan2014very}}, and Vision Transformer (ViT) {\cite{dosovitskiy2020image}}. Our experiments comprehensively cover six state-of-the-art MIAs and six state-of-the-art defenses.
Note that \textbf{\name can be viewed as a membership privacy booster that can be cascaded with other (i.e., training or post-inference) defense mechanisms}, due to its minimal deployment constraints and its plug-and-play flexibility. 
This also allows us to combine the strengths of other methods.
For instance, \name can be combined with defenses that address label-only attacks, enhancing their effectiveness against other stronger attacks. While \name does not directly counter label-only threats, this combination creates a comprehensive defense strategy effective against all types of attacks.
% For instance, although \name boosts existing defenses on stronger attacks, it does not directly counter label-only attacks. However, it can be combined with methods that address such attacks, thereby forming a comprehensive defense strategy effective against all types of attacks.
Our empirical results validate \name's effectiveness in enhancing membership privacy without decreasing the model utility, irrespective of the model's operational context—be it a baseline (vanilla) or a defended setting. 
For example, we decreased attack accuracy for SELENA, Relaxloss and HAMP by 9.3\%, 12.2\%, and 14.4\%, and attack AUC by 10.0\%, 11.4\%, and 14.4\%, respectively, on average across three benchmark datasets. \name also reduced the true positive rate (TPR) under 0.1\% false positive rate (FPR) by 52.8\% on average across six tested defenses on the SVHN dataset. All of this was achieved without any loss in model utility, with an average additional inference overhead of only 57ms. 

In this context of utility, unlike previous studies on MIAs that focused solely on model accuracy, we recognize model utility encompasses not just accuracy, but also the meaningfulness of the output confidence scores or how well the confidences are calirbated \cite{naeini2015obtaining,guo2017calibration}. Through our evaluation, we have also verified that \name effectively maintains the meaningfulness of confidence scores, which indicates that it indeed preserves the overall utility.

% \amir{give more numbers on results; also numbers on added overhead. you did not talk about overhead. YUEFENG: I have included more experimental results, including overhead}

\paragraphb{Summary of contributions:} Our contributions are as follows:
\begin{enumerate}
    \item We propose a novel diffusion model-based membership inference defense framework called \name, which can boost the membership privacy of pre-existing (both undefended and defended) models without compromising the model utility.
    \item We propose a new defense pipeline, which for the first time combines defenses deployed at different stages to achieve better defense performance.
    \item We implemented the prototype of the \name. Our extensive experiments show that \name can effectively improve the membership privacy of existing models without utility loss. Furthermore, we show that in certain settings \name can enhance both the model accuracy and privacy of the model.
\end{enumerate}

% \begin{figure*}
%   \centering
%   \begin{subfigure}{0.68\linewidth}
%     \fbox{\rule{0pt}{2in} \rule{.9\linewidth}{0pt}}
%     \caption{An example of a subfigure.}
%     \label{fig:short-a}
%   \end{subfigure}
%   \hfill
%   \begin{subfigure}{0.28\linewidth}
%     \fbox{\rule{0pt}{2in} \rule{.9\linewidth}{0pt}}
%     \caption{Another example of a subfigure.}
%     \label{fig:short-b}
%   \end{subfigure}
%   \caption{Example of a short caption, which should be centered.}
%   \label{fig:short}
% \end{figure*}

\section{Background and Preliminaries}
\label{sec:preliminaries}

\subsection{Diffusion Models}

Generative image modeling has seen remarkable advancements in recent years, with Generative Adversarial Networks (GANs)~\cite{goodfellow2014generative} and Variational Autoencoders (VAEs)~\cite{kingma2013auto} standing out as the pioneering architectures for synthesizing realistic images~\cite{brock2018large, karras2020analyzing, kingma2018glow, vahdat2020nvae}. While these frameworks have laid the groundwork and achieved significant successes, diffusion models~\cite{sohl2015deep} have recently emerged, surpassing their predecessors in terms of performance and establishing themselves as a leading approach in the domain of image synthesis~\cite{dhariwal2021diffusion}.

\emph{Denoising Diffusion Probabilistic Models (DDPMs)}~\cite{ho2020denoising} have emerged as a significant advancement in generative image modeling and operate by reversing a diffusion process. This diffusion mechanism can be expressed as 
\begin{equation}
    x_t = \sqrt{1-\beta_t} x_{t-1} + \sqrt{\beta_t} \epsilon_t
    \label{eq1}
\end{equation}
\noindent where \( \epsilon_t \) represents Gaussian noise, and \( \beta_t \) dictates the magnitude of noise introduced at each iteration. After a set number of timesteps, typically denoted as \( T \), the data \( x_T \) becomes predominantly noise-infused.

Reconstructing meaningful samples involves reversing this process. A denoising function is trained by DDPM, which, when given a noised sample \( x_t \), predicts the less-noisy version from the preceding timestep \( x_{t-1} \). The optimization goal is to minimize the difference between the true and the predicted data, typically through a mean squared error loss. During the sampling phase, DDPM initiates with a noise distribution sample and employs the trained denoising function to reverse the diffusion, yielding samples that closely mirror the original data distribution.

% \ancomment{Shouldn't you explain shortly why you listed three different sets of assumptions? Is it because you are using different attacks and each has different assumption? These are not different assumptions, but rather three aspects of our overall threat model. I adopted this format from the SELENA paper }
\subsection{Threat Model}
\subsubsection{Attacker} 
\noindent{\textbf{Black-box Access:}} Following prior works~\cite{jia2019memguard,chen2023overconfidence}, we assume the attacker has black-box access to the target model. This means the attacker can query the target model using a black-box API and receive corresponding prediction vectors, while direct access to any intermediate results, including the reconstructions from the diffusion model and intermediate outputs from the target model remains restricted. Therefore, we assume there is no privacy leakage through the diffusion model or the intermediate results of the target model.

\noindent{\textbf{Partial Knowledge of Membership for Members:}} Like previous defenses~\cite{tang2022mitigating,chen2023overconfidence}, we assume a strong attacker that knows a limited number of samples from both the training and test sets, i.e., they know the membership status of some members and some non-members. Therefore, the attacker can directly train their attack classifier using the known members and non-members of the target model without the need to train any shadow models. The goal of the attack is to infer the membership status of any other unexposed sample.

\noindent{\textbf{Full Knowledge of Defense Technique:}} We assume the attacker is fully aware of the deployment of the defense technique and the architecture of the target model. In the context of \name, we assume the attacker has full knowledge of how the diffusion model is deployed. All attacks against \name in our experiments can be considered adaptive attacks, as we assume the attacker can process samples in the same manner as \name when training their attack classifier. This means the attacker can use reconstructions from the same diffusion model to train the attack classifier instead of using the original samples.

\subsubsection{Defender}
We assume the defender possesses a private dataset and uses it to train a target model. The defender's objective is to securely release the target model for user accessibility. The defender aims to strike a balance between achieving minimal membership privacy leakage and maintaining high classification accuracy.

\section{Methodology}
\label{sec:methodology}

% The model's memorization of training data can lead to privacy leakage \cite{Carlini2019secret,Leino2020Stolen}. During the inference stage, this memorization results in distinctive model behaviors for training data compared to unseen data, such as higher prediction confidence for members. While the techniques behind membership inference attacks may vary, they fundamentally exploit these discrepancies in model behavior to infer samples' membership. Our proposed methodology aims to mitigate such prediction inconsistencies stemming from the model's recall of training samples, addressing the root cause of membership privacy vulnerabilities through the model's prediction interface. We commence by discussing how a model's memorization of training samples can compromise membership privacy and subsequently describe the utility of diffusion models in countering this threat.

Our main idea is to eliminate the differences in predictive behavior between members and non-members, thereby fundamentally mitigating membership privacy leakage. As described in Section \ref{sec:intro}, we categorize these disparities into the train-to-test accuracy gap and the prediction distribution gap. \name primarily targets the elimination of the prediction distribution gap. 
% However, by combining our approach with other defenses, we can simultaneously address both gaps, achieving state-of-the-art defense performance. 
However, \name can seamlessly integrate with other defenses that reduce the train-to-test accuracy gap, such as SELENA {\cite{tang2022mitigating}}, enabling a comprehensive defense strategy that simultaneously addresses both gaps without sacrificing the model's utility, achieving state-of-the-art defense performance.
In this section, we first explain how differences in predictive behavior can lead to privacy leakage. We then introduce how we design \name to narrow these gaps.

\subsection{Intuition Behind \name}

Prior studies on MIAs have highlighted a distinct model behavior concerning training and test data, pinpointing it as a key contributor to membership privacy leakage~\cite{carlini2022membership}. The gap in prediction distributions between members and non-members can be reflected in various features. These may include assigning higher confidence levels, lower loss, and reduced prediction entropy to members. The disparities in these features are key factors leading to membership privacy leakage, with almost all effective MIAs exploiting one or several of these feature discrepancies \cite{yeom2018privacy,shokri2017membership,song2021systematic,carlini2022membership}. Moreover, these features are interrelated; for instance, lower loss often implies reduced prediction entropy and higher confidence levels.

Carlini et al.~\cite{carlini2022membership} proposed parametric modeling of prediction confidence to achieve a more Gaussian-like distribution. Such a distribution more effectively distinguishes the prediction distribution gap between members and non-members. In our subsequent discussions, we adopt this parametric modeling approach to represent the prediction gap. The only difference is that we use the maximum confidence from the output vector rather than the confidence of the correct class, as our focus is on the prediction distribution gap rather than the train-to-test accuracy gap. The parametric modeling function is shown in Equation \ref{eq:pm}.

\begin{equation}
    \phi(p) = \log\left(\frac{p}{1-p}\right), \text{ for } p = \max(f(x))
\label{eq:pm}
\end{equation}
where \(f(x)\) is the model's output vector for input \(x\).
\begin{figure*}[htbp]
  \centering
  \includegraphics[width=0.6\columnwidth]{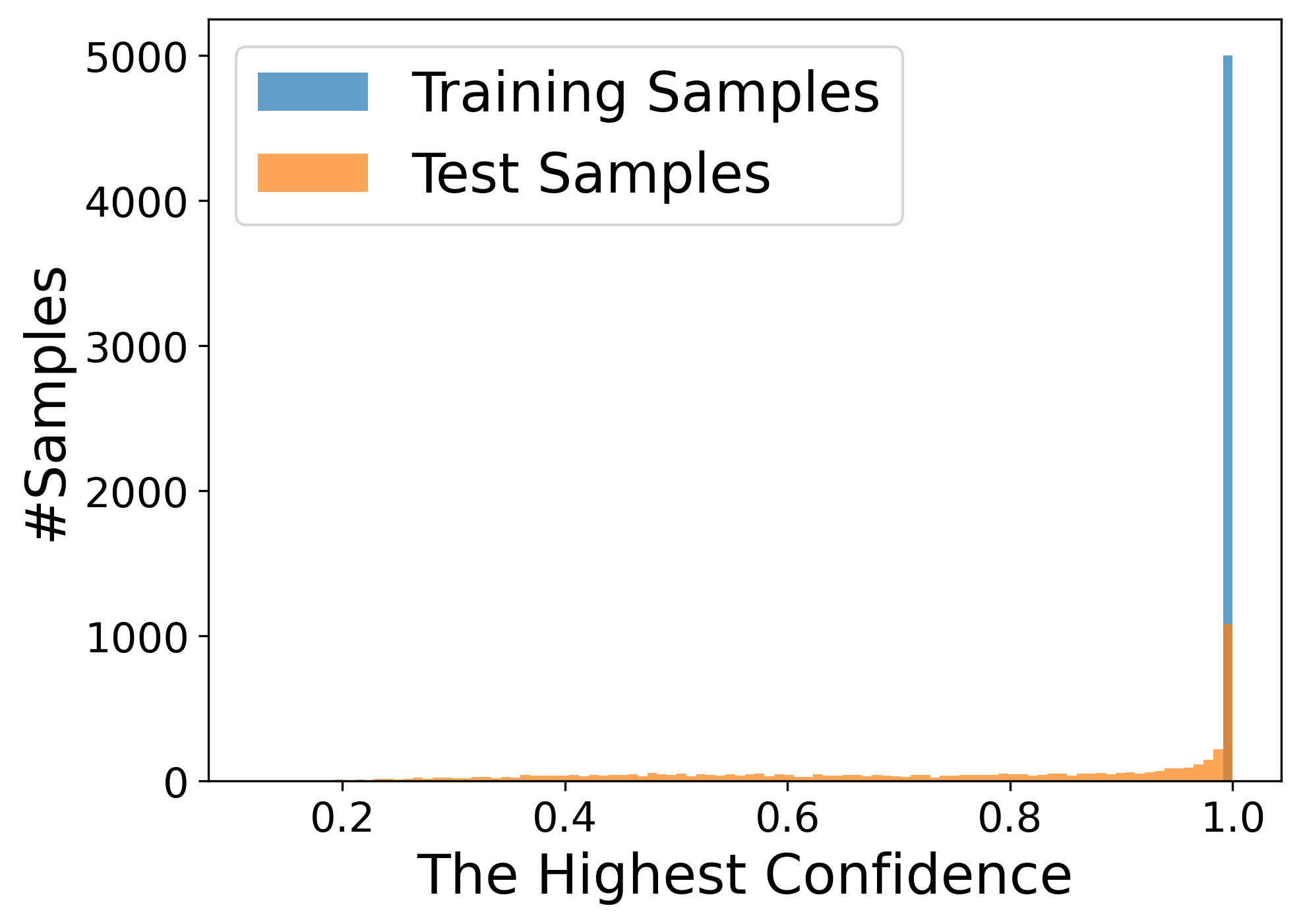}
  \hfill
  \includegraphics[width=0.6\columnwidth]{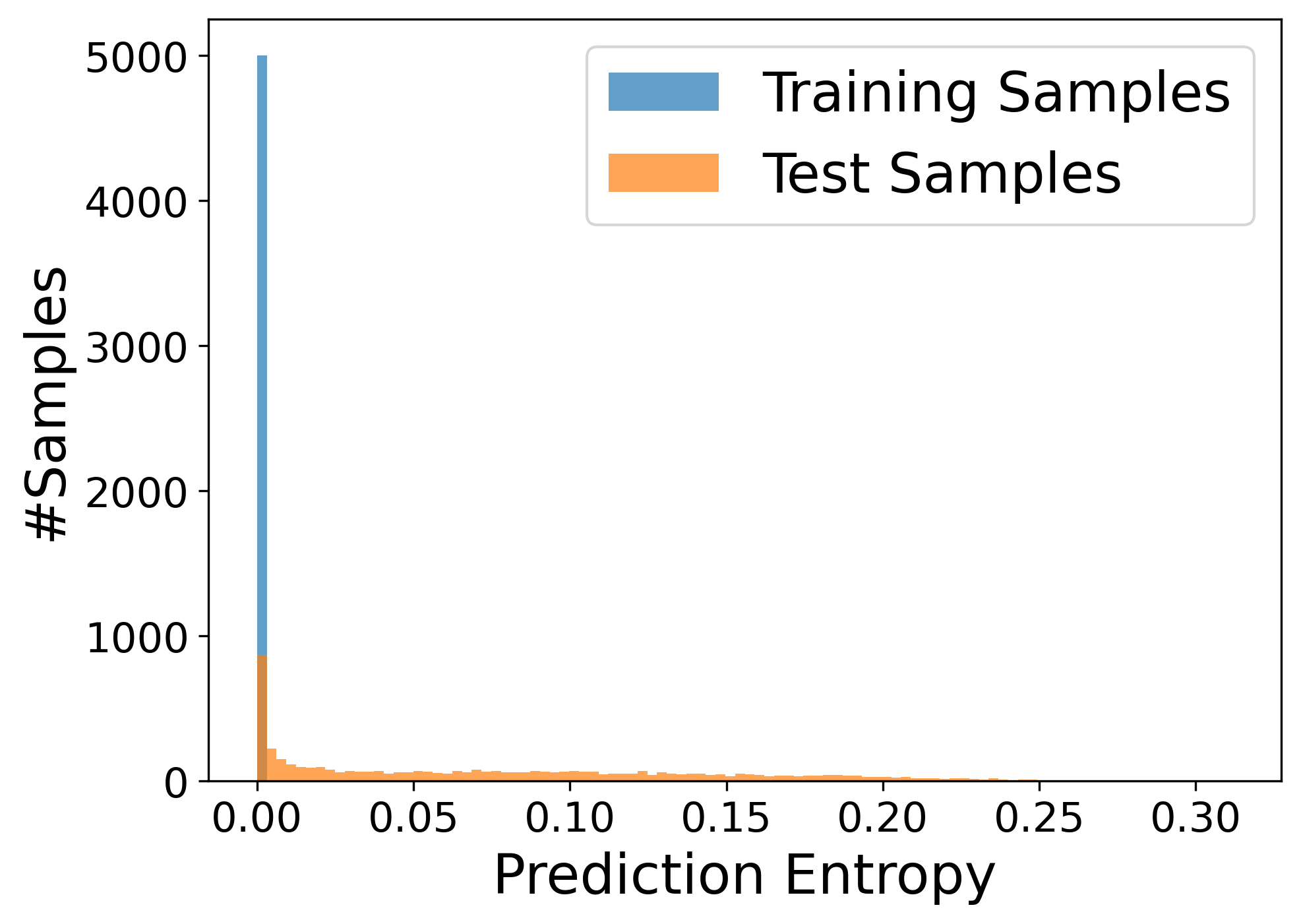}
  \hfill
  \includegraphics[width=0.59\columnwidth]{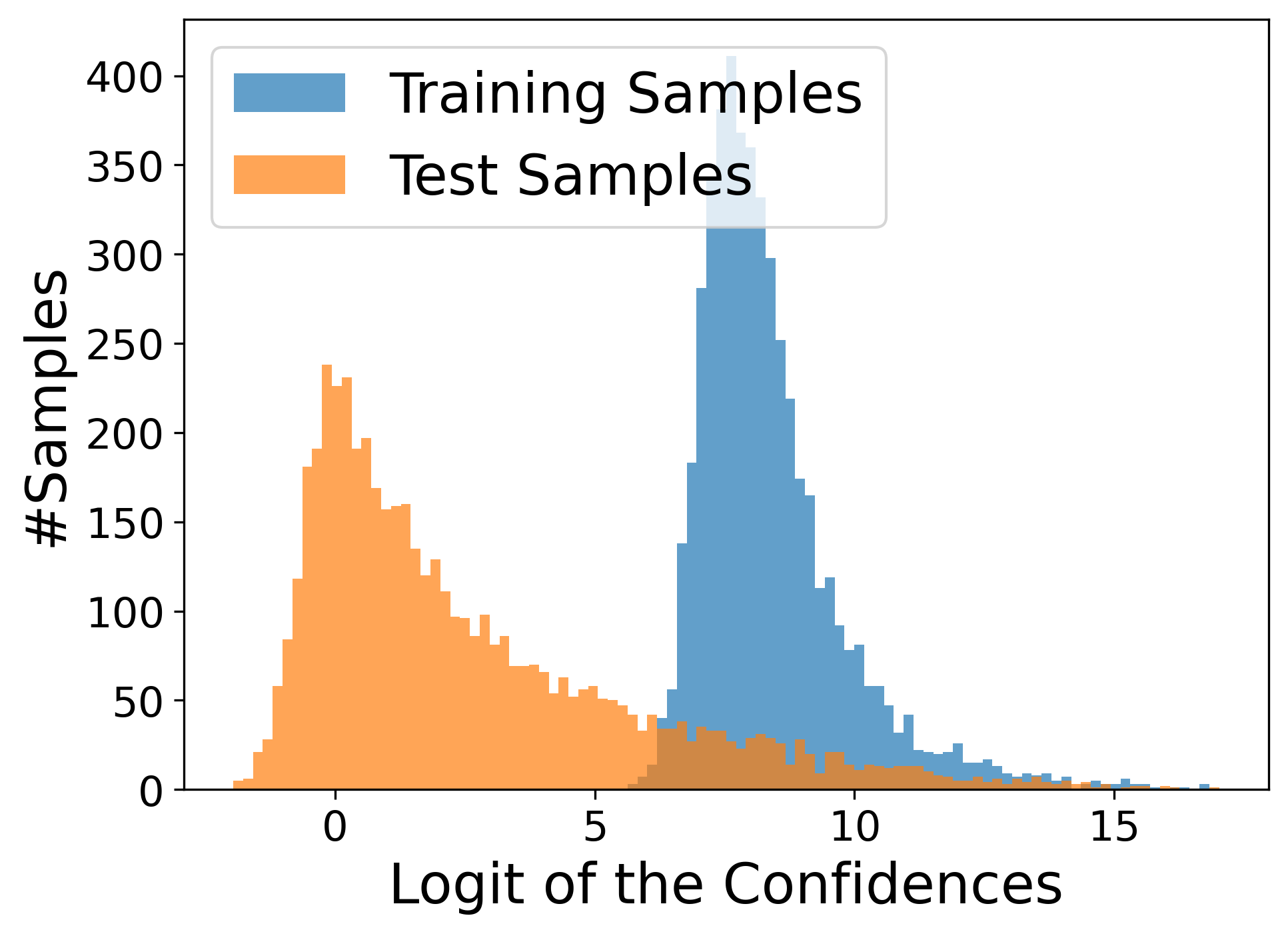}
  \caption{\textbf{Prediction distribution gap between members and non-members:} The figure illustrates the disparities in prediction distributions with respect to confidence levels, prediction entropy, and parametrically modeled confidence (predicted logits).}
  \label{fig:dist_gap}
\end{figure*}
Figure \ref{fig:dist_gap} plots histograms illustrating the differences in predictions for members and non-members on CIFAR-100 without defenses. It can be observed that there are significant disparities in the output distributions between members and non-members, especially in the parametrically modeled confidence, which we refer to as \textit{logit} in this paper. Attackers can easily exploit these differences to distinguish between members and non-members.

To address these issues, we propose a novel method that leverages input reconstruction to eliminate the prediction disparity between members and non-members. Our core idea is that by reconstructing samples, the model, during classification, encounters samples that are distinct from those it has seen during the training phase, irrespective of whether they are from members or non-members, thereby reducing inconsistencies in predictions. Our proposed method, \name, aims to reconstruct the finer details of samples without altering their semantic content. To this end, we employ diffusion models as our generative model, as they align well with our objectives.

Figure \ref{fig:scenarios} provides examples of how \name\ narrows the prediction gap. By reconstructing the input, \name\ ensures that the model's predictions for both members and non-members become more consistent, thereby mitigating the risk of membership inference attacks. This process reduces the gaps in confidence levels, prediction entropy, and the parametrically modeled confidence.

\subsection{Our Proposed \name}
\label{sec:method_diff}
\name involves generating multiple reconstructed images and selecting the best one that suits our purpose. Among the generated samples, only those with the same predicted label as the original sample are considered as candidates for selection. \textbf{This ensures that the model's predictive outcomes remain unchanged, thereby maintaining the model accuracy}. We designed different sample selection strategies for three scenarios based on the assumptions in the defender. These scenarios include a defender having access to both some member and non-member samples, having access only to members, and having access solely to a trained model without any knowledge of samples' memberships.  With more information, defenders can perform a more nuanced analysis of the prediction distribution. This enables a more strategic selection of reconstructed images, forcing the prediction distributions of members and non-members to align more closely, thereby providing robust protection. We provide more details below.
\label{sec:method}

% Recent work has shown that generative models, especially the recently rapidly developed diffusion model, can be used to purify unexpected noise in input samples and transform malicious samples into benign samples \cite{Nie2022Pure,wang2022guided}. We introduced similar ideas to the defense against privacy attacks, treating the details memorized by the model in the input sample as noise that can be purified.

\subsubsection{Sample Reconstruction}
\label{sample_re}
 For each input image, we apply a two-phase process to obscure and subsequently reconstruct its inherent details. First, we apply the forward diffusion procedure using the closed-formed expression in Equation \ref{eq2} provided by Ho et al.~\cite{ho2020denoising} to add Gaussian noise to the image.
\begin{equation}
    \bm{x}_t = \sqrt{\bar \alpha_t}\bm{x}_{0} + \sqrt{1-\bar \alpha_t}\bm{\epsilon}
    \label{eq2}
\end{equation}
% where $\alpha(t)=e^{-\int^t_0 \beta(s)ds}$ and $\bm{\epsilon}\sim \mathcal{N}(\mathbf{0},\bm{I}_d)$.
Where $\alpha_t=1-\beta_t$ and $\bar \alpha_t=\prod_{i=1}^t\alpha_i$.
Then, the reverse process is applied aiming to recover the original image from the noisified image.

The two-step reconstruction of the image aligns well with our defensive objectives. If we look into the frequency domain by using Fourier Transform to both input images $x_0$ and noise $\epsilon$, we get:
\begin{equation}
    \mathcal{F}(\bm{x}_t)= \sqrt{\bar \alpha_t}\mathcal{F}(\bm{x}_{0}) + \sqrt{1-\bar \alpha_t}\mathcal{F}(\bm{\epsilon})
\end{equation}

Typically, images display a high response to low-frequency content and a notably weaker response to high-frequency content. This is because of the inherent smoothness of most images. The dominant low-frequency components encapsulate essential visual information, while the high-frequency components, associated with fine details and edges, are comparatively subdued~\cite{kreis2022denoising}.

Considering a small \( t \), where \( \bar{\alpha}_t \) closely approximates 1, the perturbations in the frequency domain are minor. Note that the Fourier transform of a Gaussian sampling is itself Gaussian. Consequently, at small \( t \) values, the forward process washes out the high-frequency content without perturbing the low-frequency content much. This leads to a faster alteration of the high-frequency elements compared to the low-frequency ones in the forward process. This dynamic is also relevant in the context of the reverse process of denoising models. At lower \( t \) values, these models predominantly focus on reconstructing high-frequency content~\cite{Yang2023ddpmslim}.

We denote the diffusion step applied to each sample in \name as $T$. By selecting an appropriate value for $T$, we can preserve the essential semantic content of each image while altering its finer details. The impact of the diffusion step $T$ on the performance of \name is further explored in Section \ref{sec:hyper}.
 During the inference phase, the two-step reconstruction procedure reduces the gap between the prediction distributions of members and non-members. 
 
 % It achieves this alignment without prior knowledge of a sample's membership status and thus does not require external non-member references. We provide a more detailed analysis in Section \ref{sec:entropy}.

\subsubsection{Sample Selection} 

\begin{figure*}[htbp]
  \centering
  % \hspace*{\fill}%
\begin{subfigure}{0.3\linewidth}
    \includegraphics[width=1\columnwidth]{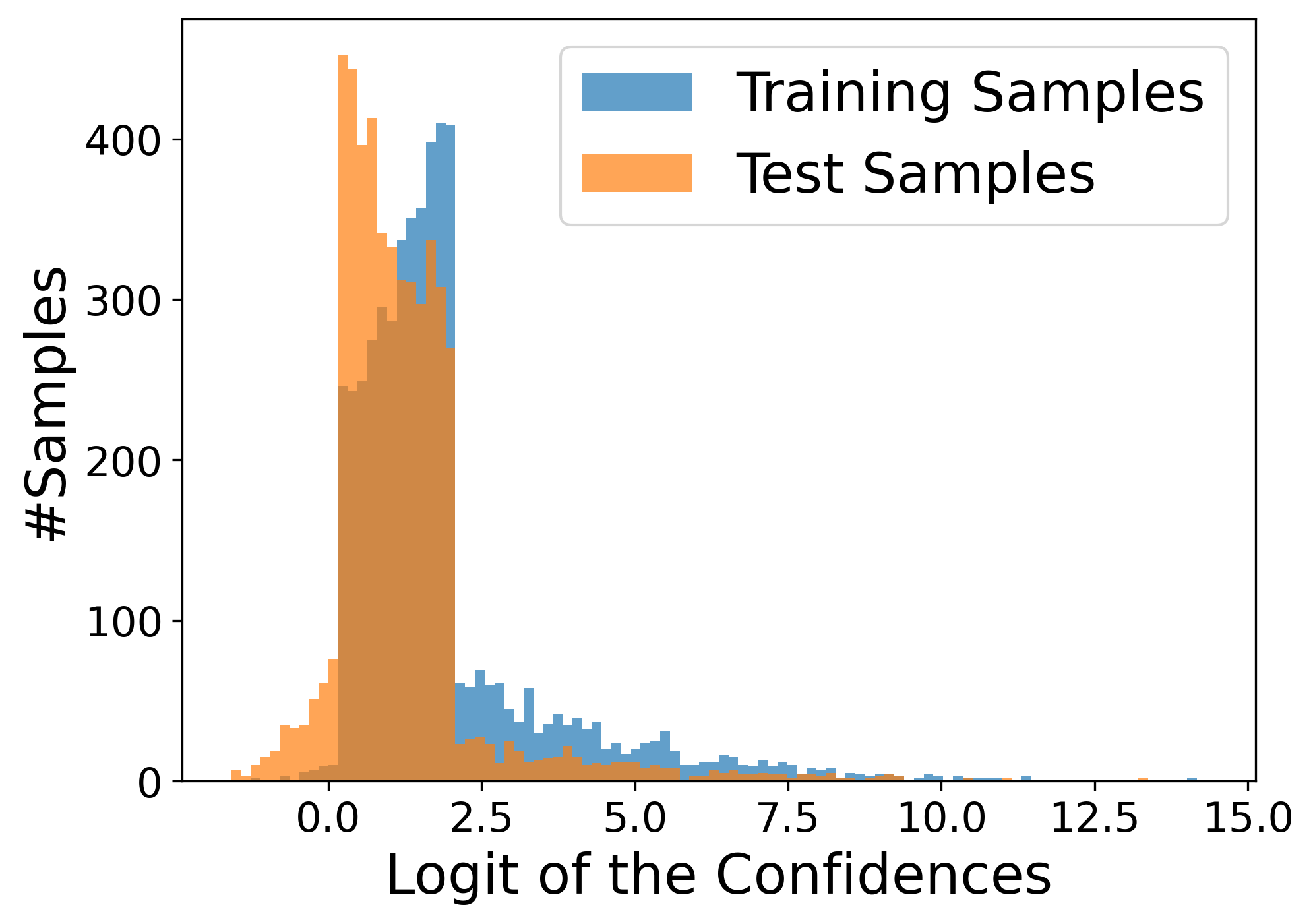}
    \label{fig:s1}
    \caption{Scenario 1}
  \end{subfigure}
     \hfill
\begin{subfigure}{0.3\linewidth}
    \includegraphics[width=1\columnwidth]{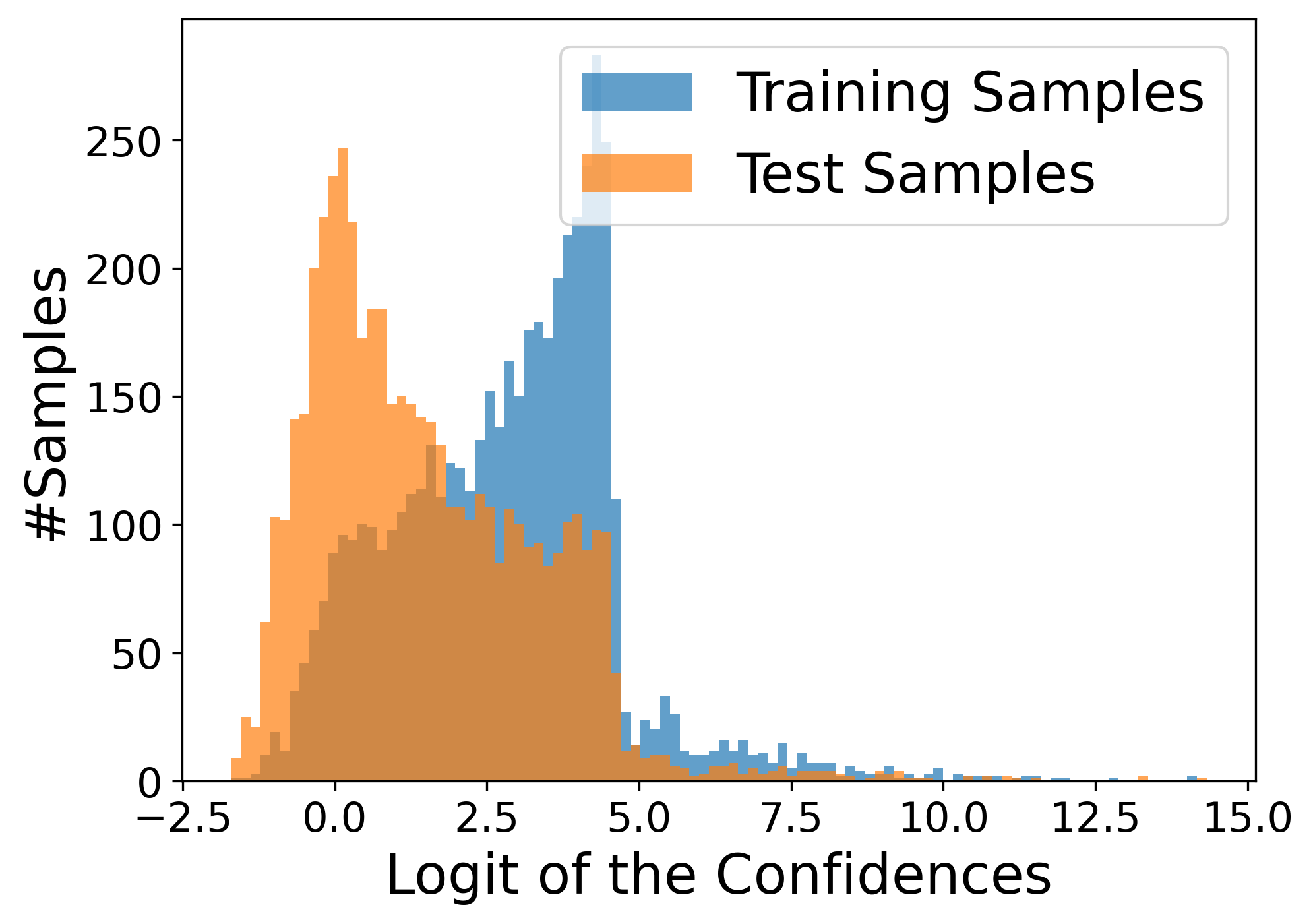}
    \label{fig:s2}
    \caption{Scenario 2 }
  \end{subfigure}
     \hfill
\begin{subfigure}{0.293\linewidth}
    \includegraphics[width=1\columnwidth]{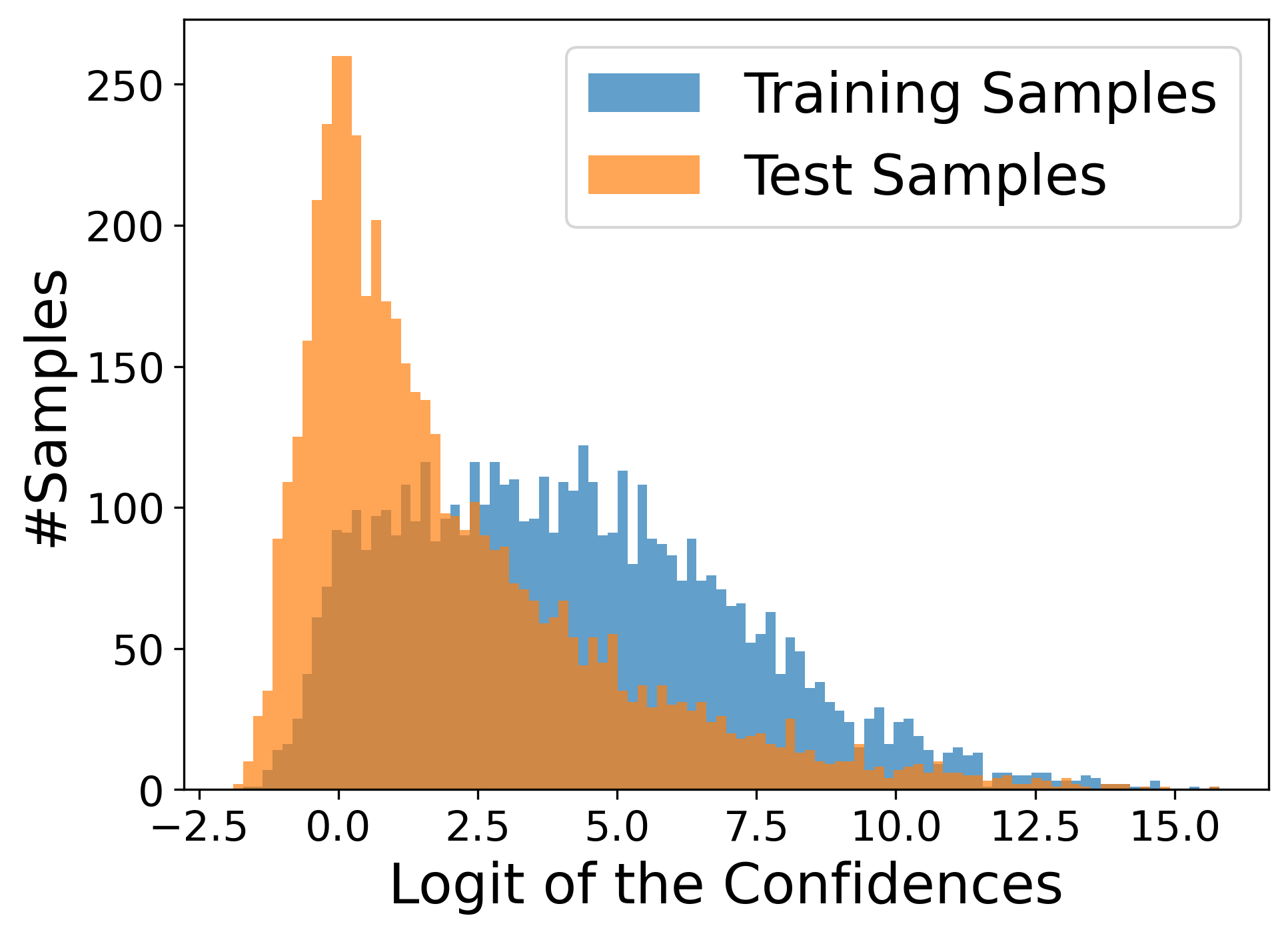}
    \label{fig:s2}
    \caption{Scenario 3 }
  \end{subfigure}
  \caption{\textbf{Prediction distribution gap between members and non-members under \name across the three scenarios.} This figure displays the distribution of predicted logits on the CIFAR-100 dataset using ResNet18 with \name in the three scenarios.}
  \label{fig:scenarios}
  % \hspace*{\fill}%
\end{figure*}

\label{sec:sample_s}
% When using the diffusion model to reconstruct the image, the diffusion step $t$ should be large enough to sufficiently mask the memorized details during forward diffuse phase and thus we can remove these memorized details during the backward reconstruction phase. However, an excessively large diffusion step 
% $t$ risks not only obfuscating the memorized noise but also undermining the global label semantics, which may hurt the model's ability to accurately classify reconstructed samples.

We observed that the stochastic nature of sample generation can lead to a decline in sample quality sometimes when only one reconstructed sample is generated for each image, potentially resulting in decreased model test accuracy. To address this issue, we propose generating multiple reconstructed images for each original image.  Then in alignment with our defense objective, we carefully select the most appropriate samples to be utilized as the final input sample.

For each original image \( I_i \) requiring privacy protection, \name generates \( N \) reconstructed versions, denoted as \( R_i = \{ R_{i1}, R_{i2}, \ldots, R_{iN} \} \). The model $f$ outputs a prediction for each reconstructed image \( R_{ij} \), i.e., \( f(R_{ij}) \). From these, we identify a subset of reconstructions whose predicted labels match the predicted label of the original image $I_i$ forming a set of candidate reconstructions, denoted as $C$. 
We then select the most appropriate reconstructed image from the candidates. 
Selecting a sample from this candidate set ensures that the original prediction label of the sample remains unchanged, guaranteeing that \name does not impact the model's accuracy.

\begin{small}
\begin{equation}
    P_i = \text{select}(\{f(R_{ij}) \mid f(R_{ij})=f(I_i), j = 1, \ldots, N\})
\end{equation}
\end{small}

Here, \( P_i \) represents the final prediction for the original image \( I_i \), derived from the model's prediction for the selected reconstruction. The selection function, denoted as \( \text{select}(\cdot) \), chooses the optimal reconstruction from \( R_i \) based on a criterion that evaluates the predictions of all \( N \) reconstructions. This criterion is aligned with the defense's objectives and may take into account the defender's prior knowledge about the prediction distributions of members and non-members. We categorize \name into three scenarios based on the information available to the defender and have accordingly designed distinct sample selection strategies for each.

\noindent \textbf{Scenario 1.} In this scenario, we assume that the defender has access to subsets of both member and non-member data. The defender initially generates \( N \) reconstructed samples for each data point and then plots the prediction distributions for all these reconstructions. Through the analysis of these distribution plots, the defender can identify an optimal interval where the predictions should ideally fall to maximize the reduction of the prediction distribution gap between members and non-members. Specifically, a grid search is conducted over the overlapping regions of member and non-member prediction distribution $[min(logit_{mem}),max(logit_{nonmem})]$ to determine this interval. The chosen interval aims to minimize the Jensen-Shannon (JS) divergence~\cite{lin1991divergence} between the prediction distributions of members and non-members within the selected range. This approach is based on the observation that a lower JS divergence between member and non-member prediction distributions typically indicates reduced membership privacy leakage \cite{he2022membershipdoctor}. This correlation has been confirmed by our experiments, as detailed in Section \ref{sec:keep generating}.

Defenders require only a small subset of samples to select the optimal interval. After the interval is determined, for each sample requiring protection, the defender randomly selects one sample from the candidate set $C$ that falls within this interval based on the prediction. If none of the candidates are within the interval, the closest one is chosen. Another possible approach involves continuously generating reconstructions until the sample falls within the defined interval. However, we found this method to be inefficient. More details on this are discussed in Section \ref{sec:keep generating}. 

Figure \ref{fig:scenarios} provides an example of \name as applied to the CIFAR-100 dataset. It illustrates the prediction distribution of reconstructed samples for 1000 data points which is used to select the optimal interval. The figure also showcases the distribution of prediction logits for both members and non-members after the selection of the optimal interval using the steps outlined in \name. This visualization demonstrates the effectiveness of \name in aligning the prediction distributions of member and non-member data, thereby minimizing the potential for privacy leakage.

\noindent \textbf{Scenario 2.} In this scenario, the defender only has access to a subset of members, the interval selection must rely solely on the prediction distribution of member reconstructions. In such cases, where only the member prediction distribution is available, we set the interval as $[min(logit_{mem}),mean(logit_{mem})]$. This is based on the observation that members typically exhibit higher confidence levels, and often, the lower half of the member distribution significantly overlaps with the upper half of the non-member distribution.  By observing Figure \ref{fig:scenarios} this approach also significantly narrows the gap between the prediction logits of members and non-members.

\noindent \textbf{Scenario 3.} In this scenario, the defender is unaware of the membership status of any sample, typically occurring when the defender is tasked with protecting an already trained model without having been involved in its training process. \name involves randomly selecting the prediction of one sample from the generated candidates as the output. Although this method does not deliberately restrict the prediction distribution, the reconstruction process inherently contributes to reducing the prediction distribution gap between members and non-members. This effect is achieved as the model encounters newly generated samples during the reconstruction phase, which are distinct from any data it was exposed to during training. This novelty in the samples ensures a more uniform response from the model, diminishing the likelihood of differential predictions between members and non-members.

% as discussed in Section \ref{sample_re}, the reconstruction process itself tends to narrow the gap in prediction distributions between members and non-members as the model is encountering a new generated sample which it has never seen during the training phase.
\subsubsection{Prediction Aggregation} 
By default, we adopt the aforementioned sample selection strategy as it consistently provides privacy protection without altering the model's accuracy. An alternative option is to aggregate the predictions of all generated samples instead of selecting one. This approach may yield superior results in specific contexts; for example, we observed that direct averaging of predictions from generated samples can \textbf{enhance both membership privacy and model accuracy} when the diffusion model is trained on sufficient data. Further details on this are provided in Appendix \ref{sec:agg}.

\subsubsection{Integration with Other Defenses in a Plug-and-Play Manner} 

\name employs off-the-shelf diffusion models, offering substantial flexibility in deployment. As outlined in Section \ref{sec:intro}, we categorize defenses into three deployment phases. Defenses deployed in different stages are compatible and can be used concurrently. \name, characterized by minimal defense assumptions, is unique in its deployment at the \textit{pre-inference} stage and offers plug-and-play capability, allowing for integration with all other existing methods. As illustrated in Figure \ref{Fig: illus}, when combined with training phase defenses, \name can integrated into the inference pipeline immediately after the model has been trained using privacy-preserving techniques.

When combined with post-inference defenses, \name can be employed to reconstruct input samples within the inference pipeline, followed by the application of the post-inference defenses, such as MemGuard, to introduce noise into the output prediction vector. Our experiments validate the effectiveness of \name in conjunction with other defenses, while also providing new perspectives for the deployment of future defense mechanisms. 

\subsection{Defending against Label-Only Attacks} 
 \name effectively reduces the prediction distribution gap while preserving the model's accuracy by not altering the predicted labels for samples. However, this design choice means that \name is not directly effective against label-only attacks, which exploit the model's predicted labels rather than the prediction confidence. Label-only attacks are generally considered weaker and less of a priority in recent defense works. For instance, RelaxLoss \cite{chen2022relaxloss} excludes label-only attacks from their evaluation, arguing that these attacks are strictly weaker than the ones used to assess model privacy. Similarly, HAMP {\cite{chen2023overconfidence}} demonstrates that existing label-only attacks are unsuccessful in a low false positive/negative regime.

Nonetheless, it's worth noting that previous defenses, such as SELENA \cite{tang2022mitigating} and HAMP {\cite{chen2023overconfidence}, have shown success in resisting label-only attacks. While \name does not directly improve defense against label-only attacks, it can be seamlessly integrated with these existing defenses, enhancing overall protection without sacrificing the model's utility. Although \name does not specifically target label-only attacks, it offers a flexible and non-intrusive approach to enhancing membership privacy that can be effectively combined with other defenses to achieve comprehensive protection.
\section{Evaluations}

\subsection{Experimental Setup}

\noindent \textbf{Datasets.} We consider three benchmark datasets and two high-resolution datasets : 
\begin{itemize}
\item \textbf{CIFAR-10~\cite{krizhevsky2009learning}}: CIFAR-10 Comprising 60,000 32x32 color images across 10 classes, Each class contains 6,000 images. 
\item \textbf{CIFAR-100~\cite{krizhevsky2009learning}}: CIFAR-100 have the same data format as CIFAR-10, but it has 100 classes, so each class has only 600 images.
\item \textbf{SVHN~\cite{netzer2011reading}}: SVHN contains 99,289 digit images of house numbers collected from Google Street View. Each image has a resolution of 32x32 and is labeled with the integer value of the digit it represents, from 0 to 9.
\item \textbf{CelebA~\cite{liu2015deep}}: CelebA contains over 200,000 high-resolution celebrity images (178x218 pixels), each annotated with 40 attributes. Following prior work \cite{liu2022midoctor}, we selected 3 attributes to create 8-class labels.
\item \textbf{UTKFace~\cite{zhang2017age}}: A large-scale face dataset with over 20,000 images labeled by age, gender, and ethnicity (resolutions up to 200x200 pixels). We used images from the four largest races (White, Black, Asian, Indian) for race-based labels.
\end{itemize}
For the CIFAR-10 and CIFAR-100 datasets, we used 25,000 samples each for training the target models. For the SVHN dataset, we used 5,000 samples. For the CelebA and UTKFace datasets, we used 10,000 and 5,000 samples, respectively. The remaining samples from each dataset were reserved as non-members or references used by defenses or attacks. 

\noindent \textbf{Models.} 
For target classifiers, we consider the widely used ResNet18~\cite{he2016deep} as the target model. Appendix \ref{sec:densenet} provides experimental results on additional model architectures, including DenseNet121~\cite{huang2017densely}, VGG16 \cite{simonyan2014very}, and ViT \cite{dosovitskiy2020image}.

In our default setup, each target model is trained for 100 epochs using the Adam optimizer with a learning rate of 0.001. We apply an L2 weight decay coefficient of $10^{-6}$ and use a batch size of 128. For ViT, we use a learning rate of 0.0001. As we found that the performance of some recent works can be affected by the training configurations of the target models, \textit{\textbf{we also employ their training settings to demonstrate that \name enhances their best outcomes}}.

For the diffusion models integrated into \name, we train standard DDPMs from scratch using the default hyperparameters from the original DDPM paper \cite{ho2020denoising}. Unless otherwise mentioned, \name uses a diffusion model trained on the same training data as the associated defended target classifier, and generates $N = 30$ reconstructed images for each sample, with the number of diffusion steps $T$ set to 160. Detailed discussions on the choice of these hyperparameters are provided in Section \ref{sec:hyper}. 

In addition to training diffusion models on the same distribution as the target classifier, we also demonstrate the effectiveness of \name using publicly available pre-trained diffusion models trained on ImageNet \cite{dhariwal2021diffusion}. This illustrates that the diffusion models in \name and the target classifier's training data do not need to be from the same distribution. It also shows that when an off-the-shelf diffusion model for the same dataset is not available, using a pre-trained diffusion model on a different dataset can still provide effective protection. More details are provided in Section \ref{sec:imagenetDM}.

\noindent \textbf{Attack and Defense Methods.} 
For evaluation, we consider six state-of-the-art attack methods: NN-based attacks \cite{nasr2018machine, salem2018ml}, four threshold-based attacks (loss, confidence, entropy, M-entropy), and the recent LiRA attack \cite{carlini2022membership}.

We consider six major defenses: AdvReg \cite{nasr2018machine}, MemGuard \cite{jia2019memguard}, SELENA \cite{tang2022mitigating}, RelaxLoss \cite{chen2022relaxloss}, HAMP \cite{chen2023overconfidence}, and DPSGD \cite{abadi2016deep}. For all attack and defense methods, we follow the original papers and setups unless otherwise stated. Detailed information is provided in Appendix \ref{sec:exp_details}.

Following previous practice \cite{chen2022relaxloss}, we exclude label-only attacks in our comparative experiments. \name primarily targets attacks beyond label-only attacks, aiming to avoid accuracy loss by not altering predicted labels, and thus does not directly defend against label-only attacks (more discussions are provided in Appendix \ref{sec:label-only}).
However, \textit{\textbf{label-only attacks} can be effectively countered by existing defenses such as SELENA {\cite{tang2022mitigating}} and HAMP {\cite{chen2023overconfidence}}, but defenses against other attacks still need improvement. Our approach is designed to complement these existing defenses on the stronger attacks (i.e., non label-only MIAs), enabling a comprehensive defense strategy that addresses all types of attacks.}
% In line with previous practice, we exclude attacks using only partial model output as they are strictly weaker than attacks above. \ancomment{Don't forget to talk about the label-only attack and cite the related papers here}

% In line with previous practice, we exclude attacks using only partial model output as they are strictly weaker than attacks above. \ancomment{Don't forget to talk about the label-only attack and cite the related papers here}

\noindent \textbf{Evaluation metrics.} 
% We use Top-1 accuracy on validation dataset to quantify model's utility. 
For privacy, we default to using five of the described attacks (excluding LiRA) to evaluate two common used metric: (i) attack accuracy and (ii) attack AUC. Additionally, we report the TPR (True Positive Rate) at 0.1\% FPR  and the TNR (True Negative Rate) at 0.1\% FNR on SVHN dataset, using LiRA which is specifically designed to achieve superior results on this metric. We exclusively applied the LiRA attack on the SVHN dataset due to its high computational demands, as it requires the training of over 100 shadow models.

\begin{figure*}[htbp]
  \centering
  % \hspace*{\fill}%
  \begin{subfigure}{1\linewidth}
    \includegraphics[width=0.3\columnwidth]{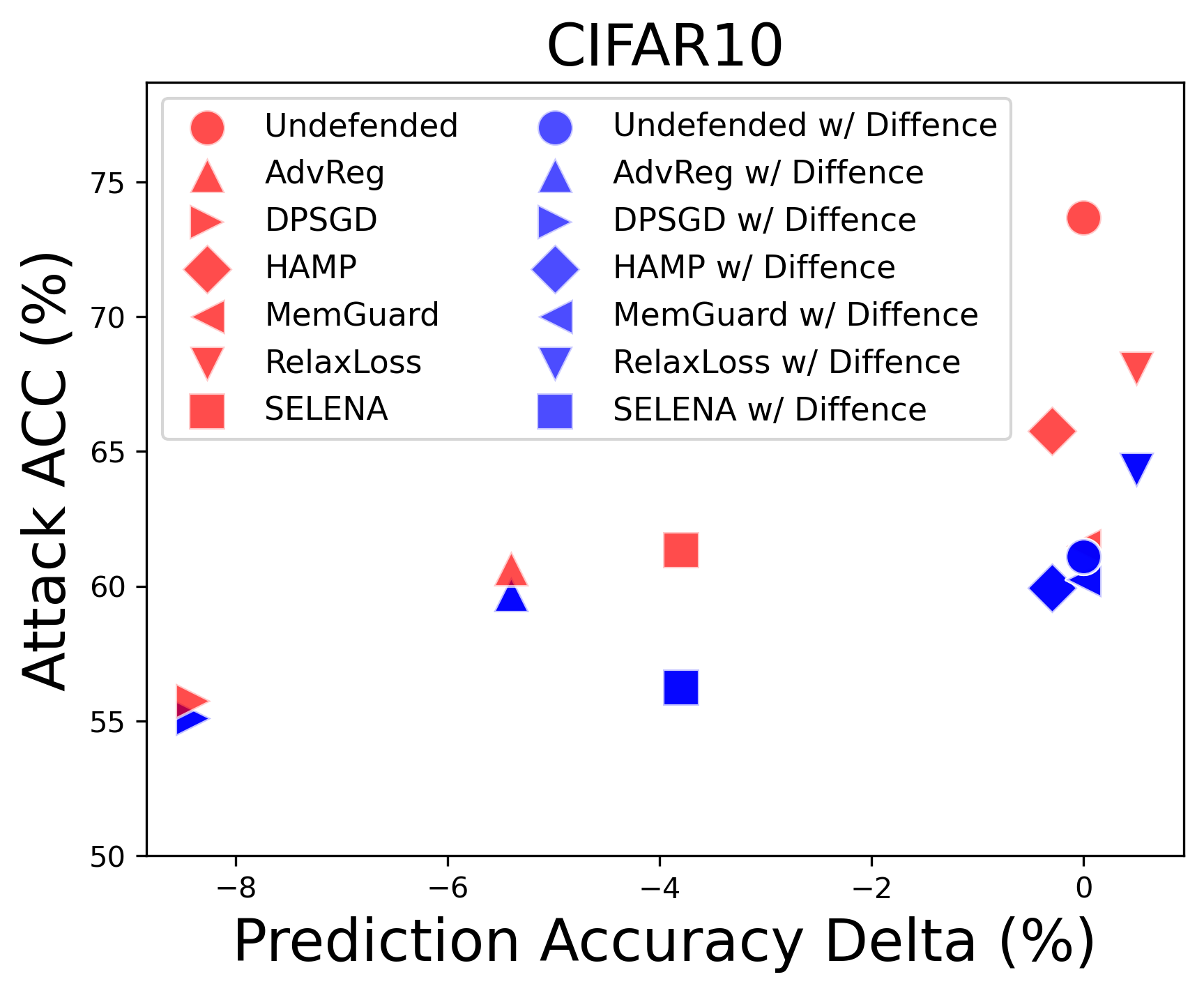}
     \hfill
    \includegraphics[width=0.3\columnwidth]{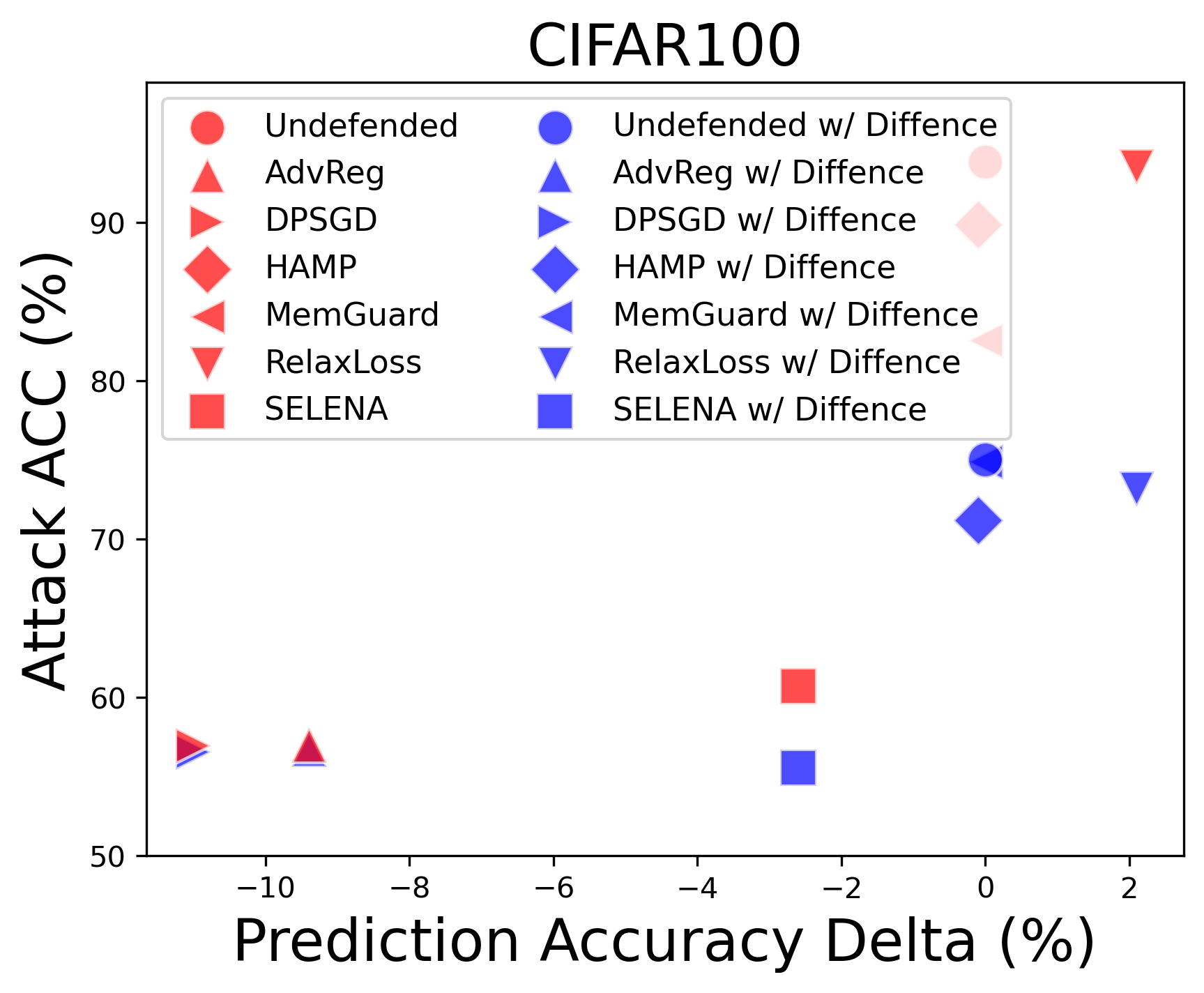}
    \hfill
    \includegraphics[width=0.3\columnwidth]{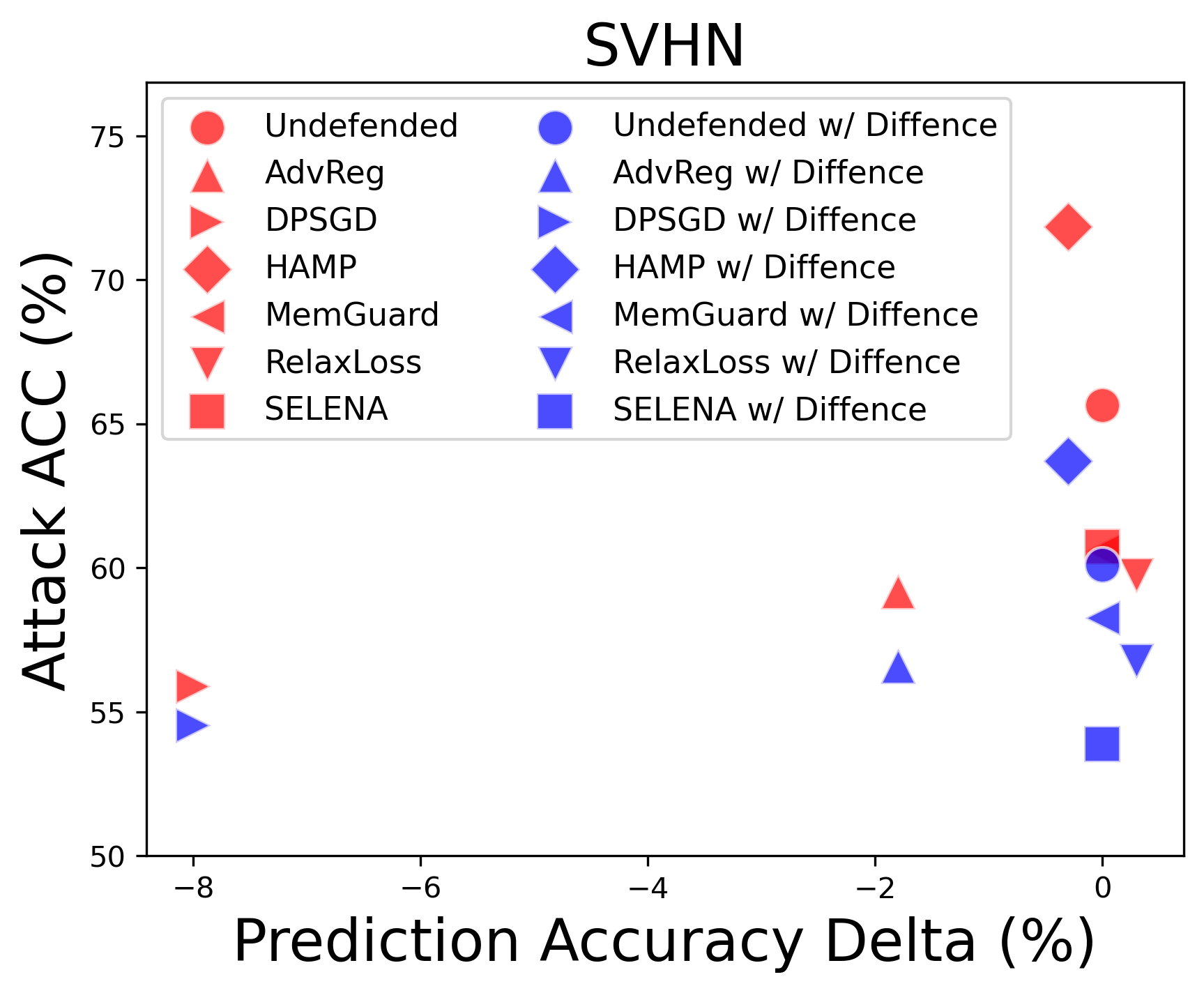}

    \label{fig:short-a}
    \caption{ Attack Accuracy}
  \end{subfigure}

  \hspace{1em}%
    \begin{subfigure}{1\linewidth}
    \includegraphics[width=0.3\columnwidth]{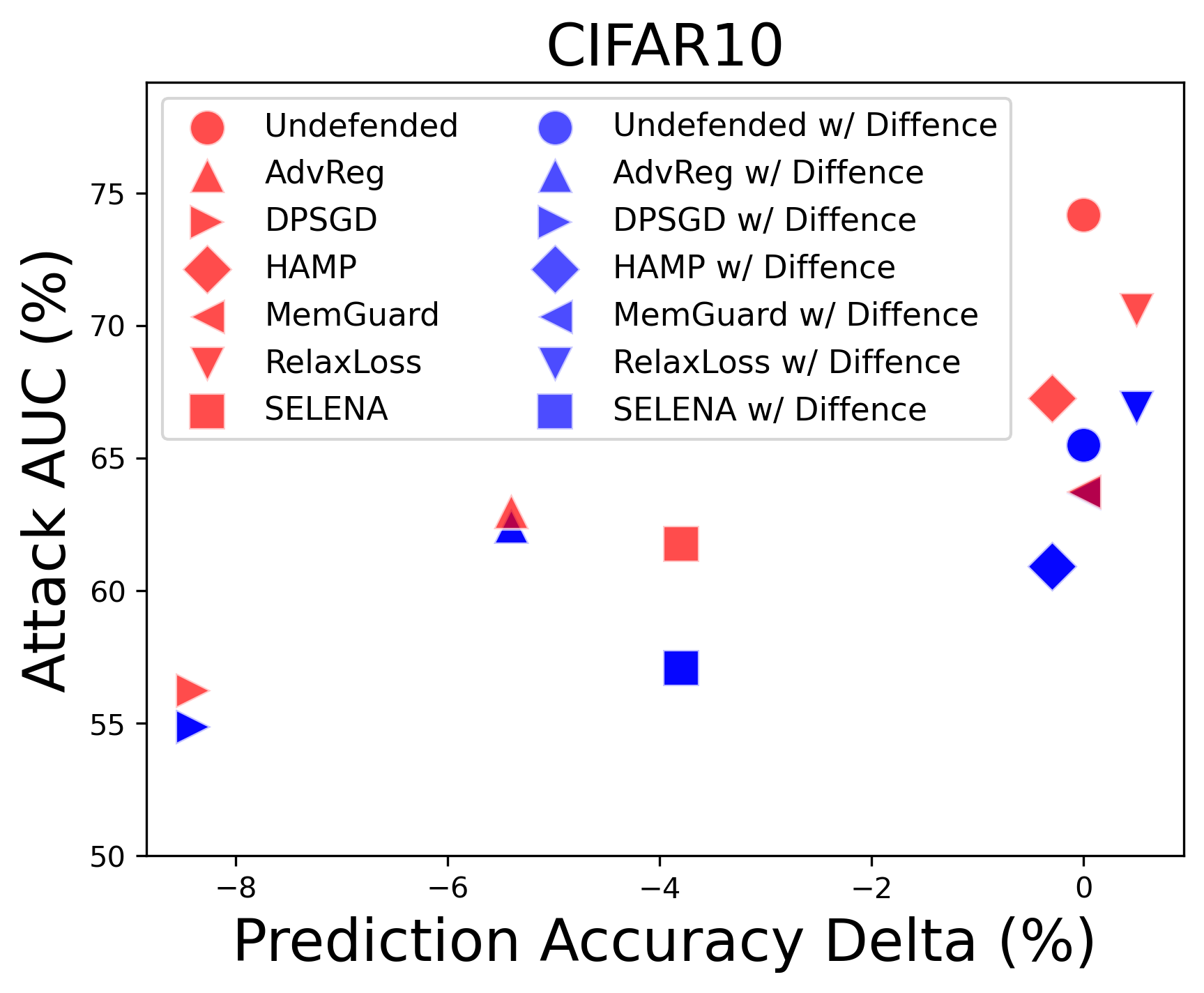}
     \hfill
    \includegraphics[width=0.3\columnwidth]{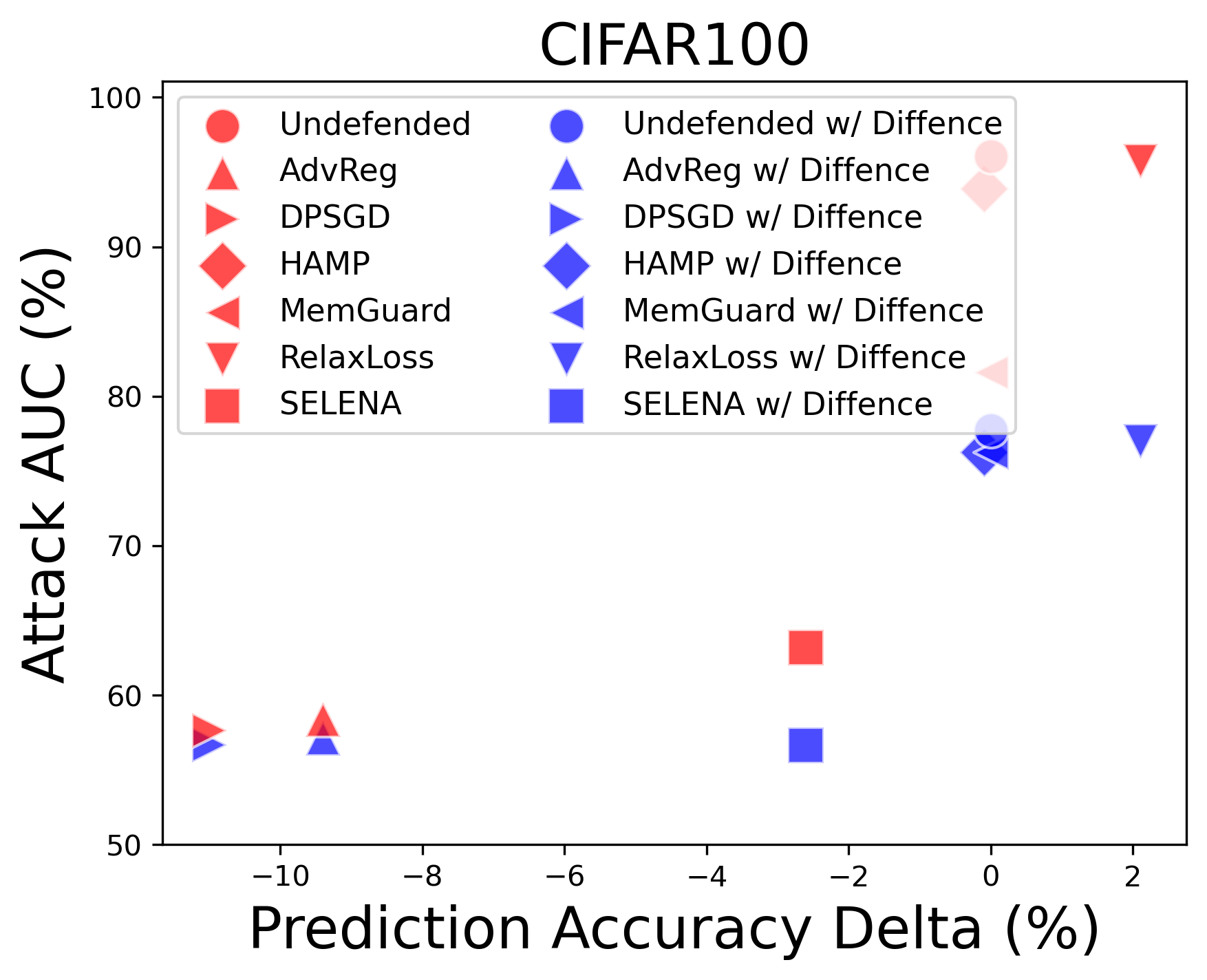}
    \hfill
    \includegraphics[width=0.3\columnwidth]{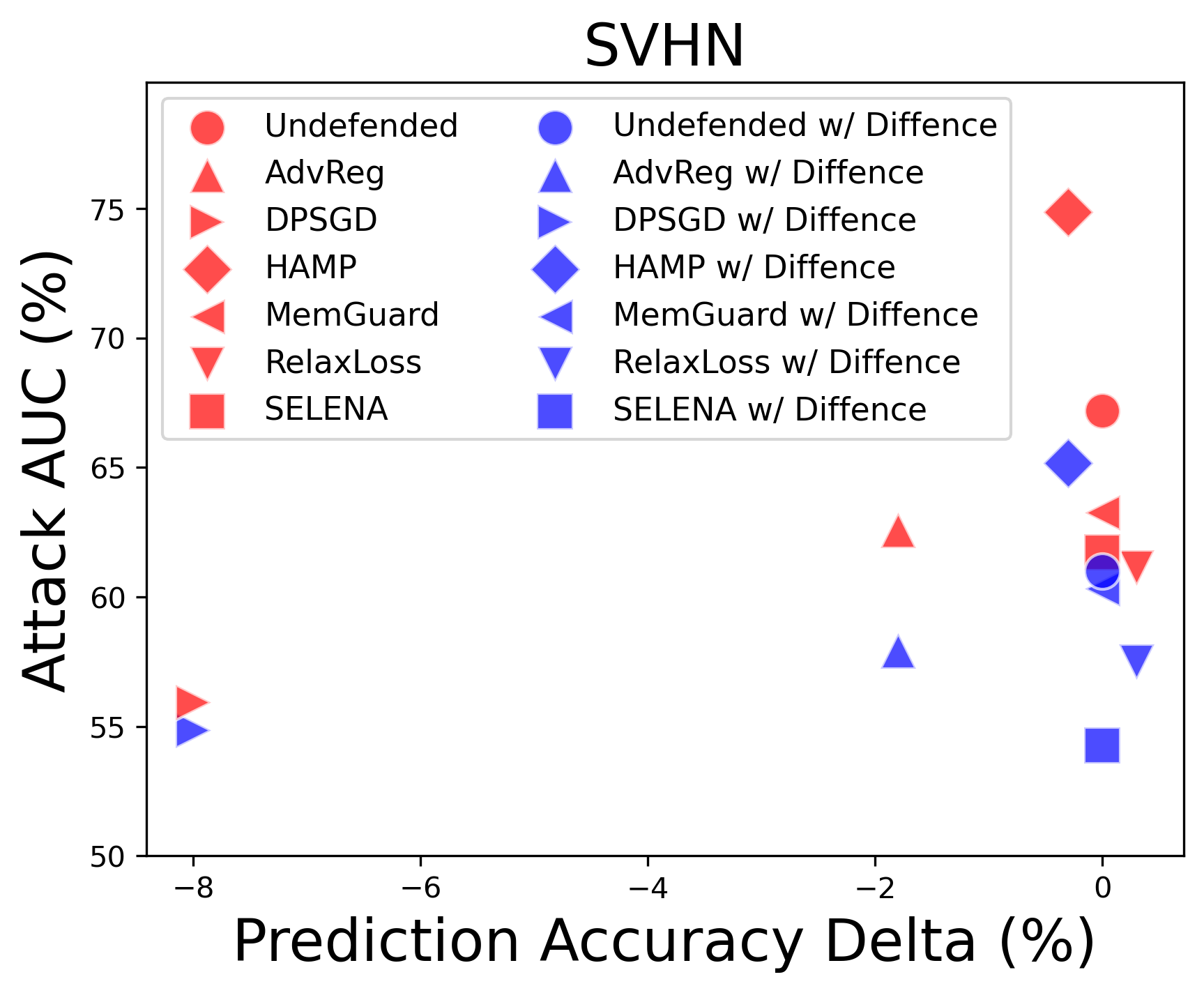}

    \label{fig:short-a}
    \caption{ Attack AUC}
  \end{subfigure}
  \caption{\textbf{Attack accuracy and AUC on three datasets against ResNet18 under Scenario 1.} We report the highest attack accuracy and attack AUC across all attacks. The prediction accuracy delta indicates the prediction accuracy gap compared to the undefended models, with negative numbers indicating a decrease in model accuracy.}
  \label{fig:all}
  % \hspace*{\fill}%
\end{figure*}

\begin{table*}
\large
    \caption{\textbf{Average attack accuracy and AUC across three datasets under three scenarios on ResNet18.} The best defense results in each case are highlighted in bold.}
    \centering
    \scalebox{0.63}{
    \begin{tabular}{m{3cm}<{\centering}m{4cm}<{\centering}m{2cm}<{\centering}m{2cm}<{\centering}m{2cm}<{\centering}m{2cm}<{\centering}m{2cm}<{\centering}m{2cm}<{\centering}m{2cm}<{\centering}<{\centering}m{2cm}<{\centering}<{\centering}m{2cm}<{\centering}}
    \toprule
    \multirow{2}*{ \textbf{Defenses}} & \multirow{2}*{\makecell{\textbf{Prediction} \\ \textbf{Accuracy Delta (\%)}}} & \multicolumn{2}{c}{\textbf{w/o \name}} & \multicolumn{2}{c}{\textbf{w/ \name (Scenario 1)}} & \multicolumn{2}{c}{\textbf{w/ \name (Scenario 2)}} & \multicolumn{2}{c}{\textbf{w/ \name (Scenario 3)}}\\
    \cmidrule(lr){3-10} & & \textbf{Attack AUC (\%)} & \textbf{Attack Accuracy (\%)} & \textbf{Attack AUC (\%)} & \textbf{Attack Accuracy (\%)} & \textbf{Attack AUC (\%)} & \textbf{Attack Accuracy (\%)} & \textbf{Attack AUC (\%)} & \textbf{Attack Accuracy (\%)} \\
    \midrule
    Undefended & 0 & 79.14 & 77.73 & \textbf{68.08} & \textbf{65.41} & 70.79 & 67.12 & 69.12 & 67.02 \\
    SELENA     & -2.13 & 62.22 & 60.92 & \textbf{56.00} & \textbf{55.23} & 60.30 & 58.58 & 57.81 & 57.16 \\
    AdvReg     & -5.53 & 61.32 & 58.94 & \textbf{59.17} & \textbf{57.68} & 61.33 & 58.58 & 60.87 & 58.62 \\
    HAMP       & -0.23 & 78.96 & 76.08 & \textbf{67.60} & \textbf{65.10} & 71.23 & 66.61 & 69.18 & 66.31 \\
    RelaxLoss  & 0.97  & 75.81 & 73.78 & \textbf{67.13} & \textbf{64.75} & 69.56 & 66.19 & 68.60 & 66.07 \\
    DP-SGD     & -9.13 & 56.61 & 56.19 & \textbf{55.47} & \textbf{55.40} & 58.40 & 56.92 & 56.60 & 56.35 \\
    Memguard   & 0     & 69.53 & 68.21 & \textbf{66.76} & \textbf{64.47} & 67.23 & 65.30 & 67.48 & 65.51 \\
    \bottomrule
    \end{tabular}}
    \label{table:scen}
\end{table*}

\subsection{Experimental Results}

\subsubsection{Comparison to Baselines}

We first conducted an extensive evaluation of \name across three benchmark datasets. For each dataset, we evaluated the performance of models across seven different cases including one without any protection and six others, each employing a different defense mechanism. We tested the impact of \name on the models' test accuracy and the changes in attack accuracy and attack AUC under three scenarios introduced in Section \ref{sec:sample_s}. Our results show that \textit{\textbf{\name consistently enhanced privacy protection across all settings without compromising the model accuracy}.} We discuss the specific effects of \name below. 

Figure \ref{fig:all} shows the performance of \name against ResNet18 under Scenario 1. We can observe that \name consistently reduces the attack's AUC and accuracy across all cases, particularly against those methods that preserve model usability but offer only limited protection. For instance, when applied to ResNet18, \name managed to decrease the attack accuracy for undefended, HAMP, RelaxLoss, and SELENA models by 15.8\%, 14.4\%, 12.2\%, and 9.3\%, respectively, on average across three datasets. Similarly, the attack AUCs showed reductions of 14.0\%, 14.4\%, 11.4\%, and 10.0\% after employing \name. The experimental results demonstrate that \name can significantly reduce privacy threats, even with the reliance on only a small amount of additional data (1000 non-members in our experiments).
Besides ResNet18, we also tested \name on DenseNet121, VGG16, and ViT. The experiments show that \name can effectively enhance membership privacy in all these settings (more details in Appendix \ref{sec:densenet}).

Our experiments suggest that the optimal defense practice is combining \name with recent utility-preserving defenses. This combination leverages their advantages in maintaining utility while enhancing privacy protection. For instance, the best privacy-utility trade-offs achieved on VGG16 were from the integrations of \name with SELENA and HAMP (Figure \ref{fig:all-dn}).

Table \ref{table:scen} presents the experimental results using ResNet18 under all three different scenarios. We show the average attack AUC and accuracy reduction of \name across three datasets in each scenario. The average defense effectiveness in Scenarios 2 and 3 is weaker than in Scenario 1, due to stronger assumptions placed on the defender in these scenarios. However, it is observable that they still achieve significant privacy enhancements across all settings without compromising the model's accuracy.
We observed that some recent defenses, such as HAMP and RelaxLoss, have not achieved satisfactory overall effectiveness. We found their performance heavily depends on the manual selection of hyperparameters. Following their claims about the impact on utility, we selected hyperparameters that matched the test accuracy of undefended models. However, this sometimes resulted in poor defense and even worsened privacy. \name can address these privacy issues while maintaining their high utility.

In the research of MIA defense, balancing sufficient privacy protection with model utility has always been one of the most important objectives. Our results suggest that \name represents a significant step forward towards the ideal defense.

\subsubsection{Evaluation of Confidence Calibration in Model Predictions}

\name preserves the predicted label of the sample, thereby not affecting the model's accuracy. Nonetheless, the utility of a model also involves the value of information provided by the output confidence vector. As discussed in Section \ref{sec:method}, \name exploits diffusion model to generate reconstructions that closely resemble the original samples, differing only in subtle details (see Figure \ref{Fig:examples} for examples). We propose that the confidence levels of these reconstructions are reliable indicators and provide a dependable measure of the original sample's characteristics.

% Previous works \cite{guo2017calibration, karras2020analyzing, wang2021beconf} highlighted the necessity for model outputs to be well-calibrated, meaning that the probability associated with a predicted class label should correspond to its actual likelihood of correctness. 

Previous research \cite{guo2017calibration, karras2020analyzing, wang2021beconf} emphasizes the importance of well-calibrated model outputs, where the predicted class probability should match its actual correctness likelihood. 
Expected Calibration Error (ECE) \cite{naeini2015obtaining, guo2017calibration, wang2021beconf} is a key metric in this context, quantifying the alignment between predicted probabilities and empirical accuracies. Essentially, ECE measures the average discrepancy between the confidence of a model's predictions and the true correctness of those predictions, providing a critical assessment of the reliability of the model's confidence estimates. A lower ECE suggests better alignment of confidence scores with true probabilities, indicating more reliable predictions. ECE is calculated as in equation \ref{eq:ece}:
\begin{equation}
    \text{ECE} = \sum_{m=1}^{M} \frac{|B_m|}{n} \left| \text{acc}(B_m) - \text{conf}(B_m) \right|
\label{eq:ece}
\end{equation}
where \( M \) denotes the number of bins into which the predictions are grouped. \( B_m \) is the set of samples in the \(m\)-th bin. \( n \) represents the total number of samples. \( \text{acc}(B_m) \) is the accuracy within bin \( B_m \). \( \text{conf}(B_m) \) is the average predicted confidence for samples in bin \( B_m \).

Figure \ref{fig:ece} shows the ECE of the confidence vectors output by both undefended models and models employing various defenses across three datasets. It is observed that the ECE values of \name are very similar to those of the undefended models. For instance, the ECE values for the undefended model and the three \name scenarios are 0.139, 0.142, 0.126, and 0.131 on average, respectively. This suggests that \textit{\textbf{\name not only maintains the accuracy of the model but also provides meaningful confidence scores.}}

However, it was observed that some defenses could lead to a significant increase in ECE. For example, compared to the undefended model, the average ECE of the HAMP is 222.3\% higher. This could be attributed to HAMP's strategy of promoting high-entropy confidence vectors as a defense against MIAs, leading to more flattened confidence distributions. Such alterations, while serving a defensive purpose, can yield confidence scores that lack meaningful interpretation, potentially compromising the utility of the model.

In summary, our results validate the effectiveness of \name in preserving the utility of confidence scores, and also highlight the importance of evaluating the meaningfulness of confidence scores in the field of MIA defenses.

\subsubsection{Effectiveness on Attack TPR and TNR Metrics}

Following the methodology of Carlini et al.~\cite{carlini2022membership}, we evaluate the reliability with which an adversary can compromise the privacy of even a few users in a sensitive dataset, using metrics such as TPR at 0.1\% FPR and TNR at 0.1\% FNR. We assess the defensive effectiveness of \name under the third scenario against these metrics on the SVHN dataset, using six different attacks including LiRA.

As illustrated in Figure \ref{fig:tpr}, \name significantly reduces both the attack TPR and TNR in most cases. For example, \name decreased the attack TPR under 0.1\% FPR and TNR under 0.1\% FNR against the undefended model by 63.9\% and 56.8\%, respectively. When cascaded with other defenses, \name reduced the TPR under 0.1\% FPR by 52.8\% on average across six tested defenses.
\begin{figure}[tbp]
  \centering
  % \hspace*{\fill}%
    \includegraphics[width=0.8\columnwidth]{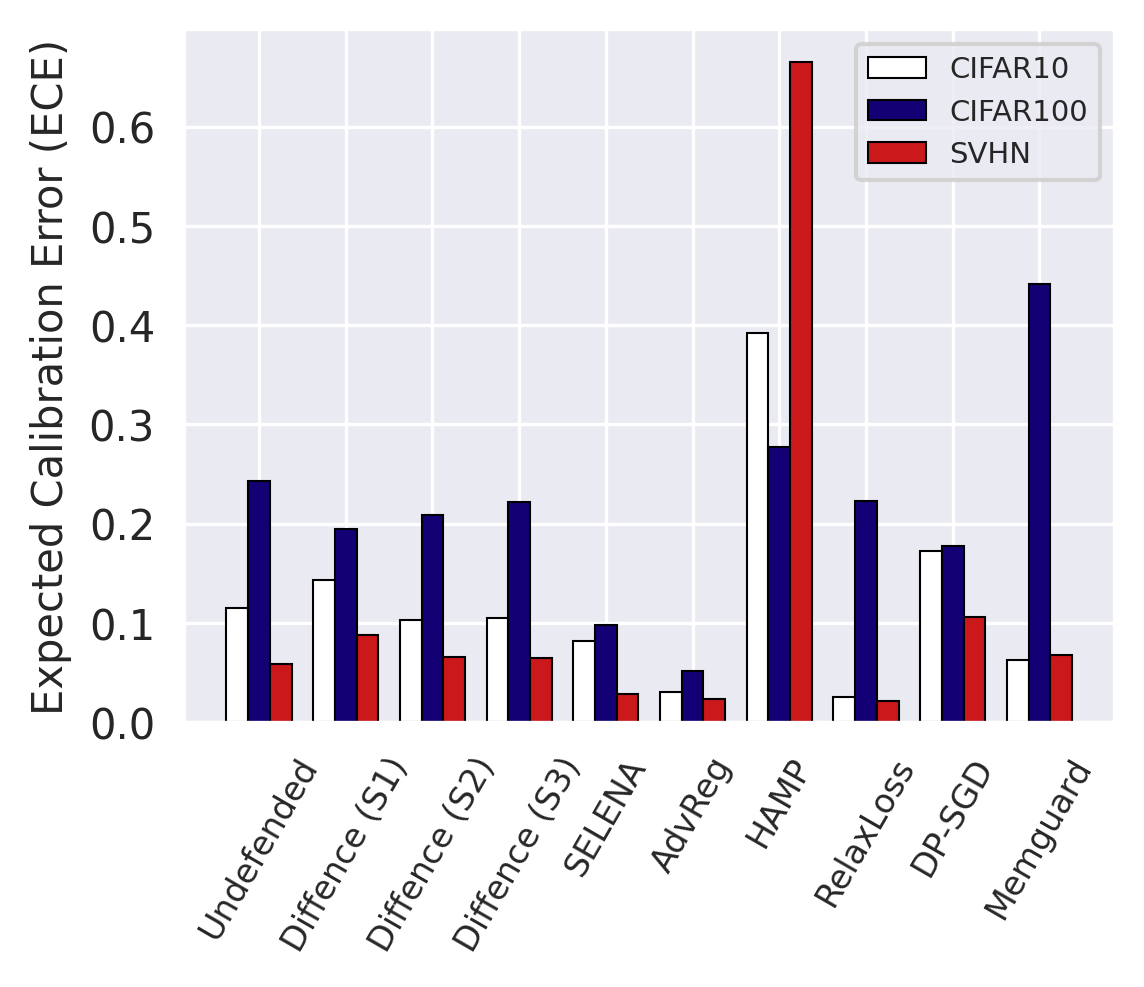}
     \hfill
  \caption{\textbf{Expected Calibration Error (ECE) of various defenses across three datasets.} A lower ECE value indicates better-calibrated confidence scores.}
  \label{fig:ece}
  % \hspace*{\fill}%
\end{figure}

\begin{figure}[htbp]
  \centering
  % \hspace*{\fill}%
    \includegraphics[width=0.7\columnwidth]{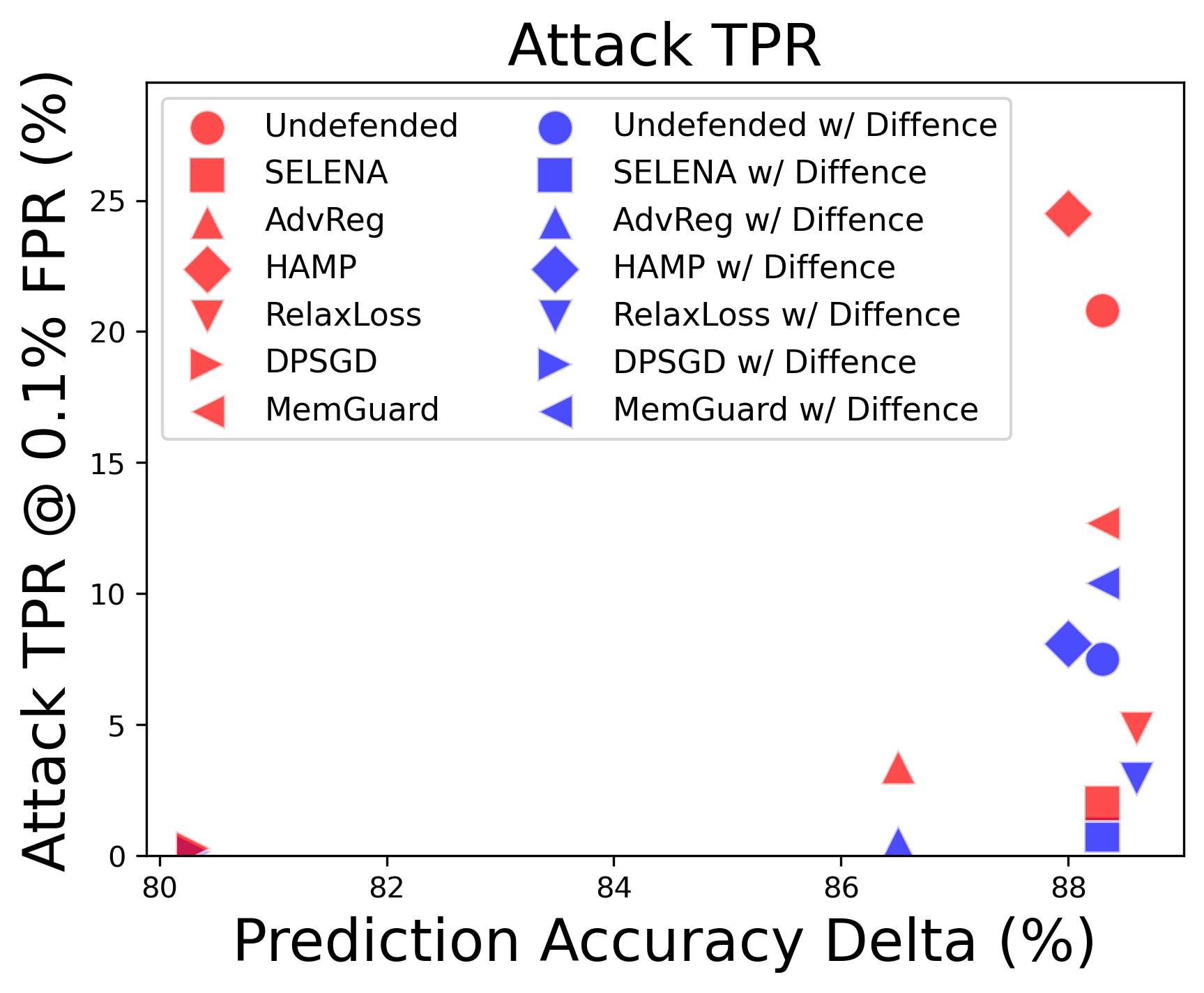}
     \hfill
    \includegraphics[width=0.7\columnwidth]{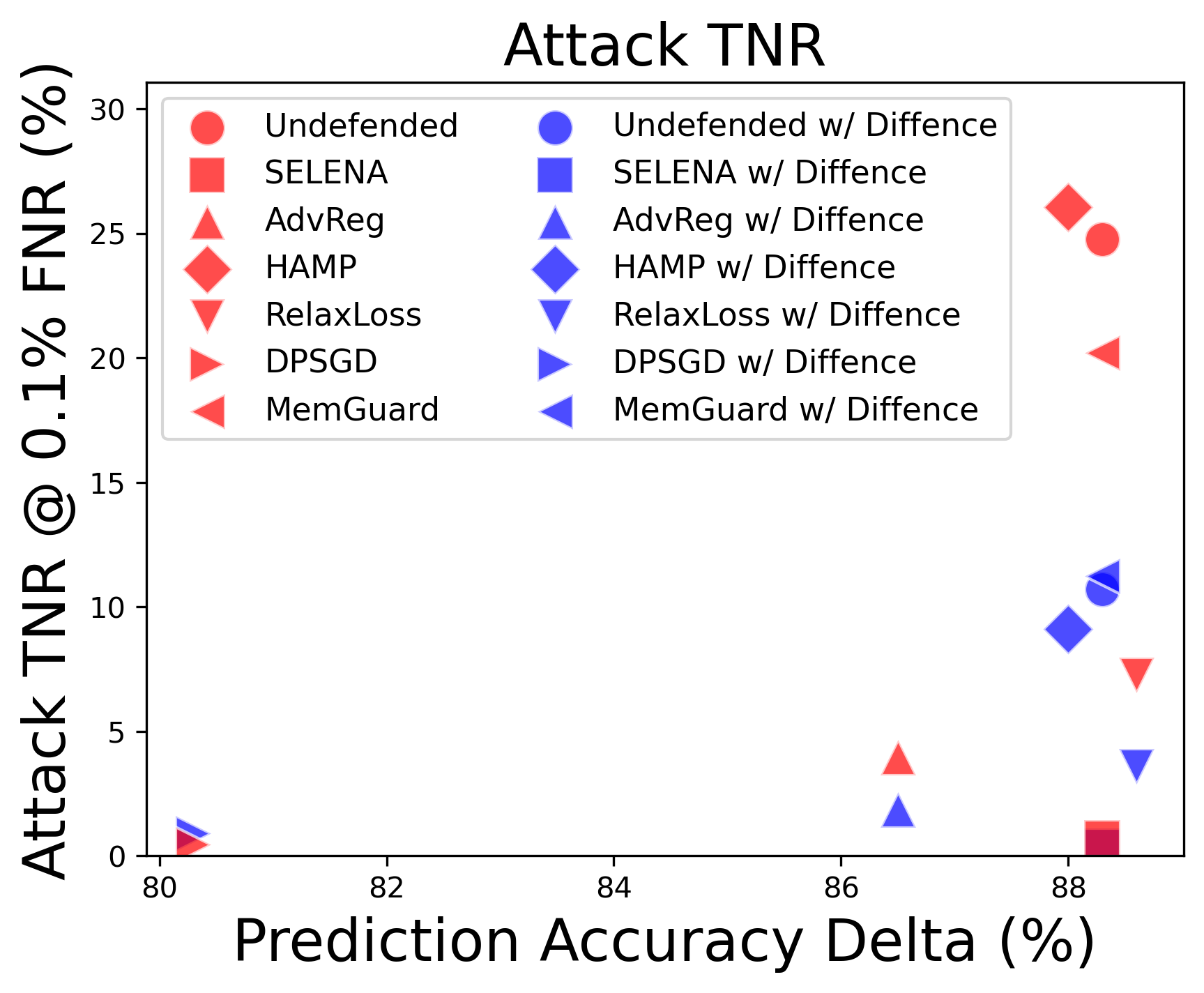}

  \caption{\textbf{Attack TPR and FPR on SVHN dataset.} We report the highest attack TPR and attack FPR across all attacks.}
  \label{fig:tpr}
  % \hspace*{\fill}%
\end{figure}
\name has a smaller impact on methods that already exhibit low TPR; however, these methods often compromise the model's utility. Consequently, \name can be advantageously combined with other defenses that preserve model utility but are relatively weaker in defense effectiveness, to enhance overall protection without significantly sacrificing utility.

\subsubsection{Improving Performance Beyond Previous State-of-the-Art}
\label{sec:best_settings}

\begin{figure*}[htbp]
  \centering

  \begin{subfigure}{1\linewidth}
    \centering
    \includegraphics[width=0.3\columnwidth]{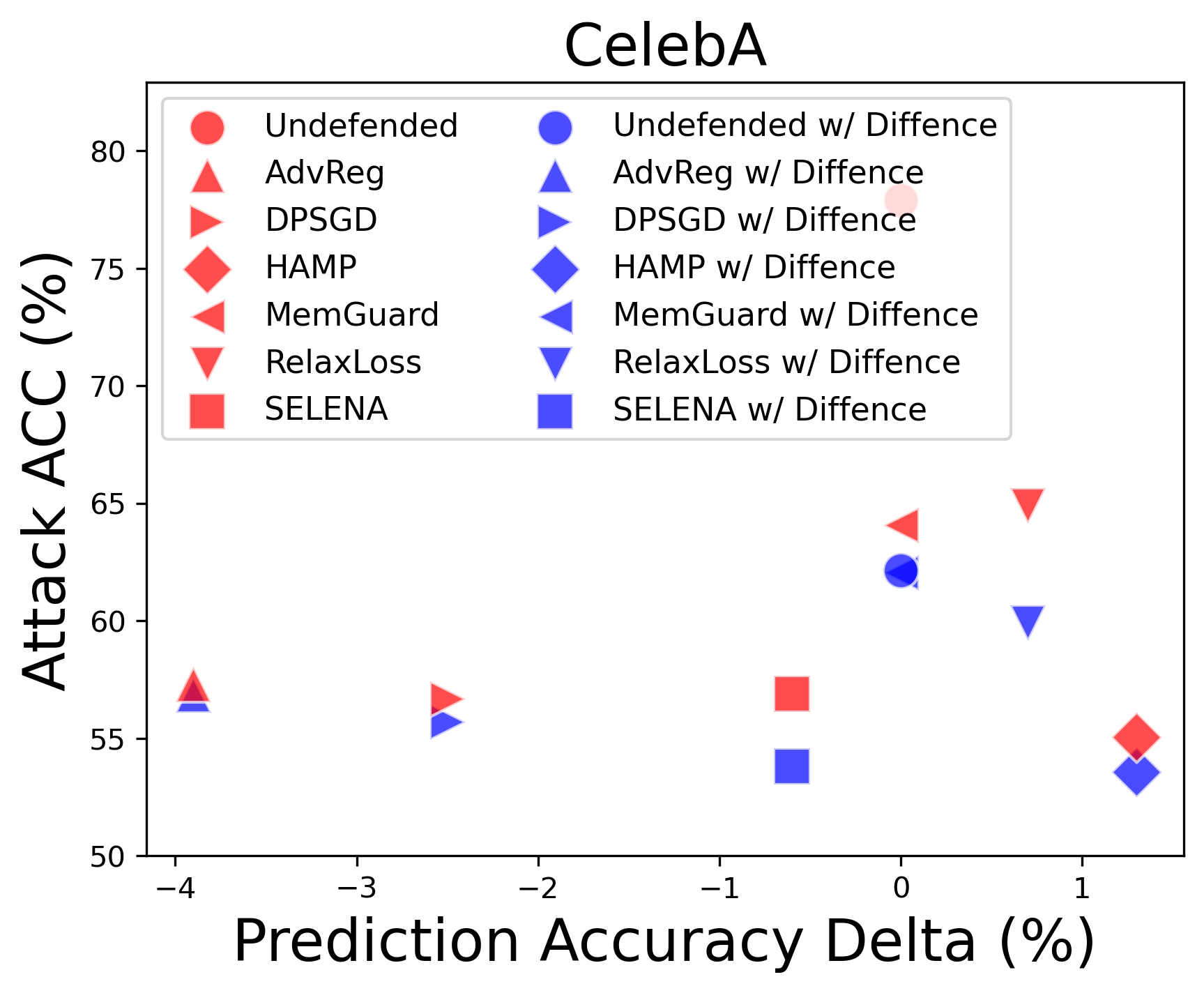}
    \hspace{2em}
    \includegraphics[width=0.3\columnwidth]{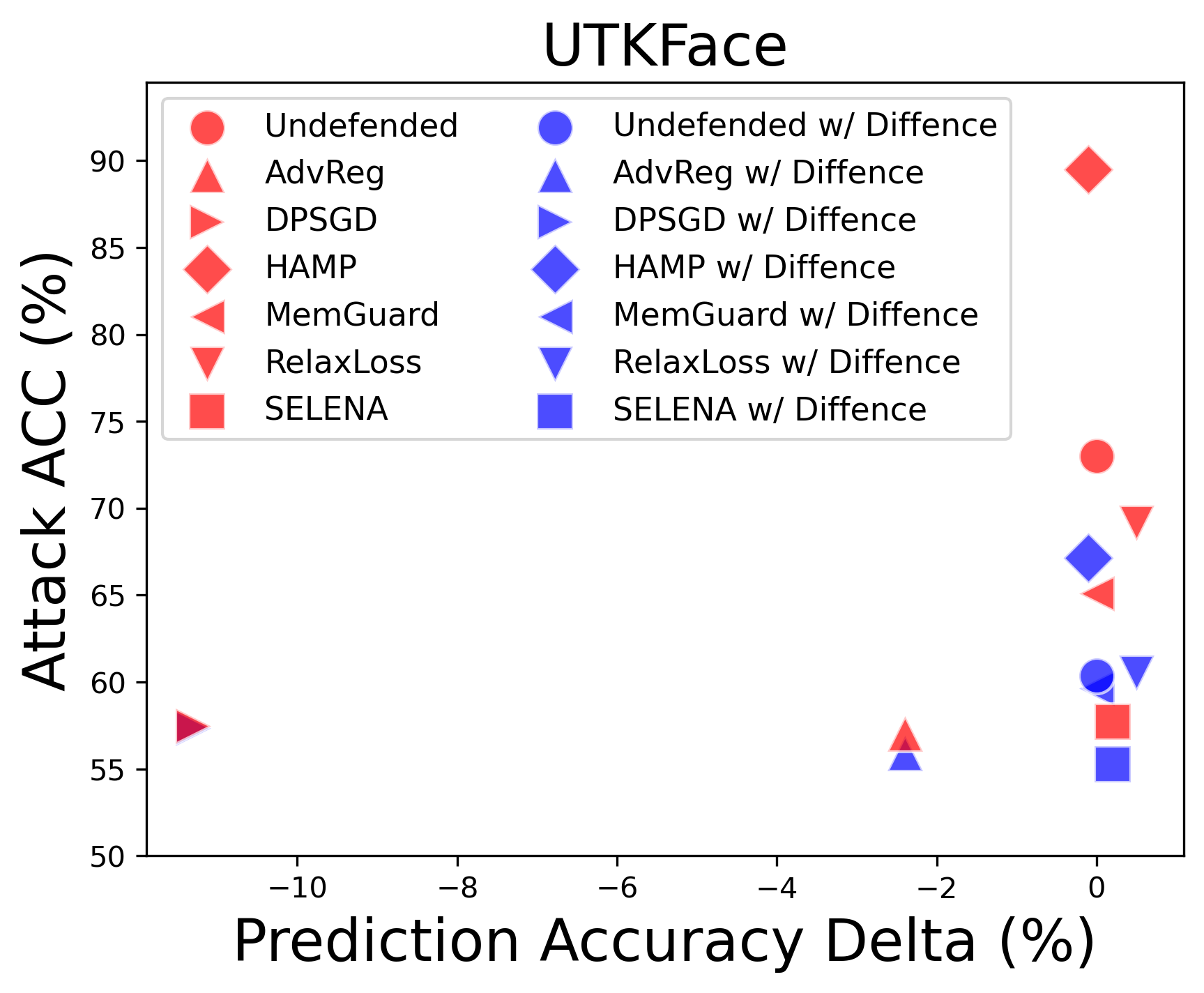}
    \caption{Attack Accuracy}
  \end{subfigure}

  \vspace{1em}

  \begin{subfigure}{1\linewidth}
    \centering
    \includegraphics[width=0.3\columnwidth]{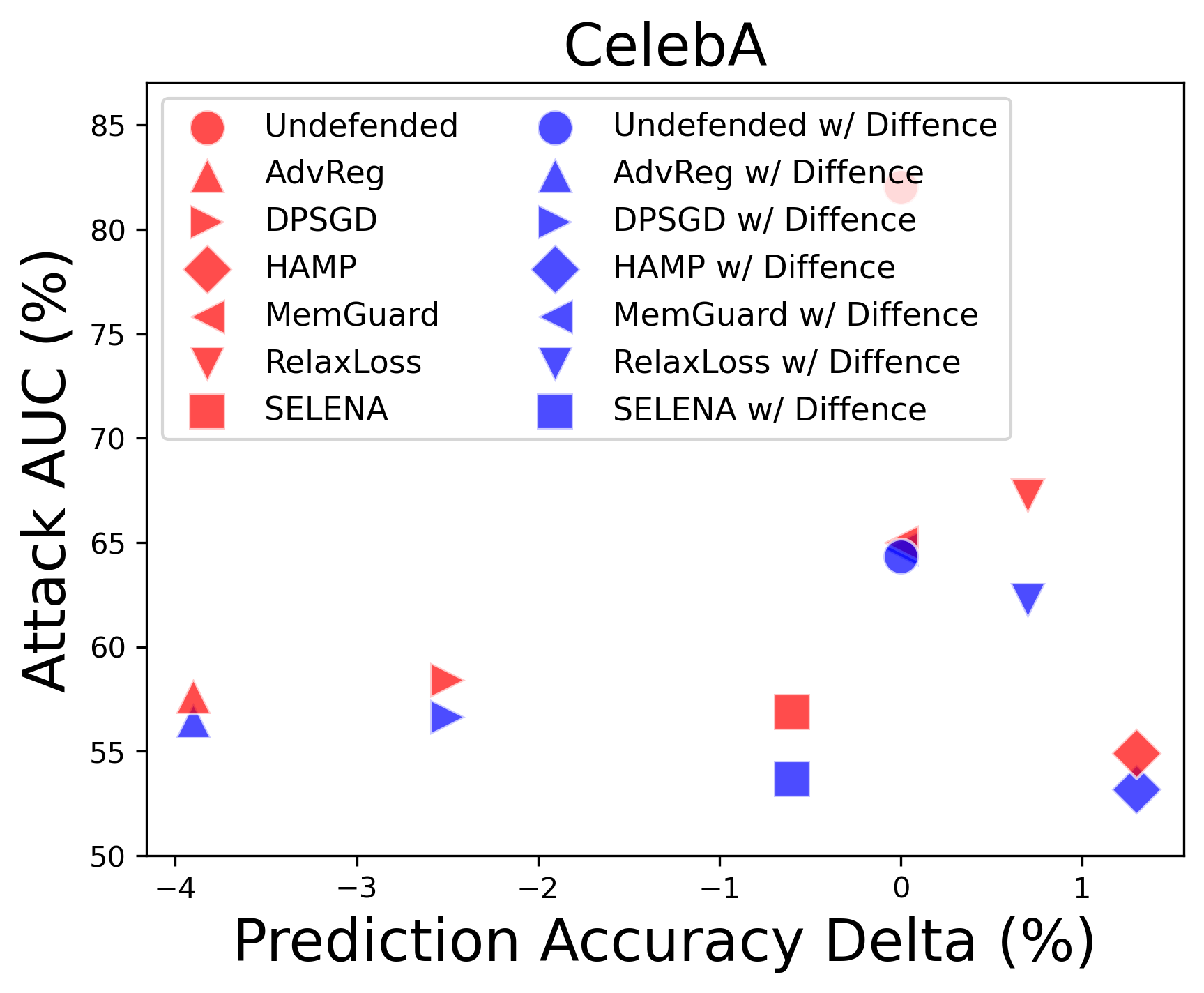}
    \hspace{2em}
    \includegraphics[width=0.3\columnwidth]{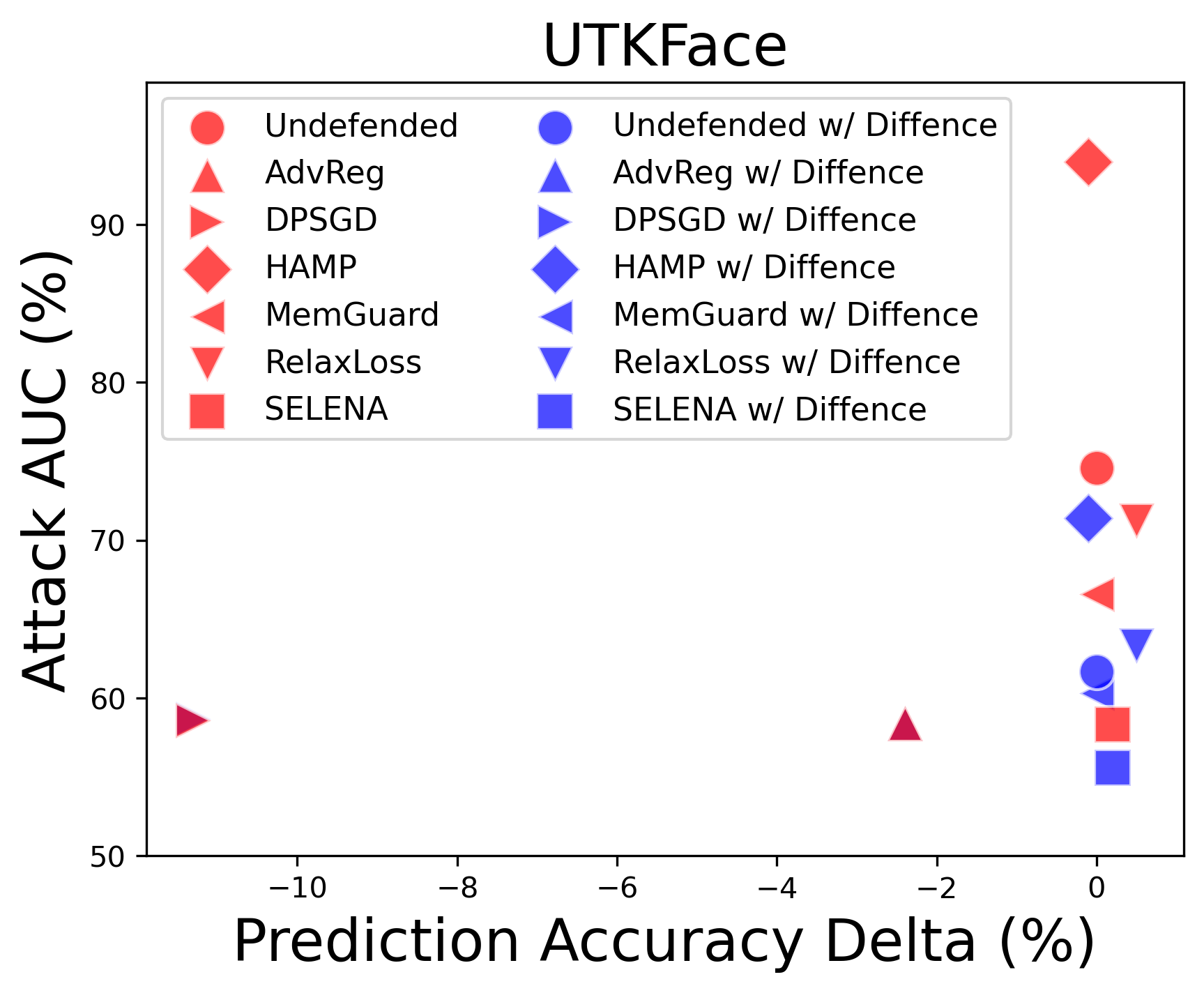}
    \caption{Attack AUC}
  \end{subfigure}

  \caption{\textbf{Attack accuracy and AUC on two high-resolution datasets against ResNet18, where \name employs a publicly released diffusion model trained on ImageNet.}}
  \label{fig:high-reso}
\end{figure*}

Recent advanced defenses like SELENA, RelaxLoss, and HAMP have been reported state-of-the-art defense performance in terms of privacy-utility trade-off in their respective papers. However, our experments suggest that the defensive robustness of RelaxLoss and HAMP is contingent upon the choice of training parameters, leading to variability in their effectiveness.

For a fair and accurate comparison, we adopted the experimental settings deployed in their original papers to assess the effectiveness of \name. In our evaluations, \name was configured with $N=10$ and $T=50$ in Scenario 3. The results are shown in Table \ref{table:sota}. Note that \name maintains the original predicted labels of the model. While this does not eliminate the privacy risks posed by the accuracy gap between training and test sets, it significantly mitigates the risk of privacy leakage arising from other inconsistent prediction behaviors.

We observed that \name can effectively enhance membership privacy in most settings. For instance, for HAMP, \name successfully reduced attack AUC and accuracy from 67.4\% and 65.9\% to 64.7\% and 61.5\%, respectively. \name did not show improvement in the RelaxLoss setting, as there was no room for privacy enhancement by reducing the prediction distribution gap. In this case, a simple attack assuming all correctly classified samples as members achieved the highest attack accuracy. Our experiments demonstrate that \name can be applied to the best defense models proposed in prior works, further enhancing privacy by reducing the prediction distribution gap.

% N=10, T=50
\begin{table}[htbp]
\centering
\large
\caption{\textbf{Performance of \name on CIFAR-100.} Here we directly adopted settings from previous papers.}
\scalebox{0.65}{
    \begin{tabular}{m{3cm}<{\centering}m{1.8cm}<{\centering}m{1.8cm}<{\centering}m{1.8cm}<{\centering}m{1.8cm}<{\centering}m{1.8cm}<{\centering}}
        \toprule
        \multicolumn{1}{c}{\textbf{Defenses}} & \textbf{Training Accuracy (\%)} & \textbf{Test Accuracy (\%)} & \textbf{Attack AUC (\%)} & \textbf{Attack Accuracy (\%)} \\
        \midrule
        \multicolumn{5}{c}{\textbf{SELENA}} \\
        \midrule
        Undefended / +\name & 100.0 / 100.0 & 78.4 / 78.4 & 76.2 / \textbf{64.8} & 74.5 / \textbf{63.0} \\
        SELENA / +\name     & 77.3 / 77.3   & 75.0 / 75.0 & 55.5 / \textbf{51.8} & 55.1 / \textbf{52.7} \\
        \midrule
        \multicolumn{5}{c}{\textbf{RelaxLoss}} \\
        \midrule
        Undefended / +\name & 67.3 / 67.3   & 34.1 / 34.1 & 72.4 / \textbf{69.5} & 67.9 / \textbf{67.0} \\
        RelaxLoss / +\name  & 52.9 / 52.9   & 36.6 / 36.6 & \textbf{59.5} / 59.7 & \textbf{57.6} / 57.7 \\
        \midrule
        \multicolumn{5}{c}{\textbf{HAMP}} \\
        \midrule
        Undefended / +\name & 91.4 / 91.4   & 58.7 / 58.7 & 70.8 / \textbf{67.9} & 67.5 / \textbf{66.1} \\
        HAMP / +\name       & 78.6 / 78.6   & 55.6 / 55.6 & 67.4 / \textbf{64.7} & 65.9 / \textbf{61.5} \\
        \bottomrule
    \end{tabular}
}
\label{table:sota}
\end{table}

\subsubsection{Evaluating \name with Publicly Available Diffusion Models }
% \label{sec:high-reso}
\label{sec:imagenetDM}

Although \name requires deploying a diffusion model, the model does not need to be trained on the same dataset as the target model. Therefore, defenders can leverage off-the-shelf diffusion models, including publicly available models trained on different datasets.

To demonstrate this, we evaluated \name on two high-resolution datasets: CelebA \cite{liu2015deep} and UTKFace \cite{zhang2017age} in Scenario 3 using a publicly available diffusion model trained on ImageNet by Dhariwal et al. \cite{dhariwal2021diffusion} with settings of $N=10$ and $T=200$. To the best of our knowledge, we are the first to conduct comprehensive MIA tests on these datasets using the original resolution without resizing, employing five attacks and seven defenses. The experimental results are shown in Figure \ref{fig:high-reso}. It can be observed that \name effectively enhances membership privacy in all cases without compromising model accuracy. On CelebA, \name decreases attack AUC by an average of 6.2\% and attack accuracy by 6.0\%. On UTKFace, \name decreases attack AUC by 9.5\% and attack accuracy by 9.9\% on average. On CelebA and UTKFace, the combinations of HAMP + \name and SELENA + \name achieved the best utility-privacy trade-offs, respectively. 

We also conducted experiments on three benchmark datasets—CIFAR-10, CIFAR-100, and SVHN-using the same public diffusion model. The results demonstrate that \name with the pretrained diffusion model provides effective defense across all three datasets. On average, \name reduces the attack AUC by 4.8\% and attack accuracy by 5.3\%. For more details, please refer to Figure \ref{fig:all-imagenet} in the Appendix. Additionally, the reconstructions generated by the diffusion model are nearly identical to the original samples, with only imperceptible visual differences, as shown in Figure \ref{Fig:examples-imagenet}.

\begin{figure*}[htbp]
  \centering
  % \hspace*{\fill}%
\begin{subfigure}{0.3\linewidth}
    \includegraphics[width=1\columnwidth]{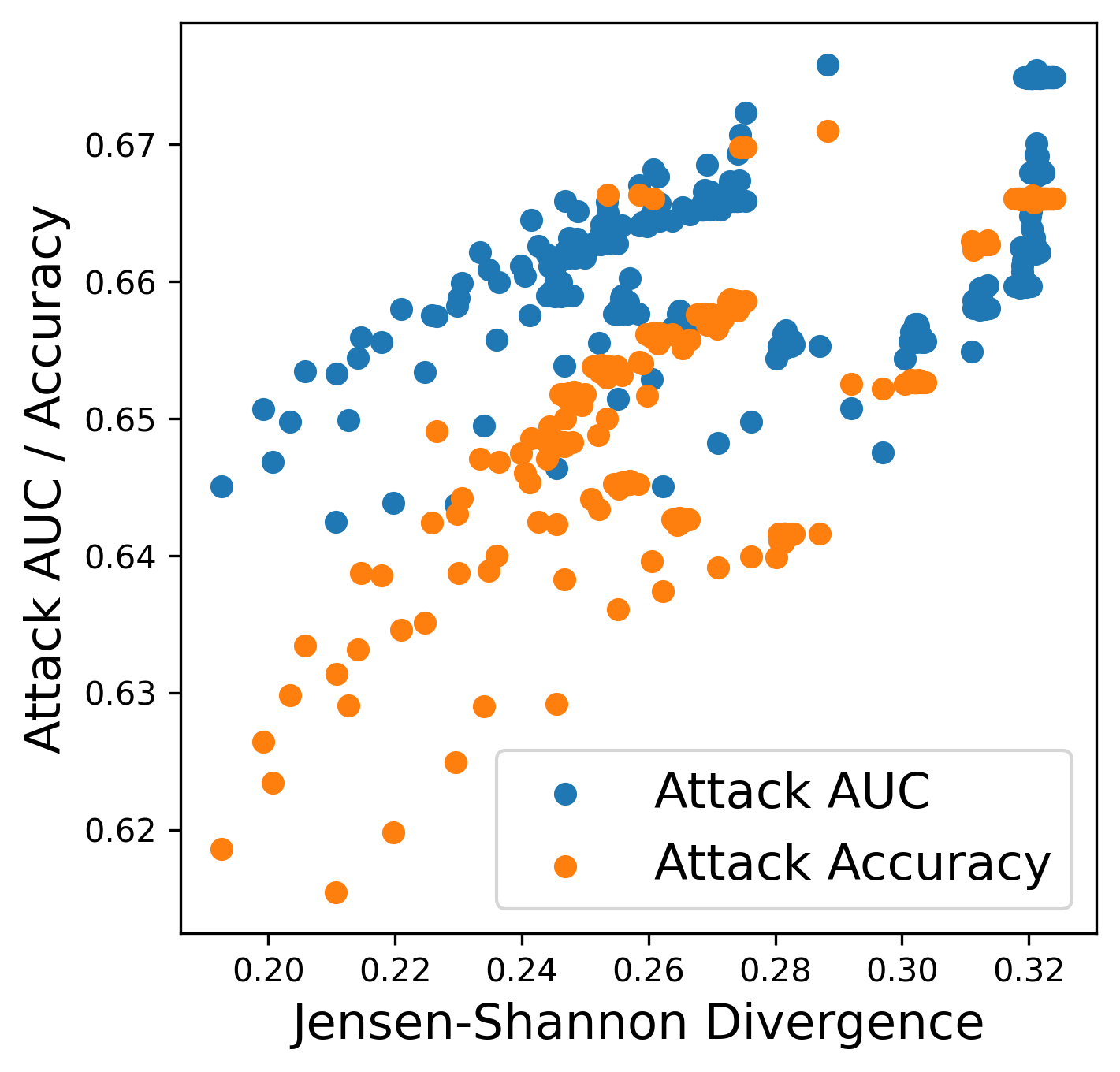}
    \label{fig:s1}
    \caption{CIFAR-10}
  \end{subfigure}
     \hfill
\begin{subfigure}{0.3\linewidth}
    \includegraphics[width=1\columnwidth]{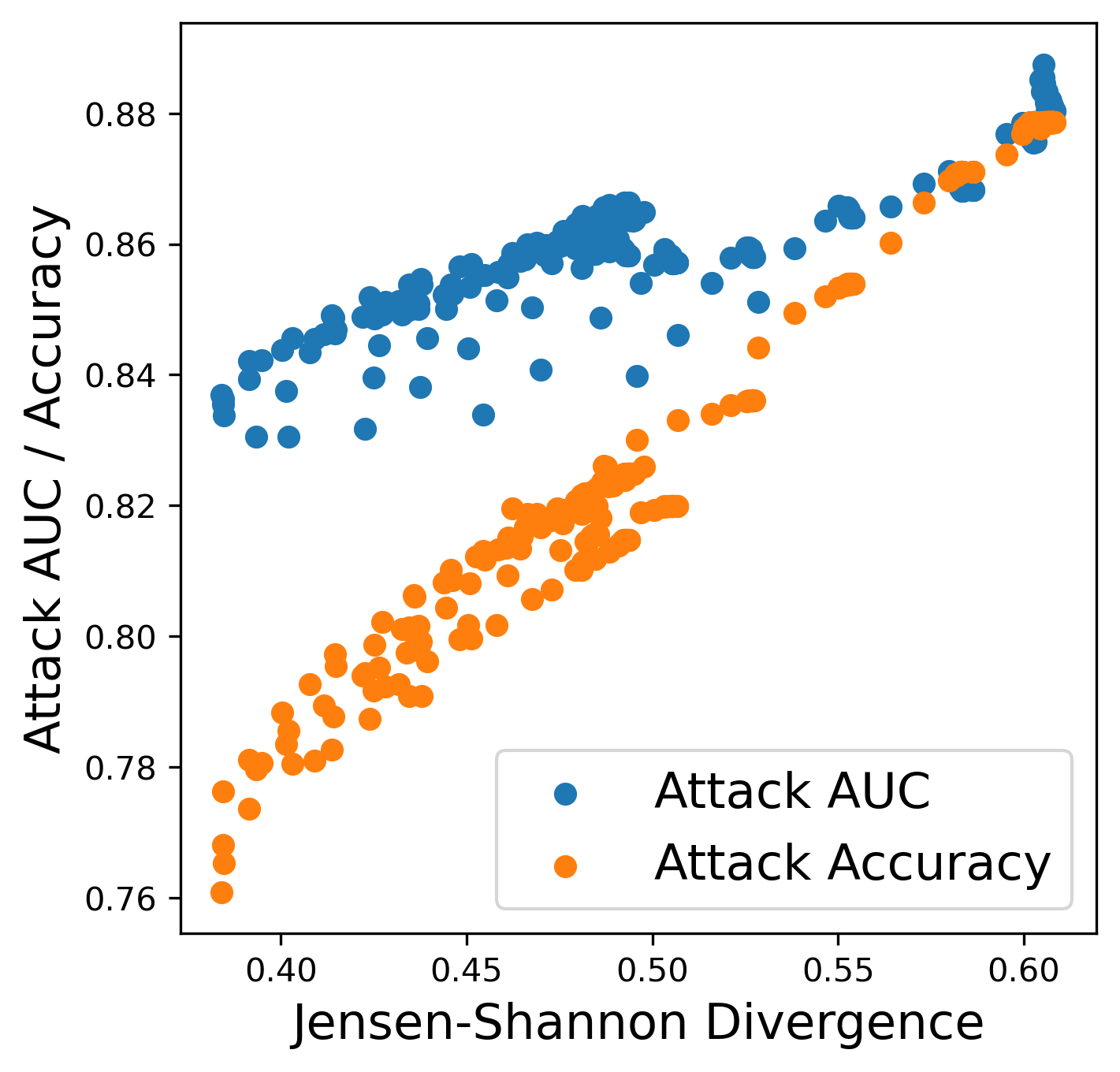}
    \label{fig:s2}
    \caption{CIFAR-100 }
  \end{subfigure}
     \hfill
\begin{subfigure}{0.31\linewidth}
    \includegraphics[width=1\columnwidth]{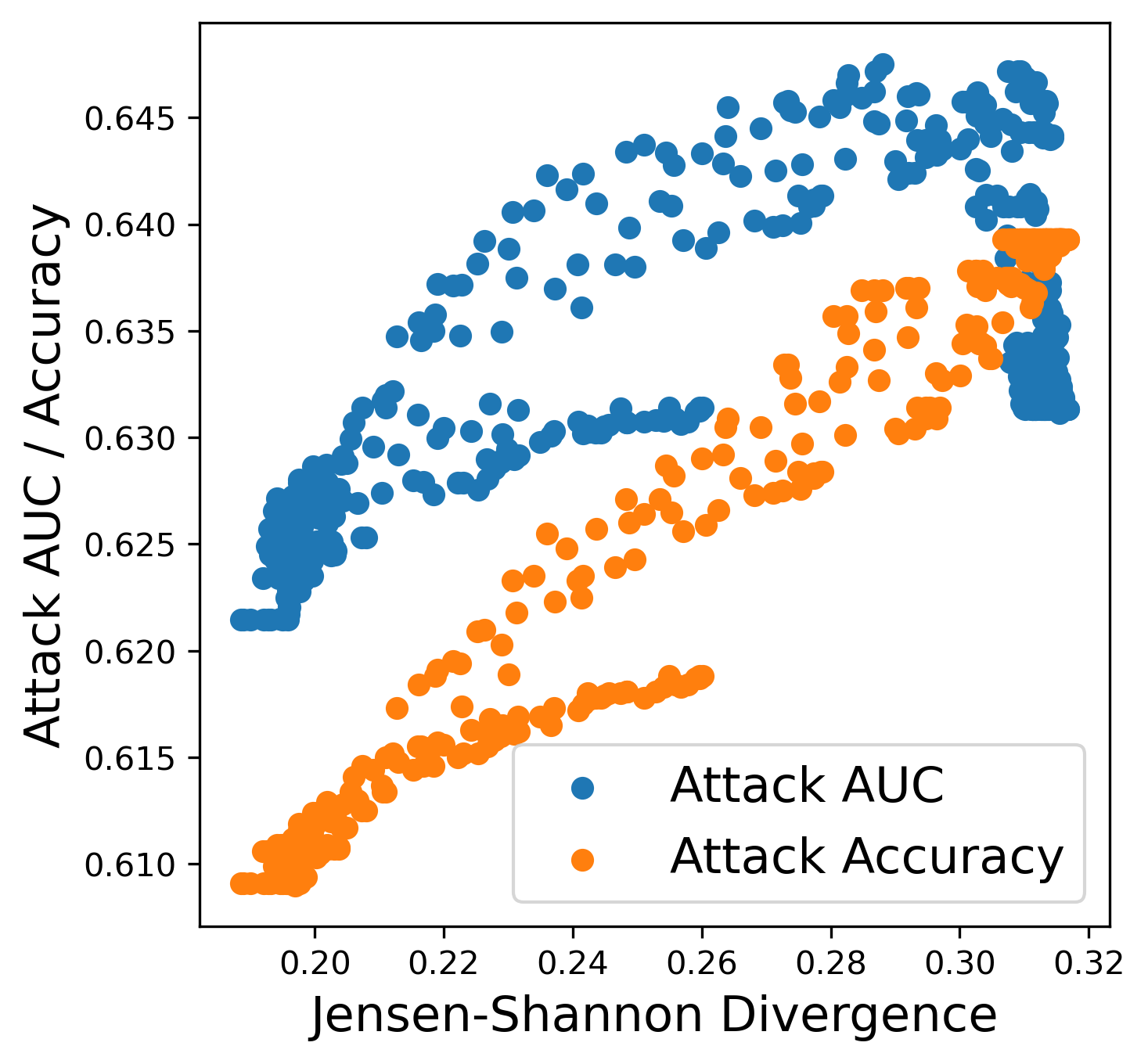}
    \label{fig:s2}
    \caption{SVHN}
  \end{subfigure}
  \caption{\textbf{Attack AUC and accuracy under different levels of Jensen-Shannon (JS) divergence between member and non-member prediction distributions.} We tested it on three dataset using ResNet18 and set diffusion steps $T=40$ and the number of reconstructions $N$ to 50.}
  \label{fig:js_attack}
  % \hspace*{\fill}%
\end{figure*}

\begin{figure}[htbp]
  \centering
  % \hspace*{\fill}%
    \includegraphics[width=0.8\columnwidth]{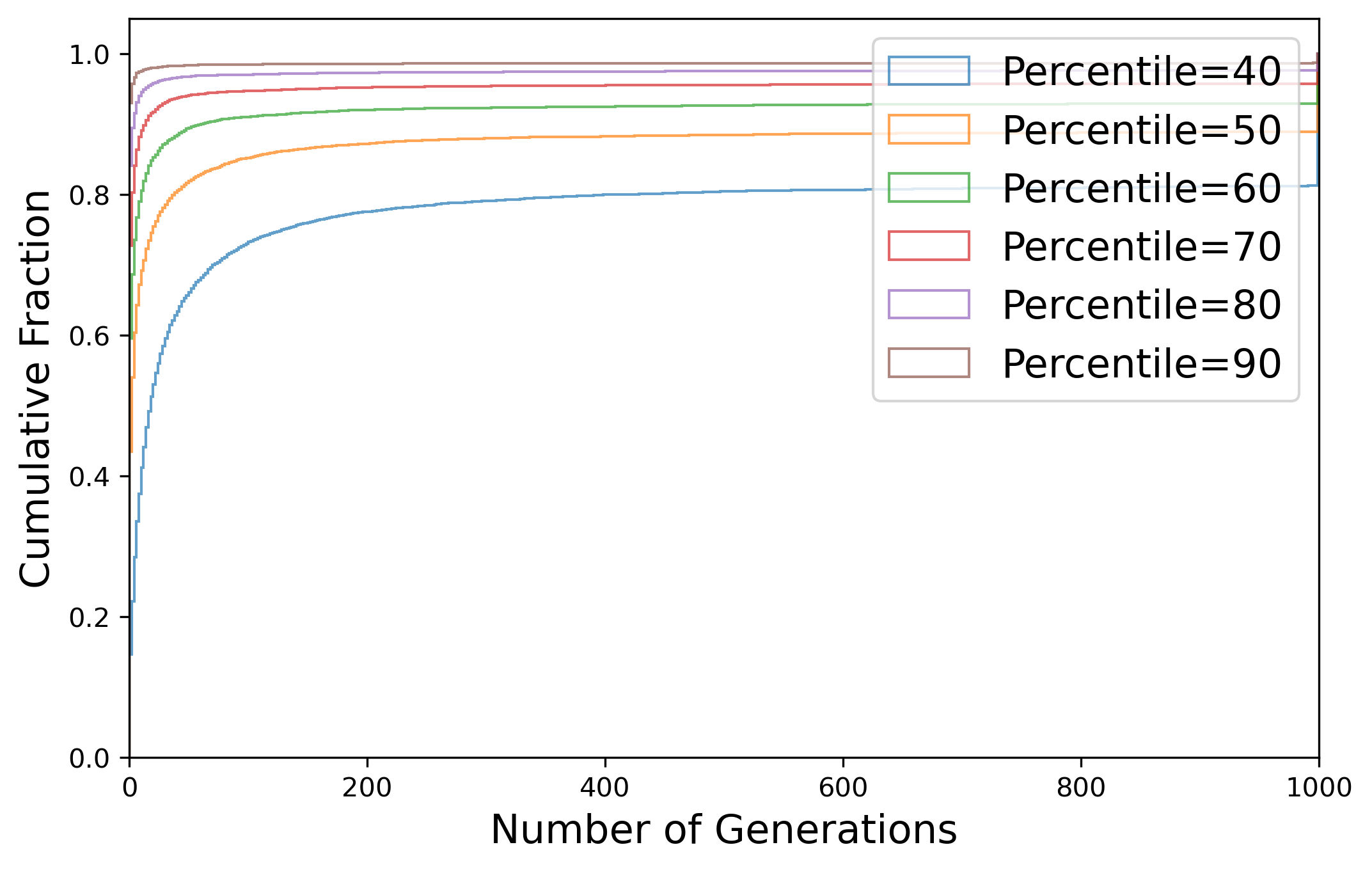}
     \hfill
  \caption{\textbf{Cumulative Distribution Function (CDF) of the number of generations required for reconstructions to fall within different intervals when $T=100$ on CIFAR-100.} The 'percentile' indicates the anticipated probability of reconstructions falling within the given interval, as determined by analyzing the samples available to the defender.}
  \label{fig:interv}
  % \hspace*{\fill}%
\end{figure}

\subsection{Ablation study} 
\label{section:ablation study}
As introduced in Section \ref{sec:methodology}, \name employs a diffusion model to reconstruct samples, protecting the membership privacy of the original samples. We generate multiple samples and select the prediction of the most suitable sample as the final output based on the defense objectives and the information available to the defender. This process involves considering several factors, including the choice of interval, the number of reconstructions $N$ generated for each sample, and the diffusion step $T$. In this section, we delve further into discussing the impact of the parameters and sample selection strategy.

\subsubsection{Sample Selection Strategies}
\label{sec:keep generating}

In Scenario 1, the main idea of \name involves setting an interval within which the prediction logits of both members and non-members are encouraged to fall. This strategy aims to align the prediction distributions of members and non-members towards a range in the middle. It is important to note that \name uses a fixed number of reconstructions $N$ and diffusion steps $T$ as the basis for selecting the optimal interval. Given the fixed $N$, not all samples can fall within the chosen interval. For these outliers, we select the sample that is closest to the interval.

An alternative approach might involve continuously generating reconstructions for each sample after setting the interval, to force their predictions into this range. However, we found this method to be highly inefficient. Figure \ref{fig:interv} illustrates the cumulative distribution of samples falling within various intervals as the number of reconstructions increases. It reveals that if a sample does not fall within the interval in its initial generation, it is unlikely to do so in subsequent iterations. Therefore, continuously generating new samples is not a practical solution.

In terms of interval selection, we tested various intervals and opted for the one where the JS divergence between members' and non-members' prediction logits is minimized. This approach is predicated on our observation that the JS divergence of prediction logits between member and non-member samples correlates significantly with attack performance. This correlation has also been noted by He et al.~\cite{he2022membershipdoctor} in their research. 

Figure \ref{fig:js_attack} depicts attack AUC and accuracy on three datasets in relation to the different JS divergence between member and non-member prediction distributions when \name is solely applied. Each point in the figure corresponds to a selectable interval, where choosing different intervals results in varying JS divergences. A key observation is the negative correlation between JS divergence and privacy protection – smaller JS divergences typically indicate stronger membership privacy protection. These results substantiate the effectiveness of our interval selection strategy.

\subsubsection{Effect of Hyperparameters}
\label{sec:hyper}

\begin{figure}[htbp]
  \centering
  % \hspace*{\fill}%
  \begin{subfigure}{0.8\linewidth}
    \includegraphics[width=1\columnwidth]{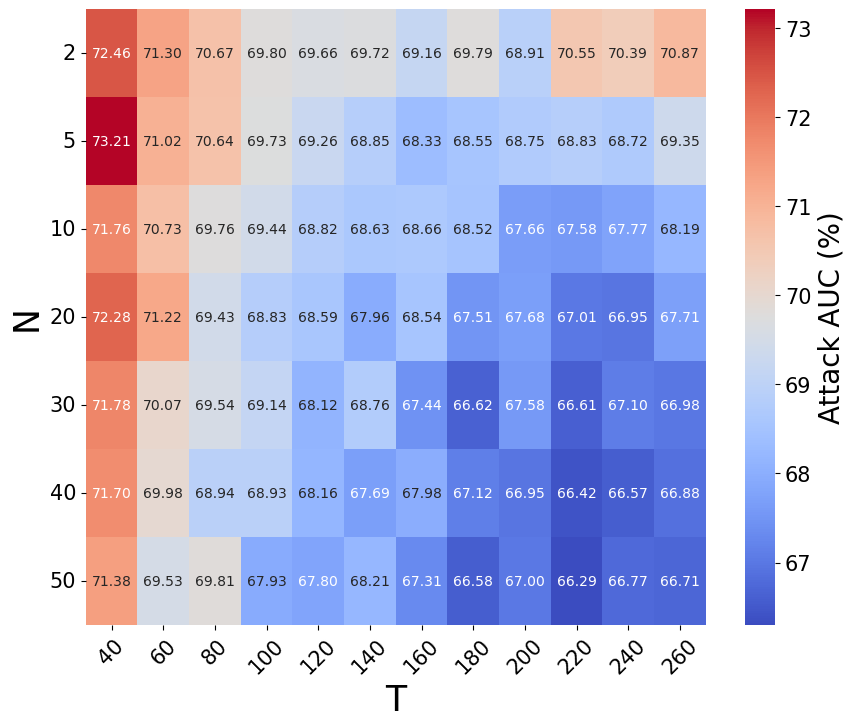}
    \label{fig:short-a}
    \caption{Attack AUC}
  \end{subfigure}
    \hfill
  \hspace{1em}%
    \begin{subfigure}{0.8\linewidth}
    \includegraphics[width=1\columnwidth]{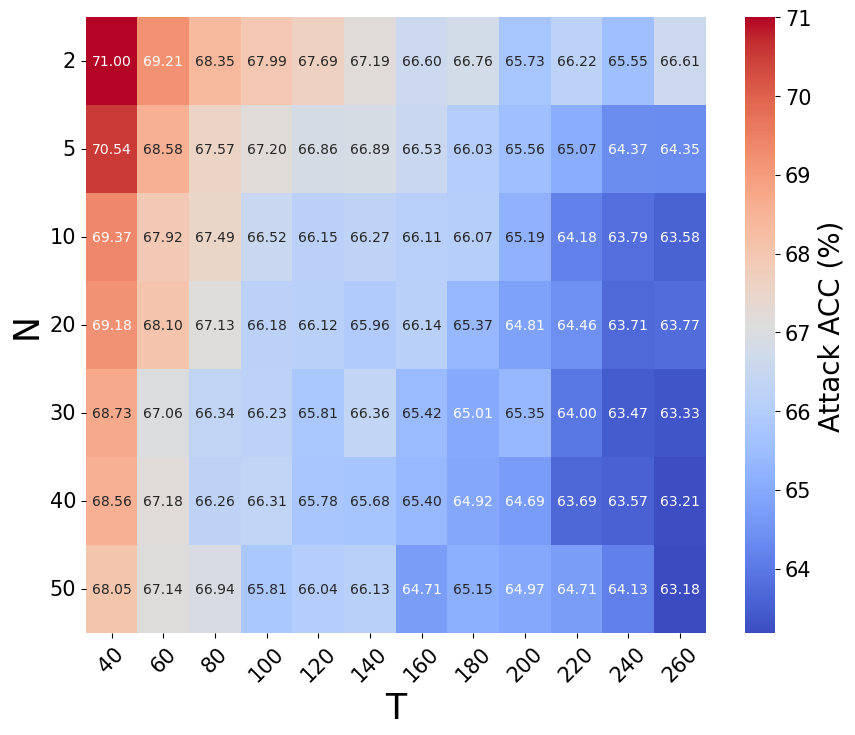}
    \label{fig:short-a}
    \caption{Attack Accuracy}
  \end{subfigure}
  \caption{\textbf{Attack AUC and attack accuracy against \name with different numbers of generated samples \( N \) and Diffusion Steps \( T \) on average across three datasets.} We varies \( N \) from 2 to 50 and \( T \) from 40 to 260.}
  \label{fig:heatmap}
  % \hspace*{\fill}%
\end{figure}

We then varied the values of $N$ and $T$ to observe the changes in defense performance. Note that \name does not alter the prediction labels, and therefore, does not impact the model's test accuracy.

An increase in $T$ leads to a more substantial alteration of the original sample, thereby enlarging the divergence between the generated samples and the original ones. A higher $N$ enhances the pool of alternative images, enriching the selection process for our sample selection strategy. This, in turn, mitigates randomness and enables a more consistent selection of suitable replacement samples.

As depicted in Figure \ref{fig:heatmap}, both the attack AUC and accuracy decrease as $T$ and $N$ increase and a more pronounced decrease is observed with larger $T$ values. However, increasing $T$ and $N$ will also increase the time of sample generation, potentially increasing the latency of inference. This introduces a trade-off: optimizing the defense's effectiveness versus minimizing inference delay. The balance between improved privacy and reduced latency should be tailored to the requirements of the specific task in practice.

\section{Discussions}
\label{sec:dis}

\subsection{Overhead of Different Defenses}
\label{sec:infer_overhead}

Our experiments include seven defense methods, where HAMP, SELENA, RelaxLoss, AdvReg, and DPSGD are training-phase defenses, and MemGuard and \name operate during the inference phase. We evaluated the overhead of these defenses during training and testing based on their deployment phase. The training and inference overhead were measured on a single NVIDIA RTX-8000 GPU with 40GB of memory. Here, we primarily discuss the inference overhead, with Appendix \ref{sec:training_overhead} reporting the results on training overhead comparison.

\begin{table}[htbp]
\centering
\caption{\textbf{Inference overhead comparison of different defenses.}}
\begin{tabular}{lccc}
\toprule
\textbf{Defenses} & \textbf{CIFAR-10 (ms)} & \textbf{CIFAR-100 (ms)} & \textbf{SVHN (ms)} \\
\midrule
Undefended & 26 & 24 & 24 \\
Memguard & 732 & 612 & 586 \\
\name & 82 & 82 & 81 \\
\bottomrule
\end{tabular}
\label{table:infer_overhead}
\end{table}

MemGuard generates adversarial noise for each sample's confidence vector to deceive the attacker’s model, necessitating the solving of complex optimization problems for each sample. In contrast, \name, which involves calculating optimal intervals for samples before inference, is a one-time task that does not add overhead during the inference phase. The primary overhead of \name comes from processing the original samples with a diffusion model. To enhance this process, we utilize Denoising Diffusion Implicit Models (DDIM) \cite{song2021denoising}, which streamline the diffusion process by providing a means to skip sampling steps and directly estimate the denoised image from noisy observations, significantly reducing computational demands. Table \ref{table:infer_overhead} reports the inference times for the undefended model, MemGuard, and \name on three benchmark datasets. We measured the inference overhead by performing inference on 100 random member and non-member samples and taking the average. As shown in Table \ref{table:infer_overhead}, \name outperforms MemGuard in terms of inference phase running time and only adds an average of 57ms to the inference time compared to the undefended model, highlighting \name's efficiency and low overhead upon deployment.

\subsection{Pre-training MIA Defenses}
\label{sec:defense_stage}
In Section \ref{sec:intro}, we categorized MIA defenses into three classes based on where they kick in during the ML pipeline: training phase defenses, pre-inference defenses, and post-inference defenses.
One could also consider techniques that work during ML \emph{pre-training} as another class of MIA defense, e.g., techniques like dataset condensation (DC) \cite{dong2022privacy}. 
However, we argue that these techniques tend to mitigate MIA more as a side effect than as a primary focus. Furthermore, their effectiveness in defending against MIA has been  questioned in recent work {\cite{carlini2022no}}.

Also, one could consider these pre-training techniques as a sub-class of our training phase defenses. In fact, some training stage defenses, such as SELENA and HAMP, already incorporate specifically designed dataset processing. Therefore, we did not make pre-training  a separate class of MIA defenses. 
% However, as more defenses developed, exploring a finer-grained division of defense phases is an interesting direction for future work.
\section{Related Works}
\label{sec:formatting}
A \emph{membership inference attack}~\cite{shokri2017membership} aims to determine if a specific sample was in a model's training data, posing risks of sensitive individual information leakage. This section overviews various such attacks and defenses, highlighting their diversity across scenarios.
%-------------------------------------------------------------------------
\subsection{Membership Inference Attacks}

Shokri et al.~\cite{shokri2017membership} introduced a \emph{black-box} MIA that employs a shadow training technique to train an attack model to differentiate the model's output, categorizing it as either a member or non-member. In a different approach, Salem et al.~\cite{salem2018ml} streamlined the process by training only a single shadow model, assuming the attacker lacks access to similar distribution data as the training dataset, yet still achieving notable effectiveness. Expanding on these concepts, Nasr et al.~\cite{nasr2019comprehensive} presented a \emph{white-box} MIA targeting ML models. For each data sample, they computed the corresponding gradients over the parameters of the white-box target classifier, utilized as features of the data sample for membership inference.

Choquette-Choo et al.~\cite{choquette2021label} developed a \emph{label-only} MIA concept, where the target model reveals only the predicted label. Their attack's efficacy relies on the model's increased resilience to perturbations like augmentations and noise in the training data. Complementing this, Li et al.~\cite{li2021membership} presented two specific label-only MIAs: the \emph{transfer-based} MIA and the \emph{perturbation-based} MIA. Remarkably, these label-only attacks achieve a balanced accuracy on par with that of the shadow-model strategies.

Song~\cite{song2021systematic} employs a \emph{modified entropy} measure, using shadow models to approximate the distributions of entropy values for members and non-members across each class. Essentially, the attacker conducts a hypothesis test between the distributions of (per-class) members and non-members, given a model $f$ and a target sample $(x,y)$. Yeom et al.~\cite{yeom2018privacy} proposed a \emph{loss-based} membership inference attack, leveraging the tendency of machine learning models to minimize training loss. This attack identifies training examples by observing lower loss values. Carlini et al.~\cite{carlini2022membership} forged ahead with the development of a \emph{Likelihood Ratio-based} attack (LiRA). This attack has demonstrated success in outperforming previous attacks, especially at low False Positive Rates. In the LiRA approach, an attacker trains $N$ shadow models using samples from distribution $D$. Half include the target point $(x,y)$, and half do not. Two Gaussian fits are applied to these model confidences. The membership inference is then deduced for $(x,y)$ in the target model using a Likelihood-ratio test based on these fits.

\subsection{Existing Defenses}

Initial research on defenses against MIA showed that certain \emph{regularization} techniques, like dropout~\cite{srivastava2014dropout}, can curb overfitting, resulting in modest privacy enhancements in neural networks~\cite{shokri2017membership}. Another method, \emph{early stopping}~\cite{caruana2000overfitting}, also serves to prevent model overfitting, potentially reducing MIA accuracy, albeit at the cost of compromising model utility.

Several studies have proposed defenses during the training phase. Nasr et al.~\cite{nasr2018machine} proposed an \emph{adversarial regularization} technique, a min-max game-based training algorithm aiming to reduce training loss and increase MIA loss. Shejwalkar~\cite{shejwalkar2021membership} introduced a defense against MIAs through \emph{knowledge distillation}, transferring knowledge from an undefended private-dataset-trained model to another using a public dataset. Tang et al.~\cite{tang2022mitigating} proposed a knowledge distillation-based defense balancing privacy and utility. They partitioned the training dataset into K subsets, trained K sub-models on each, and used these to train a separate public model with scores from non-training samples. Chen~\cite{chen2022relaxloss} developed \emph{RelaxLoss}, a training scheme balancing privacy and utility by minimizing loss distribution disparities to reduce membership privacy risks.

Some other defenses are applied separately from the training phase. Jia et al.~\cite{jia2019memguard} proposed \emph{MemGuard}, a method that operates in two stages. It first crafts a noise vector to transform confidence scores into adversarial examples under utility-loss constraints. Then, it integrates this noise into the confidence score vector based on a derived analytical probability. Chen~\cite{chen2023overconfidence} introduced \emph{HAMP}, a defense mechanism encompassing both training and test-time defenses. Its training component aims to lower model confidence on training samples, countering the overconfidence induced by hard labels in standard training.

\emph{Differential privacy (DP)}~\cite{dwork2006differential, dwork2006calibrating} is a prominent method extensively employed to offer theoretical privacy guarantees in ML models. Within this framework, noise can be introduced to both the objective function~\cite{iyengar2019towards} and gradients~\cite{abadi2016deep, song2013stochastic}. While DP provides robust privacy assurances, it has been observed to significantly compromise utility~\cite{nasr2018machine}.

\section{Conclusion}

In this work, we introduced \name, a novel defense method that addresses the challenge of membership inference attacks (MIAs) in machine learning (ML). \name effectively diminishes the distinction in prediction behaviors between members and non-members through input reconstruction using diffusion models. By generating multiple reconstructions and selectively utilizing predictions based on defined criteria, \name significantly narrows the prediction distribution gaps exploited in MIAs.

We categorized defenses into three deployment phases and, for the first time, proposed integrating different defenses to enhance overall protection. \name can be cascaded in a plug-and-play manner with other defenses. Although our strategy requires the use of a diffusion model, it does not necessarily require the diffusion model to be specifically trained for defense; instead, off-the-shelf diffusion models can be used—even those trained on datasets different from the target model's training dataset—thereby demonstrating considerable flexibility and effectiveness. By combining \name with others, we effectively address both the train-to-test accuracy gap and the prediction distribution gap. Our extensive experiments across multiple datasets have validated \name’s efficacy in enhancing membership privacy without sacrificing model utility—preserving both accuracy and the meaningfulness of confidence vectors.

% In conclusion, our research contributes to the field of privacy protection in ML by providing a robust and flexible solution to counter MIAs. Future work could further explore optimizing the selection process in our defense strategy and investigating its effectiveness against a wider range of privacy threats in ML.

\section*{Acknowledgements}
The work was supported in part by the NSF grant 2131910, and by DARPA under Grant DARPA-RA-21-03-09-YFA9-FP-003. 

\bibliographystyle{IEEEtran.bst}
\bibliography{main}

\appendices % Start the appendices section
% Appendix A

\section{Appendix}

\begin{figure*}[htbp]
  \centering
  % \hspace*{\fill}%
  \begin{subfigure}{1\linewidth}
    \includegraphics[width=0.3\columnwidth]{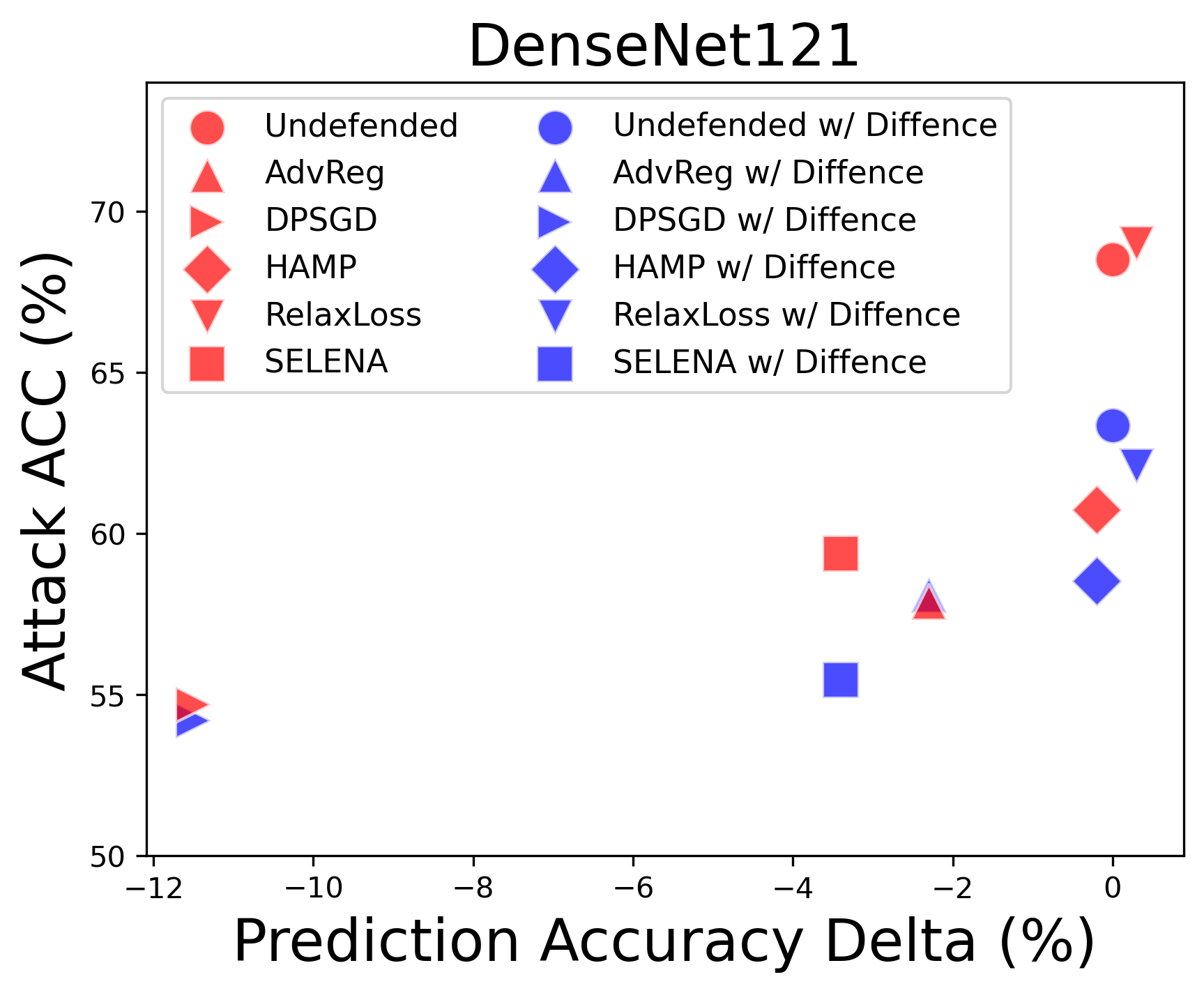}
     \hfill
    \includegraphics[width=0.3\columnwidth]{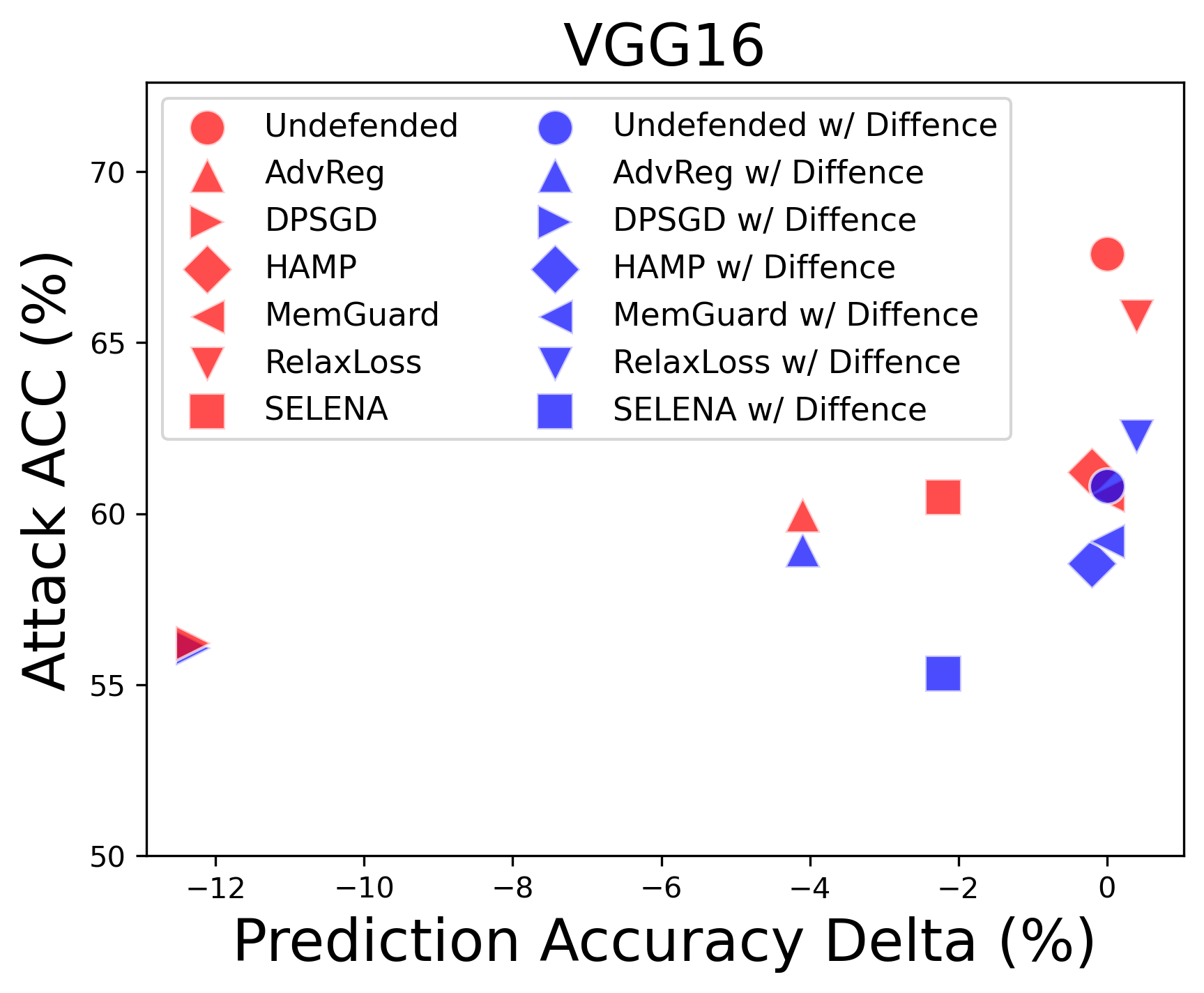}
    \hfill
    \includegraphics[width=0.3\columnwidth]{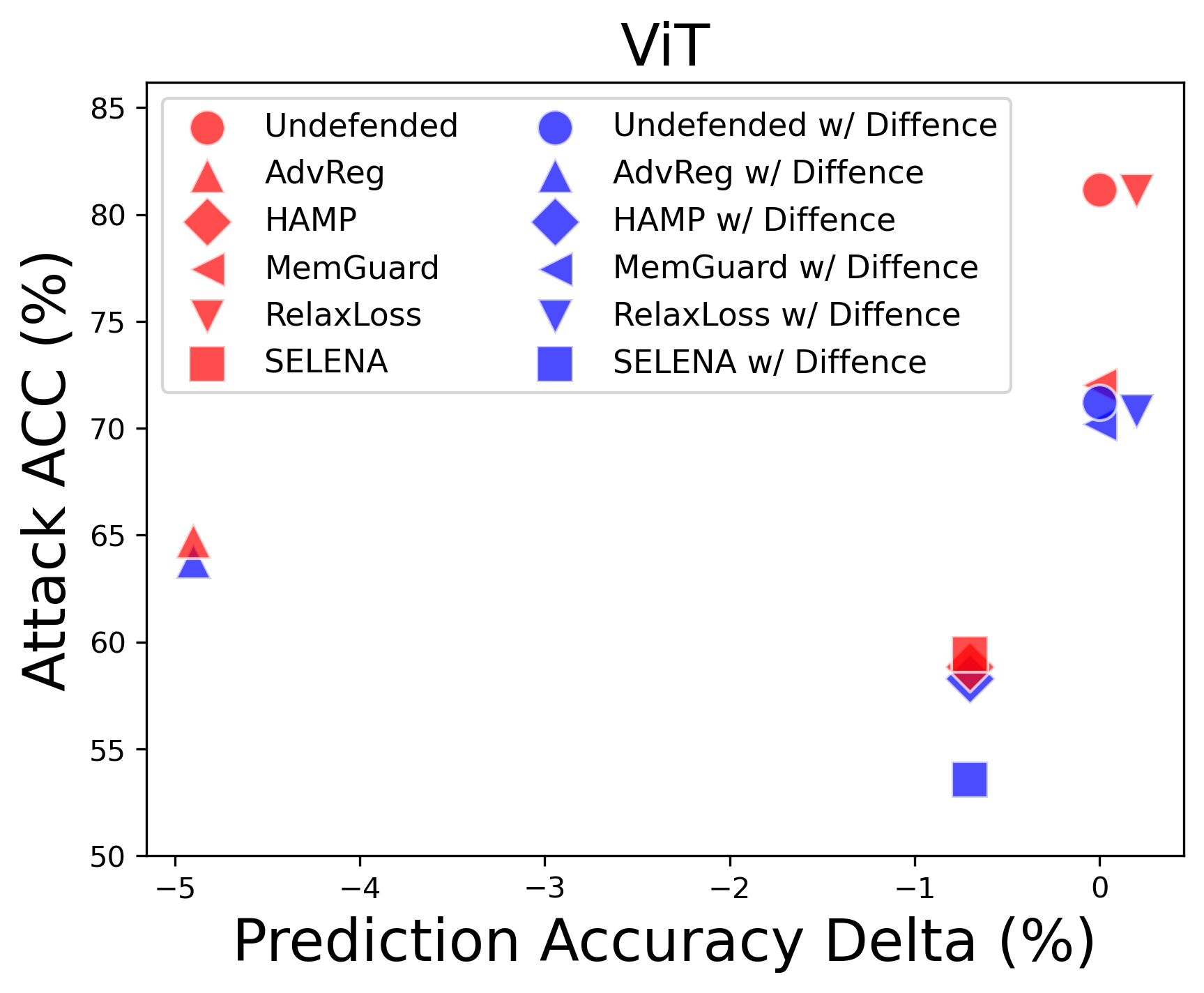}

    \label{fig:short-a}
    \caption{ Attack Accuracy}
  \end{subfigure}

  \hspace{1em}%
    \begin{subfigure}{1\linewidth}
    \includegraphics[width=0.3\columnwidth]{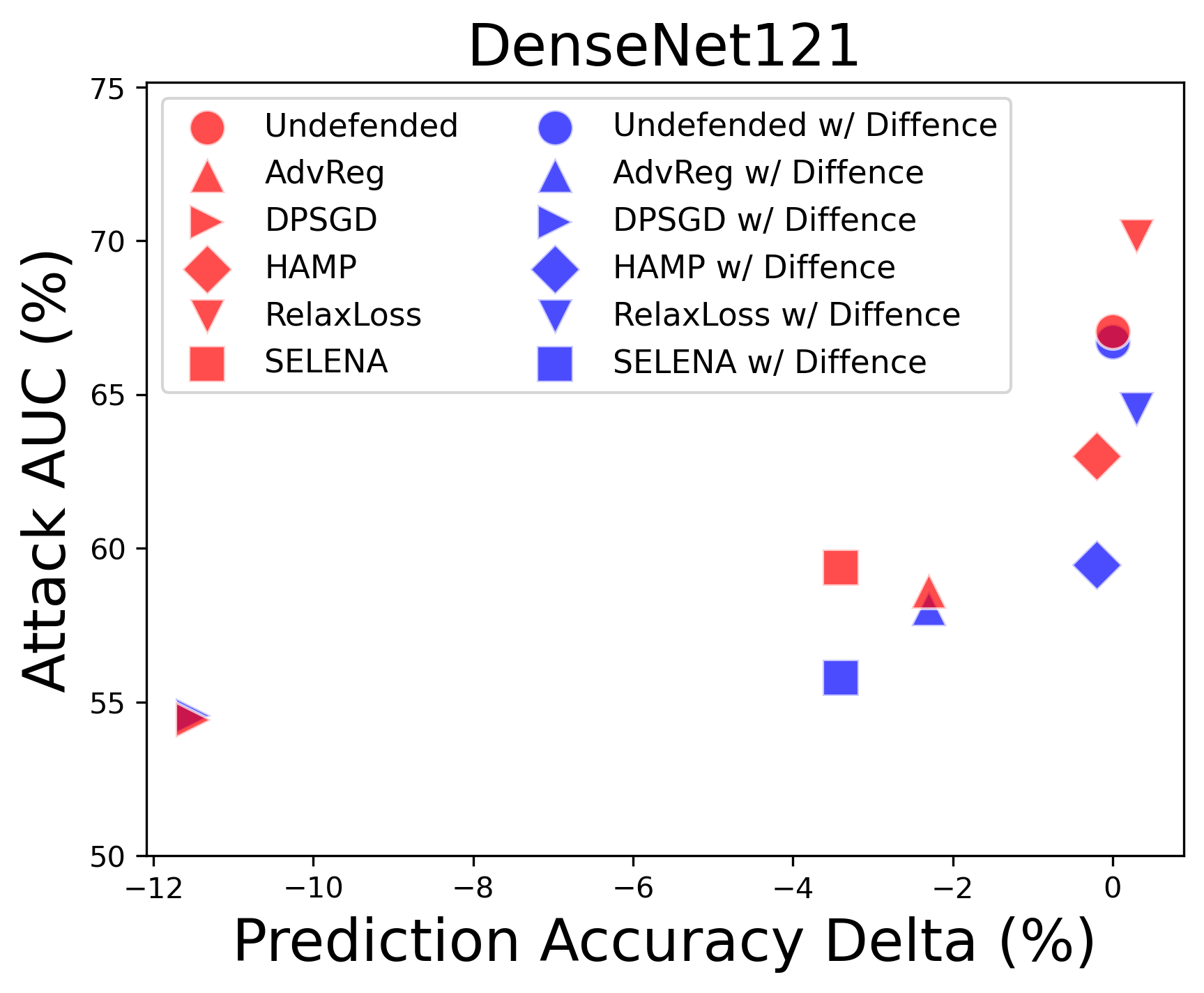}
     \hfill
    \includegraphics[width=0.3\columnwidth]{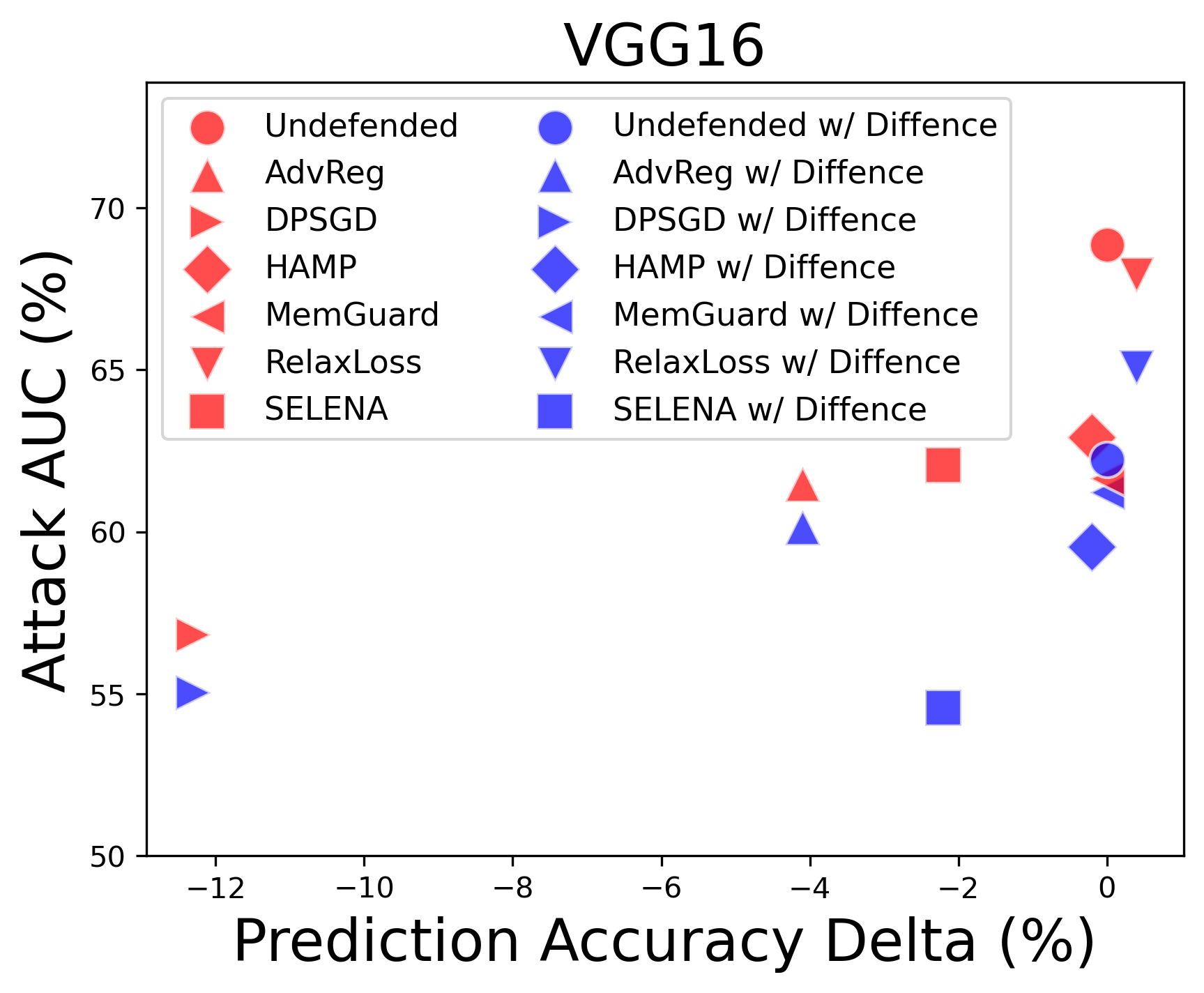}
    \hfill
    \includegraphics[width=0.3\columnwidth]{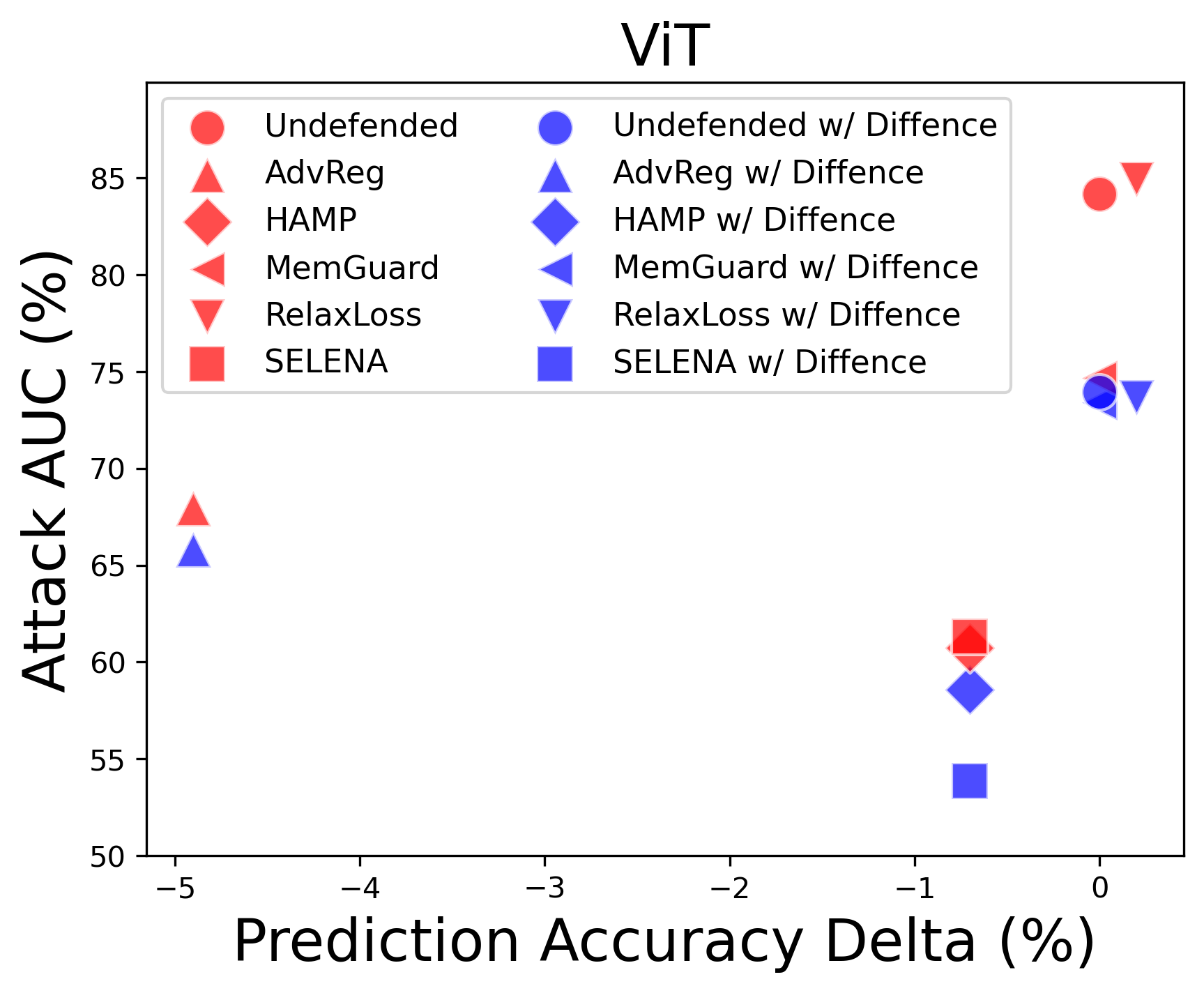}
 
    \label{fig:short-a}
    \caption{ Attack AUC}
  \end{subfigure}
  \caption{\textbf{Attack accuracy and AUC on CIFAR-10 against three models with three different architectures.} We report the highest attack accuracy and attack AUC across all attacks. The prediction accuracy delta indicates the prediction accuracy gap compared to the undefended models, with negative numbers indicating a decrease in model accuracy.}
  \label{fig:all-dn}
  % \hspace*{\fill}%
\end{figure*}

\subsection{Experimental Results on Different Model Architectures}
\label{sec:densenet}

This section presents our additional experimental results on different model architectures, including DenseNet121 \cite{huang2017densely}, VGG16 \cite{simonyan2014very}, and ViT \cite{dosovitskiy2020image}\footnote{ViT contains layers that are incompatible with the DPSGD implementation in the Opacus library, so we omitted DPSGD from the experiments on ViT.}. The results are illustrated in Figure \ref{fig:all-dn}. Consistent with our findings on ResNet18, \name effectively reduces both attack accuracy and attack AUC without impacting model accuracy. For example, \name reduces the attack AUC on average by 5.3\% and the attack accuracy by 4.6\% on average on the VGG16 model. On the ViT model, \name achieves reductions of 7.7\% in attack AUC and 6.6\% in attack accuracy on average. Additionally, it can be observed that combining \name with recent defenses such as SELENA and HAMP achieves the best privacy-utility trade-off.

\begin{figure*}[htbp]
  \centering
  % \hspace*{\fill}%
  \begin{subfigure}{1\linewidth}
    \includegraphics[width=0.3\columnwidth]{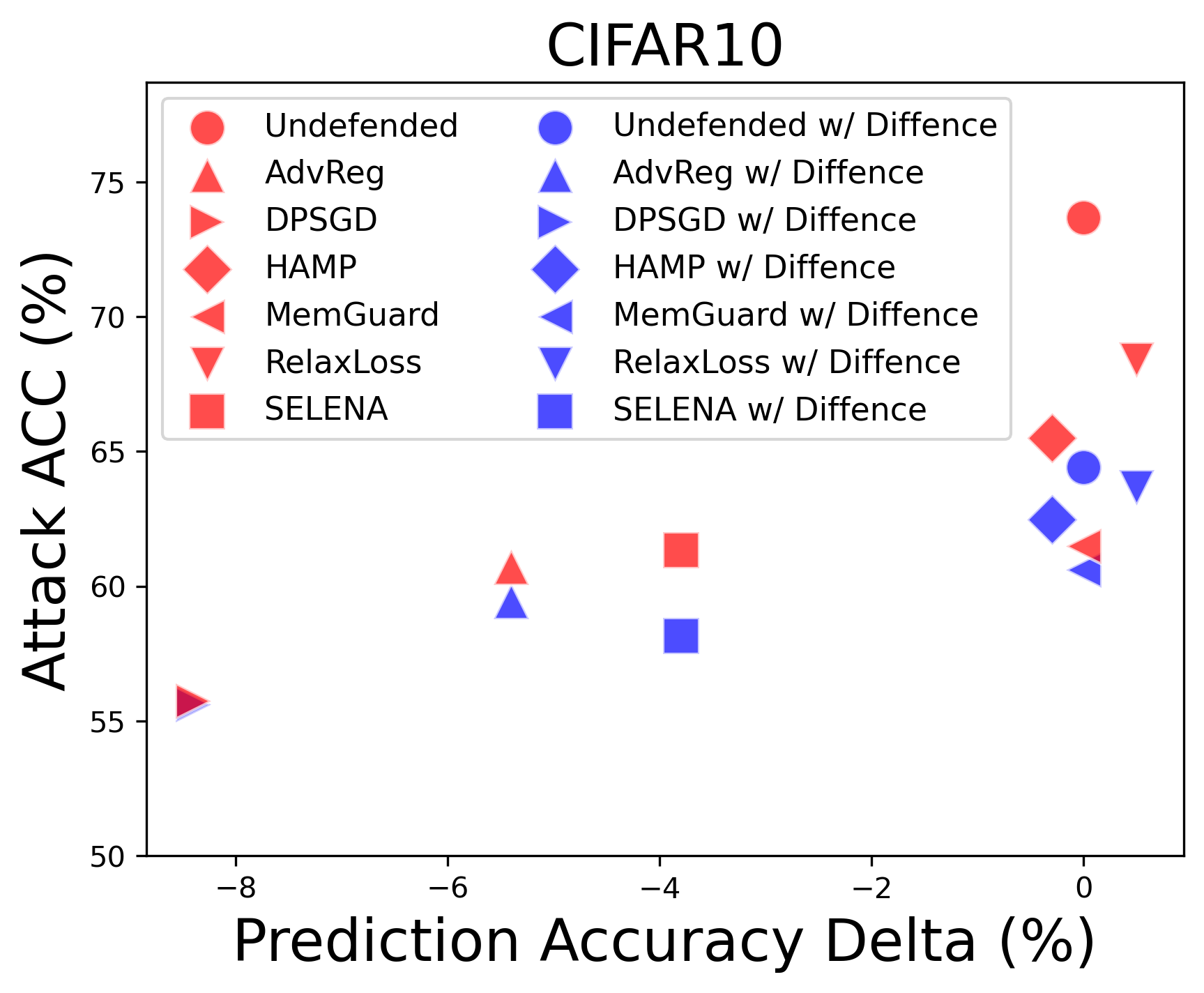}
     \hfill
    \includegraphics[width=0.3\columnwidth]{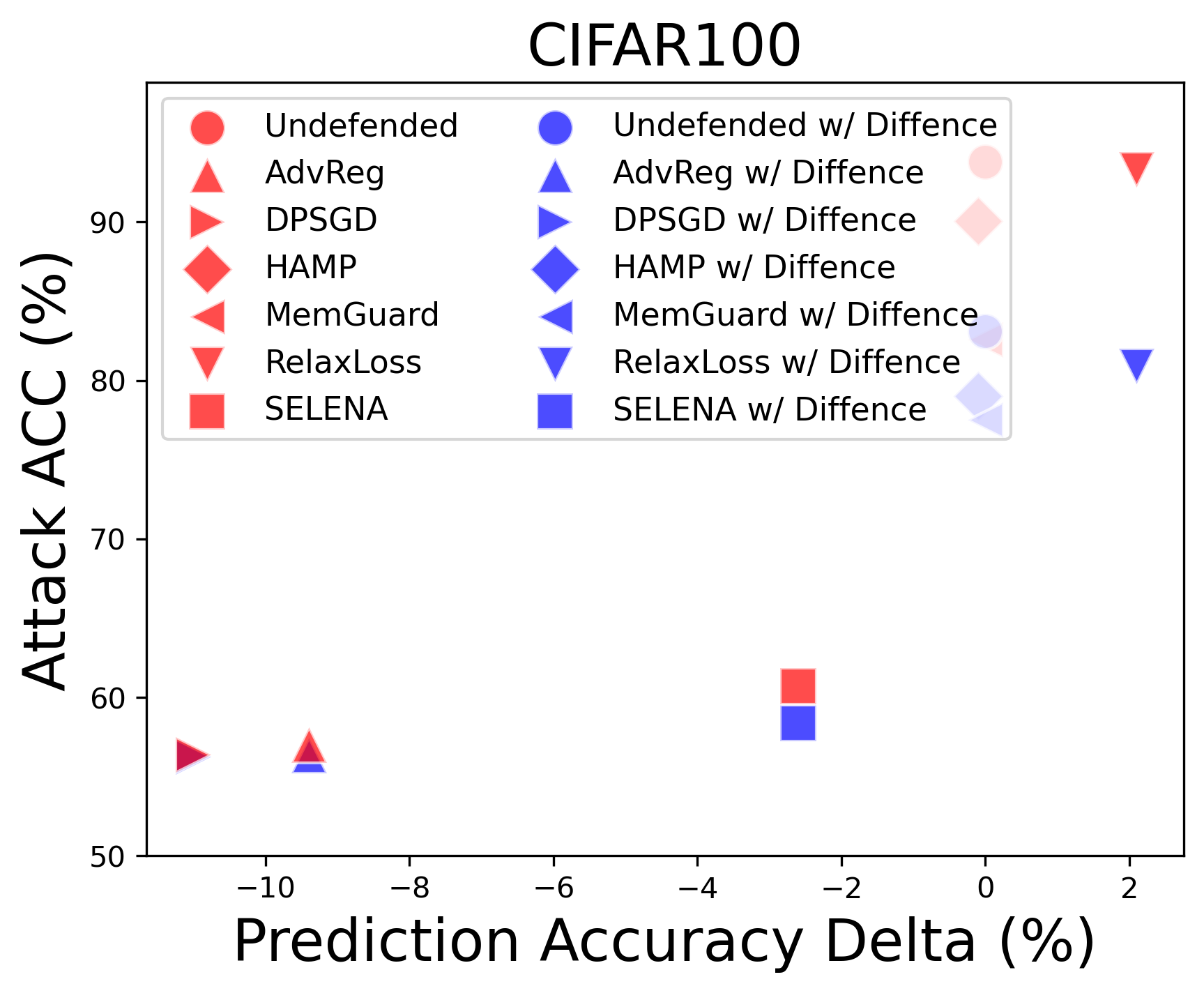}
    \hfill
    \includegraphics[width=0.3\columnwidth]{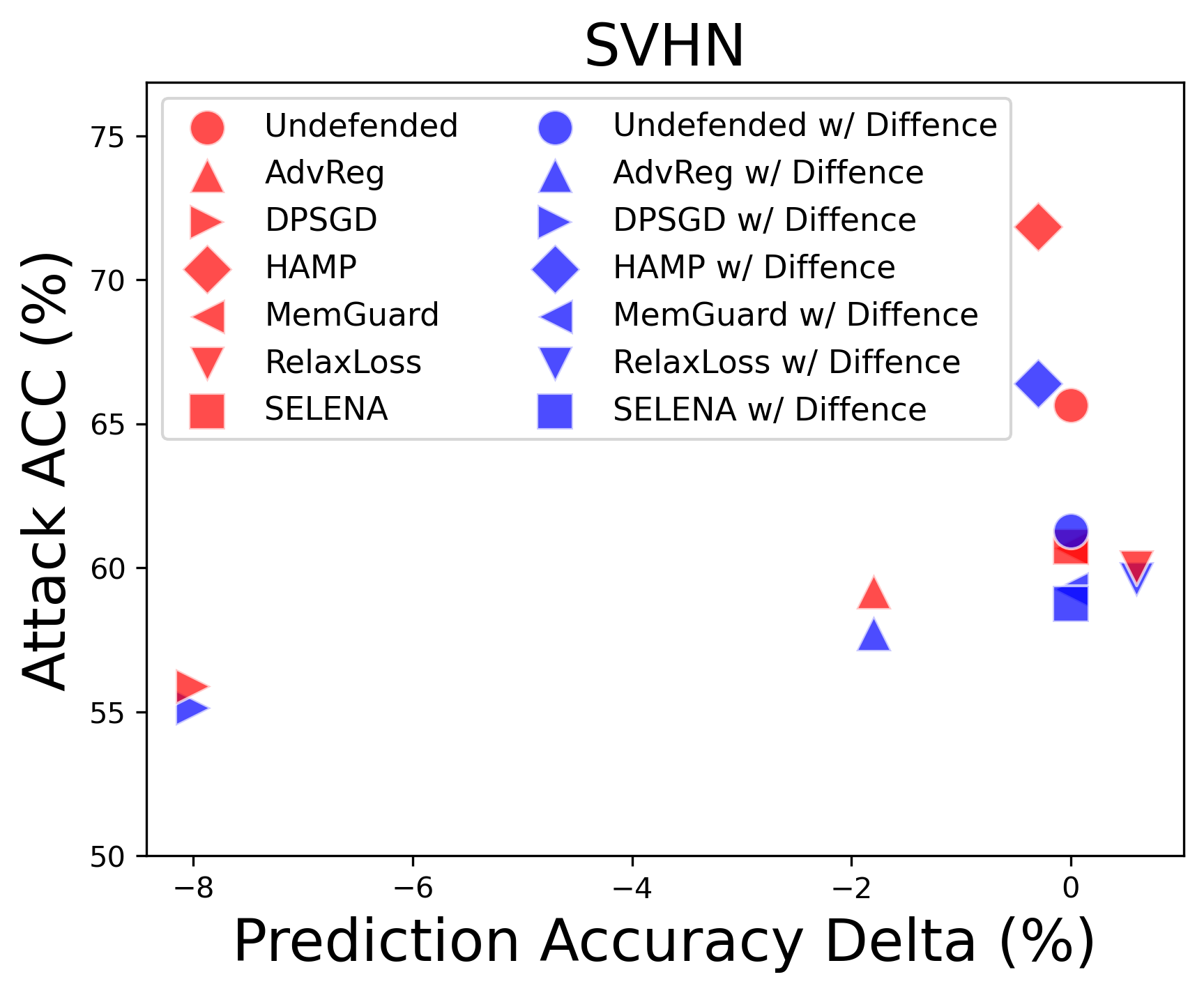}

    \label{fig:short-a}
    \caption{ Attack Accuracy}
  \end{subfigure}

  \hspace{1em}%
    \begin{subfigure}{1\linewidth}
    \includegraphics[width=0.3\columnwidth]{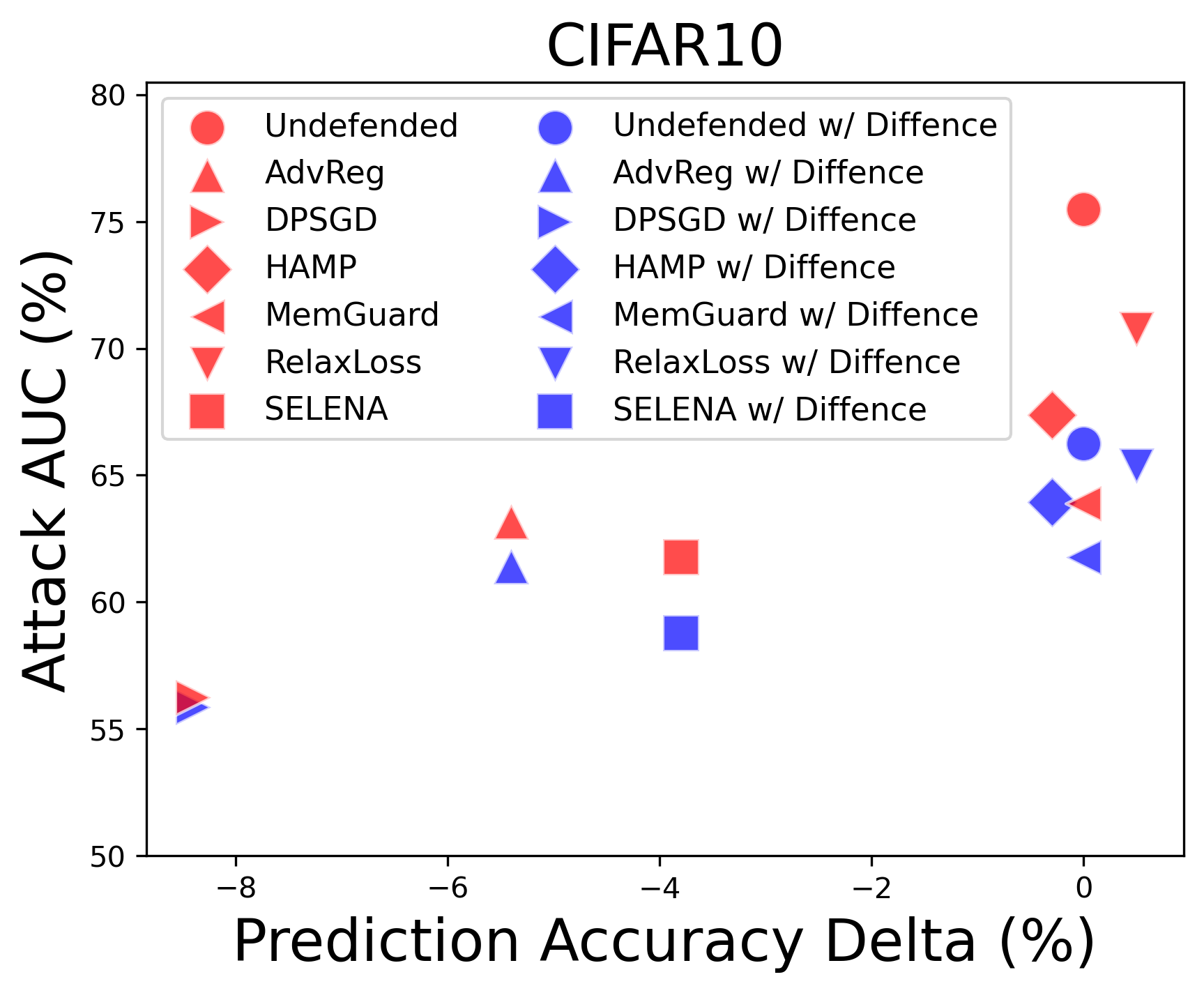}
     \hfill
    \includegraphics[width=0.3\columnwidth]{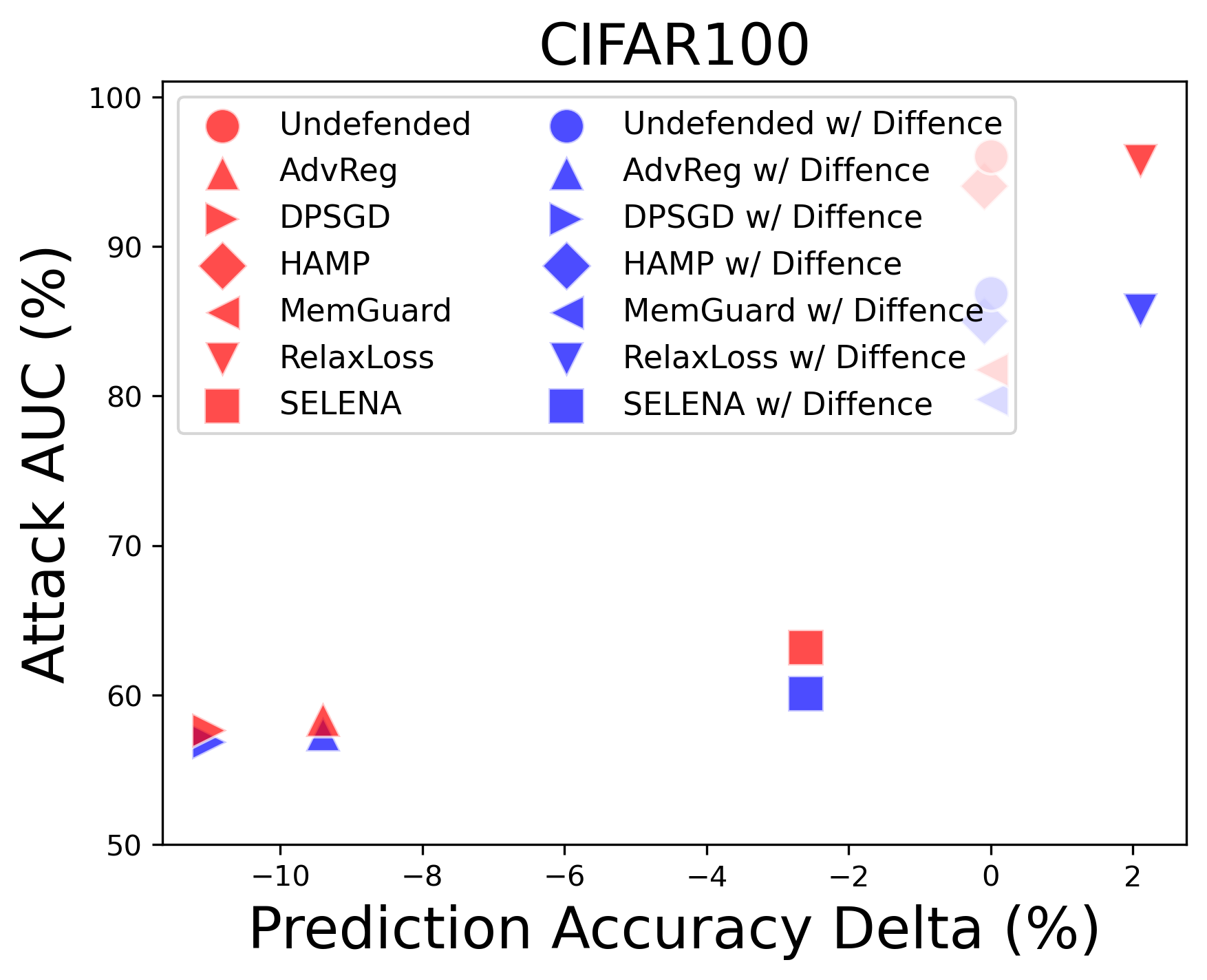}
    \hfill
    \includegraphics[width=0.3\columnwidth]{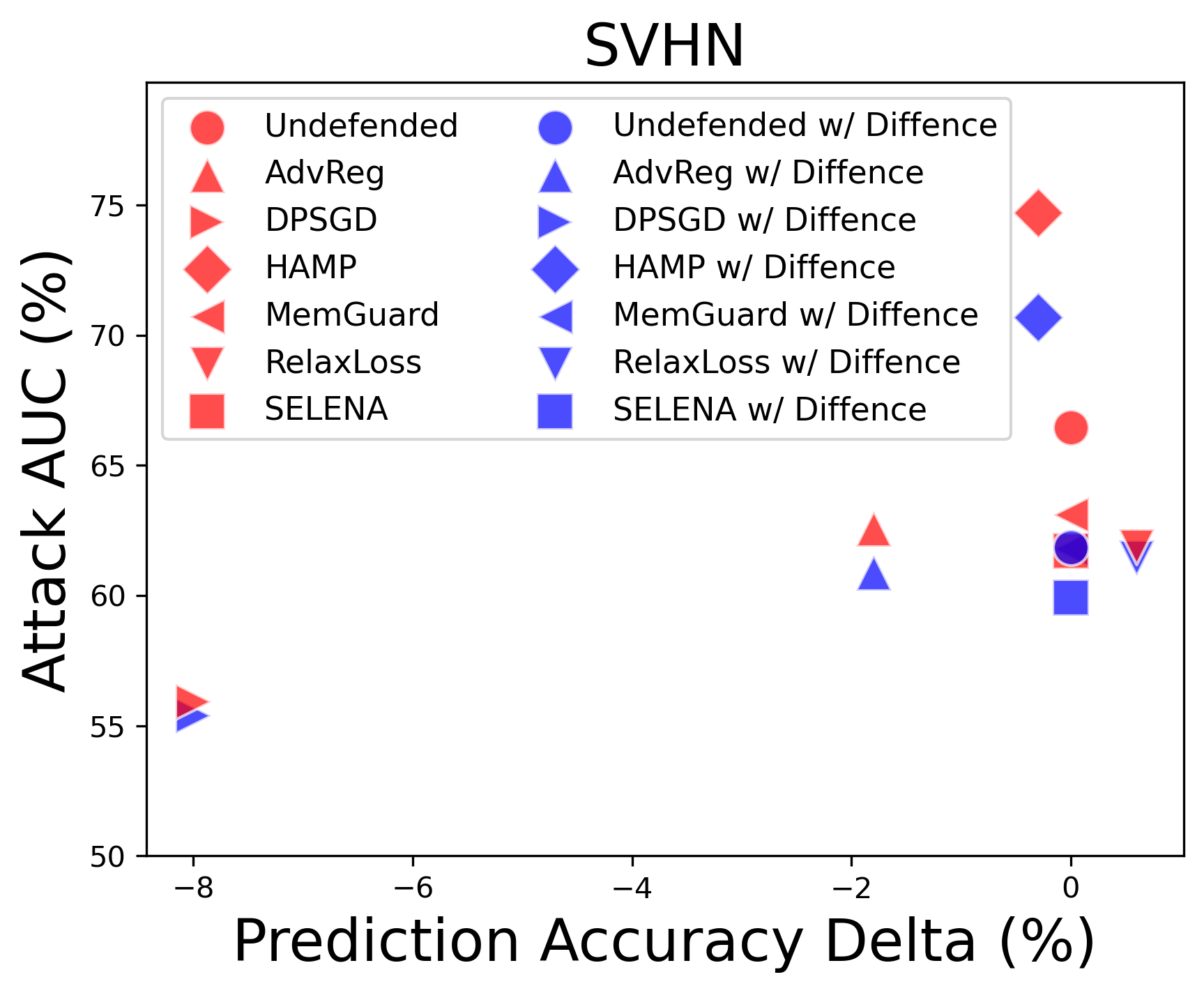}
 
    \label{fig:short-a}
    \caption{ Attack AUC}
  \end{subfigure}
  \caption{\textbf{Attack accuracy and AUC on three datasets against ResNet18, where \name employs a publicly released diffusion model trained on ImageNet.}}
  \label{fig:all-imagenet}
  % \hspace*{\fill}%
\end{figure*}

\begin{figure}[htbp]
\centering
\includegraphics[scale=0.4]{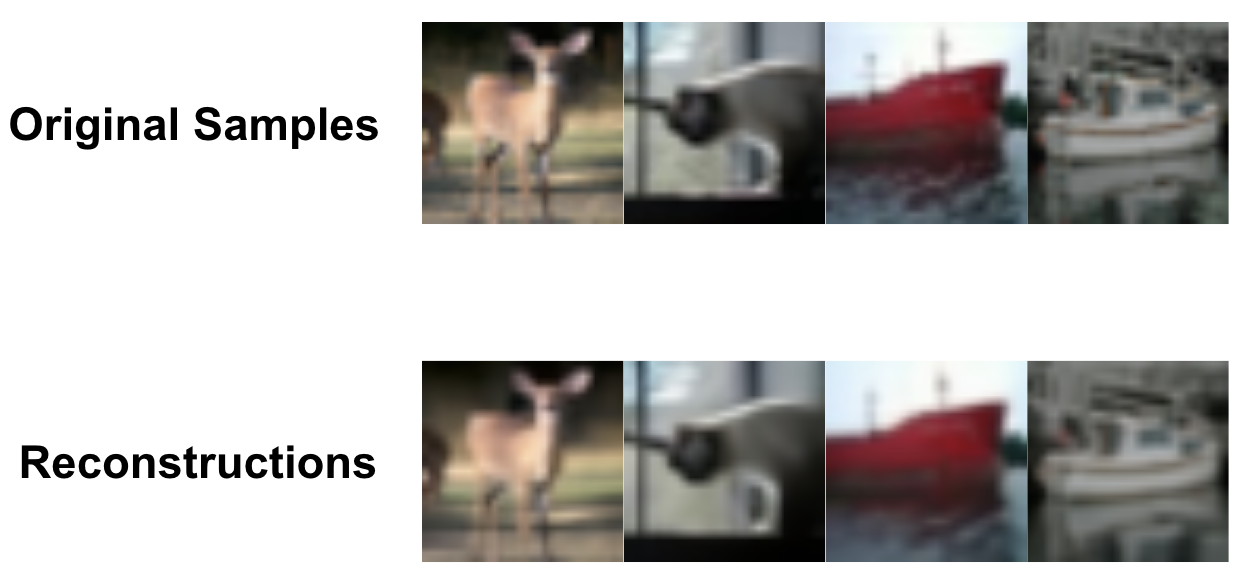}
\caption{\textbf{Examples of original samples and their reconstructions on CIFAR-10, with reconstructions generated by a diffusion model pretrained on ImageNet.} }
\label{Fig:examples-imagenet}
\end{figure}

\begin{table*}[h]
\centering
\large
\caption{\textbf{Comparison of Label-only MIAs against ResNet18 w/o and w/ \name.}}
\scalebox{0.75}{
\begin{tabular}{l m{3cm}<{\centering} m{3cm}<{\centering} m{3cm}<{\centering} m{3cm}<{\centering} m{3cm}<{\centering} m{3cm}<{\centering}}
\toprule
\multirow{2}{*}{\textbf{Defenses}} & \multicolumn{2}{c}{\textbf{Gap Attack}} & \multicolumn{4}{c}{\textbf{Boundary Attack}} \\
\cmidrule(lr){2-3} \cmidrule(lr){4-7}
 & \textbf{Accuracy w/o \name (\%)} & \textbf{Accuracy w/ \name (\%)} & \textbf{AUC w/o \name (\%)} & \textbf{Accuracy w/o \name (\%)} & \textbf{AUC w/ \name (\%)} & \textbf{Accuracy w/ \name (\%)} \\
\midrule
Undefended & 58.5 & 58.5 & 70.8 & 69.5 & 70.9 & 69.3 \\
SELENA & 52.9 & 52.9 & 59.7 & 58.9 & 59.7 & 59.4 \\
AdvReg & 57.6 & 57.6 & 61.7 & 59.7 & 60.9 & 59.4 \\
HAMP & 58.6 & 58.6 & 69.5 & 66.8 & 69.5 & 67.3 \\
RelaxLoss & 58.2 & 58.2 & 67.6 & 66.3 & 67.7 & 66.4 \\
DPSGD & 54.6 & 54.6 & 57.3 & 56.5 & 57.7 & 57.1 \\
MemGuard & 58.5 & 58.5 & 70.8 & 69.5 & 70.9 & 69.3 \\
\bottomrule
\end{tabular}
}
\label{table:label-only}
\end{table*}

\subsection{Impact of \name on label-only MIAs}
\label{sec:label-only}
In this section, we discuss the impact of \name on label-only MIAs~\cite{choquette2021label,li2021membership} and defenses. Both \name and label-only attacks rely on perturbations to the original images. Intuitively, \name modifies the input only after the target model $f$ has assigned a label to the noised inputs $x'$ created by the label-only attack. This process alters the output confidence score without changing the predicted label $f(x')$. Therefore, the target model with or without \name responds the same to label-only attack queries. Consequently, \name has no impact on label-only MIAs or defenses against them. 

We used two types of label-only MIAs: the baseline gap attack \cite{yeom2018privacy} and the state-of-the-art boundary attack \cite{li2021membership,choquette2021label} to test the impact of \name. As shown in Table~\ref{table:label-only}, as expected, \name has no effect on the performance of label-only attacks and defenses against them. In all settings, the results with and without \name were nearly identical.

\subsection{Defending with Stronger Diffusion Models}
\label{sec:agg}

\begin{table}[htbp]
\large
    \caption{\textbf{Performance of \name on CIFAR-10 when using a stronger diffusion model.}}
    \centering
    \scalebox{0.75}{
    \begin{tabular}{m{3cm}<{\centering}m{1.8cm}<{\centering}m{1.8cm}<{\centering}m{1.8cm}<{\centering}m{3cm}<{\centering}}
    \toprule
     \textbf{Defenses} & \textbf{Test Accuracy (\%)} & \textbf{Attack Accuracy (\%)} & \textbf{Attack AUC (\%)} \\
    \midrule
    Undefended / +\name      & 81.4 / \textbf{83.2}   & 73.0 / \textbf{66.2}    & 75.9 / \textbf{69.3} \\
    AdvReg / +\name      & 78.4 / \textbf{80.3}  & 60.3 / \textbf{57.5}   & 61.0 / \textbf{58.2}     \\
    DPSGD / +\name   & 74.7 / \textbf{75.7}  & 55.3 / \textbf{54.4} & 55.3 / \textbf{54.5} \\
    SELENA / +\name    & 79.2 / \textbf{79.8} & 60.4 / \textbf{57.2} & 60.1 / \textbf{57.0} \\
    RelaxLoss / +\name   &  81.2 / \textbf{82.1}    & 65.1 / \textbf{61.5}   & 69.5 / \textbf{63.7} \\
    HAMP / +\name    & 81.1 / \textbf{82.7}& 65.2 / \textbf{62.5} & 69.4 / \textbf{62.2} \\
    \bottomrule
    \end{tabular}}
    \label{table:both}
\end{table}

In our previous experiments, we assumed that the training set size of the defender's diffusion model and the size of the classifier's training set are the same. Furthermore, to ensure no decrease in model accuracy, we only select reconstructed samples that match the original sample labels.

Interestingly, however, we discovered that when a stronger diffusion model is employed in the defense—specifically, a model trained on a dataset larger than that of the classifier, directly averaging the predictions of all generated samples can enhance both the classifier's privacy and its accuracy.

We utilized a publicly available diffusion model trained on 50,000 samples from CIFAR-10 to defend our classifiers, which were trained on 25,000 samples. The results, as shown in the Table~\ref{table:both}, indicate that in all cases, the classifiers not only achieved enhanced privacy but also exhibited higher test accuracy. Given that diffusion model training does not require labeled data, it can be performed on a large, public, unlabeled dataset. We leave a more detailed discussion on how diffusion models can improve classifiers' accuracy for future work.

\subsection{Comparison of Defense Training Overhead}
\label{sec:training_overhead}

In addition to the inference overhead discussed in Section \ref{sec:infer_overhead}, we also report the overhead of training-phase defenses during the training stage. The results are shown in Table \ref{tabel:train_overhead}.

\begin{table}[htbp]
\large
\centering
\caption{\textbf{Training overhead comparison of different defenses.}}
\scalebox{0.75}{\begin{tabular}{lccc}
\toprule
\textbf{Defenses} & \textbf{CIFAR-10 (h)} & \textbf{CIFAR-100 (h)} & \textbf{SVHN (h)} \\
\midrule
Undefended & 0.6 & 0.6 & 0.4 \\
SELENA & 10.9 & 11.9 & 9.3 \\
AdvReg & 10.9 & 12.6 & 6.8 \\
HAMP & 0.8 & 0.8 & 0.5 \\
RelaxLoss & 0.5 & 0.6 & 0.4 \\
DP-SGD & 0.8 & 0.9 & 0.6 \\
\bottomrule
\end{tabular}}
\label{tabel:train_overhead}
\end{table}

\subsection{Additional Experimental Details}
\label{sec:exp_details}

For all attacks, we randomly selected 2000 members and 2000 non-members as target samples to perform MIAs. For NN-based attacks, we used the same attack model architecture and training configuration as in previous work~\cite{chen2023overconfidence}, saving training checkpoints and using the attack model that achieved the highest attack AUC on the target samples.

For defenses, HAMP~\cite{chen2023overconfidence} uses an entropy threshold $\gamma$ and a regularizer parameter $\alpha$, while RelaxLoss~\cite{chen2022relaxloss} uses $\alpha$ to balance utility and privacy. However, automated parameter selection is not provided. Following their papers, we used grid search to train multiple models with different parameters and selected the models based on their claimed impact on utility. HAMP includes both training-phase regularization and testing-time output modification (retaining predicted labels while randomizing confidence scores). In our experiments, we included only the training-phase defense of HAMP because its testing-time output modification renders the confidence scores meaningless, as the generated confidence scores are entirely derived from randomly generated images. Adversarial regularization~\cite{nasr2018machine} uses the $\alpha$ parameter to balance model accuracy and privacy protection. We set $\alpha$ to 6 for CIFAR-10, CIFAR-100, and SVHN, and 4 for CelebA and UTKFace. For DP-SGD \cite{abadi2016deep}, we used PyTorch Opacus~\footnote{\url{https://github.com/pytorch/opacus}} to train the DP-SGD model, fixing a norm clipping bound of 1.2 and setting a noise multiplier of 0.1 for CIFAR-10, 0.05 for CIFAR-100 and SVHN, and 0.01 for CelebA and UTKFace.
For SELENA~\cite{tang2022mitigating}, we followed the original paper' settings with $K=25$ and $L=10$, where $K$ is the total number of teacher models and $L$ is the number of teacher models whose training sets do not contain a given member sample.

\end{document}